\documentclass[12pt]{emulateapj}
\usepackage{epstopdf}
\usepackage{graphicx,rotating}
\usepackage{longtable}
\usepackage{amssymb}
\usepackage{amsmath}
\usepackage{mathtools}
\usepackage{natbib}
\usepackage{xcolor}
\usepackage{latexsym}
\newcommand{\beq}{\begin{equation}}
\newcommand{\eeq}{\end{equation}}
\newcommand{\etal}{{\sl et~al.~}}
\newcommand{\kms}{km s$^{-1}$}
\newcommand{\kmse}{km s$^{-1}$~}

\newcommand{\FGS}{{\it FGS~}}
\newcommand{\FGSns}{{\it FGS}}
\newcommand{\HST}{{\it HST~}}

\newcommand{\G}{{\sl Gaia}}
\newcommand{\Gs}{{\sl Gaia~}}
\newcommand{\vA}{{vA\,351}}
\newcommand{\vAs}{{vA\,351~}}

\newcommand{\vAAs}{{vA\,351A~}}

\newcommand{\vABs}{{vA\,351B~}}
\newcommand{\vAC}{{vA\,351C}}
\newcommand{\vACs}{{vA\,351C~}}

\newcommand{\m}{$\cal{M}$}
\newcommand{\mA}{$\cal{M}_{\rm A}$}
\newcommand{\mAD}{$\cal{M}_{\rm AD}$}
\newcommand{\mB}{$\cal{M}_{\rm B}$}
\newcommand{\mBC}{$\cal{M}_{\rm BC}$}
\newcommand{\mC}{$\cal{M}_{\rm C}$}
\newcommand{\mD}{$\cal{M}_{\rm D}$}

\newcommand{\msun}{$\cal{M}_{\odot}~$}
\newcommand{\msune}{$\cal{M}_{\odot}$}
\newcommand{\He}{He\,\text{\small{I}}}
\newcommand{\Hes}{He\,\text{\small{I}}~}

\def\fdg{\hbox{$.\!\!^\circ$}}
\def\arcmin{\hbox{$^\prime$}}

\def\Ha{H$\alpha$}
\def\Has{H$\alpha$~}

\begin{document}

\received{}
\revised{}
\accepted{}

\shorttitle{vA 351}
\shortauthors{Benedict \etal}

\bibliographystyle{/Active/my2}

\title{
Dissecting the Quadruple Binary Hyad \vAs -  Masses for Three M Dwarfs and a White Dwarf} \footnote{We dedicate this paper to John Stauffer, who died on 2021 January 29, in honor of his many contributions to the field.}

\email{fritz@astro.as.utexas.edu}

\author{G. Fritz Benedict}
\affiliation{McDonald Observatory, University of Texas, Austin, TX 78712}
  \author{Otto G. Franz}
  \affiliation{Lowell Observatory, 1400 West Mars Hill Rd., Flagstaff, AZ 86001}
  \author{Elliott P. Horch}
  \affiliation{Southern Connecticut State University, New Haven, CT 06515 }
  \author{L. Prato}
  \affiliation{Lowell Observatory, 1400 West Mars Hill Rd., Flagstaff, AZ 86001} 
  \author{Guillermo Torres}
  \affiliation{Center for Astrophysics $\vert$ Harvard \& Smithsonian, Cambridge MA 02138}
    \author{Barbara E.  McArthur}
  \affiliation{McDonald Observatory, University of Texas, Austin, TX 78712}
  \author{Lawrence H. Wasserman}
  \affiliation{Lowell Observatory, 1400 West Mars Hill Rd., Flagstaff, AZ 86001}
 \author{David W. Latham}
  \affiliation{Center for Astrophysics $\vert$ Harvard \& Smithsonian, Cambridge MA 02138}
  \author{Robert P. Stefanik}
  \affiliation{Center for Astrophysics $\vert$ Harvard \& Smithsonian, Cambridge MA 02138}
   \author{Christian Latham}
  \affiliation{Center for Astrophysics $\vert$ Harvard \& Smithsonian, Cambridge MA 02138}
  \author{Brian A. Skiff}
  \affiliation{Lowell Observatory, 1400 West Mars Hill Rd., Flagstaff, AZ 86001}



\begin{abstract}

We extend results first announced by Franz et al. (1998),
that identified \vAs = H346 in the Hyades as a multiple star system containing a white dwarf. 
With Hubble Space Telescope Fine Guidance Sensor fringe tracking and scanning, and more recent speckle observations, all spanning 20.7 years, 
we establish a parallax, relative orbit, 
and mass fraction for two components, with a period, $P=2.70$y and total mass 2.1\msune. With ground-based radial velocities from the 
McDonald Observatory Otto Struve 2.1m telescope Sandiford Spectrograph, and Center for Astrophysics Digital Speedometers, 
spanning 37 years, we find that 
component B consists of BC, two M dwarf stars orbiting with a very short period ($P_{\rm BC}=0.749$ days), 
having a mass ratio \mC/\mB=0.95. We confirm that the total mass of the system can only be reconciled with 
the distance and component photometry by including a fainter, higher mass component.
The quadruple system consists of three M dwarfs (A,B,C) and one white dwarf (D). 
We determine individual M dwarf masses \mA=0.53$\pm0.10$\msune, \mB=0.43$\pm0.04$\msune, and \mC=0.41$\pm0.04$\msune. 
The WD mass, 0.54$\pm0.04$\msune, comes from cooling models,  
an assumed Hyades age of 670My, and consistency with all previous and derived astrometric, photometric, and RV results. 
Velocities from \Has and \Hes emission lines confirm the BC 
 period derived from absorption lines, with similar (\He) and higher (\Ha) velocity amplitudes. We ascribe the larger
 \Has amplitude to  emission from a region each component shadows from the other,  depending on the line of sight. 

\end{abstract}




%

\section{Introduction}

The Hyades, an open cluster
whose every member contributes to our knowledge of stellar evolution, open cluster formation, and open cluster evolution, remains an important rung on the extragalactic distance scale ladder \citep{An07b,deB01}.
Much effort has resulted in an ever improving Hyades distance, from convergent point methods \citep{Bue52, Han75} using only proper motions,
to directly measured parallax \citep{Per98, McA11}, through recent efforts using a combination of parallax,  proper motion, and radial velocities \citep{Lee17}, with cluster membership determination always a critical issue. Hence the first goal of this project is an unambiguous determination of cluster membership for \vA=H346 = L42 = V805 Tau = HG7-203 = LP415-65, RA $04^{\rm h} 25^{\rm m} 13.54^{\rm s}$ DEC $+17\arcdeg 16' 05\farcs48$ (2000), V=13.27, K=8.27.

This Hyad was first angularly resolved \citep{Fra94} as a result of a blind search of faint   members of the Hyades,  using the {\it Hubble Space Telescope} (\HST) Fine Guidance Sensor \#3 (\FGSns\,3) in the transfer function or fringe scanning (TRANS) mode. \cite{Fra98} details this \FGS observation mode. Further observations  yielded a relative orbit, and, depending on an assumed parallax based on Hyades membership, a total mass for the system \citep{Fra98b} inconsistent with the colors and spectral types suggesting M dwarf components. A white dwarf component was posited to bring the total mass into agreement with the dynamical mass. Subsequently, we acquired \FGSns\,3 POS mode \nocite{Ben16} observations (the fringe tracking POS mode described in Benedict et al. 2016) with which to derive a mass fraction and independent parallax. Once \vAs was identified as a binary we added it to our then ongoing Mass-Luminosity Relation radial velocity program, also described in \cite{Ben16}. Thus our second goal: to establish the physical properties of all \vAs components, determining how many stars comprise this system, and derive their individual component masses.

Observational astronomers are hard-pressed to devise experiments. Nature provides these by presenting us with binary stars. Put one star next to another and see what happens. Extreme cases provide laboratories within which to test our knowledge of stellar magnetic fields
and mass transfer.
\vAs provides such a laboratory. Hence,  a detailed examination of the complex behavior of a subset of \vAs components provides a third goal,
that of generating a qualitative model explaining the observations. 

These observations include \FGS TRANS which use the entire fringe to
match a superposition of the fringe of each component to the fringe of a known single star, resulting in position angles and separations with which to derive a relative orbit. They also provide component 
magnitude differences. \FGS POS mode produces positions 
of the zero crossing of a fringe for each star observed. Measuring a series of \vAs  positions relative to a known reference frame provides a parallax, establishes a proper motion for the center of mass of the system, and yields a mass ratio for the supposed two components. For many of the M dwarf binary systems investigated in \cite{Ben16},
adding radial velocities (RVs) to the astrometry allowed us to
completely sample the motion of a binary system. The three
complementary data sets improve the accuracies of the final component masses.   For \vAs all assist in dissecting the system, yielding masses for the four components.

This paper summarizes our efforts to understand the \vAs system. We first present the results of our analysis of the TRANS-mode astrometric data, yielding a binary orbit and total mass (Section~\ref{AstRel}).  Next we present a radial velocity (RV)  analysis using  absorption lines, yielding  a \mC/\mB~ mass ratio (Section~\ref{spec}) and a confirmation of the component A RVs with respect to the AD-BC barycenter.  Finally we fold in the POS-mode data and derive a mass fraction, a proper motion, and a parallax for the system (Section~\ref{POSast}).  Section~\ref{allphot} presents evidence of small amplitude but precisely measured variability from photometry acquired with the POS mode observations, and recent ground-based $V-$band photometry. In Section~\ref{BS} we use to  these results to  confirm masses for the three M dwarf  components and to derive a WD mass  required to satisfy the derived total mass and the (\mA+\mD)/(\mB+\mC) mass ratio. Half our spectra include emission lines of both \Has and \Hes ($\lambda 5876.6\AA$).  We derive equivalent widths and  RVs for both \Has and \Hes emission lines (Section~\ref{specem}).  
We describe the structure of this quadruple system and posit some possible emission line formation mechanisms (Section~\ref{SciFi}), offer conclusions (Section~\ref{summ}), and summarize our results in Section~\ref{sum}.

Unless otherwise noted, times are modified Julian Date, mJD = JD-2400000.5. We abbreviate millisecond of arc as mas throughout.

\section{The \vA~ Relative Orbit}  \label{AstRel}

We  have  13 position angle and separation
TRANS mode measurements of component B relative to component A acquired over 4.7 yr. 
Five  speckle measures, taken with the DSSI speckle camera on the 4.3m Lowell Observatory Discovery Telescope (LDT; formerly DCT), extend the relative orbit astrometric coverage to 26.0 yr. The LDT speckle program \citep{Hor12,Hor15}, in operation since 2014,  obtained data for vA 351 on four separate nights in October and November of that year. A more recent observation, under very poor weather conditions, was secured in early February 2020. Analysis of these data files proceeded using the same methodology as described most recently in \cite{Hor17}, using a weighted least-squares fit to the summed spatial frequency power spectrum of the speckle data frames. 

We transformed all position angle and separation observations into x and y positions and assigned a 1.5 mas error to each x, y, this to yield a reduced $\chi^2$ near unity for the relative orbit modeling. Figure~\ref{Rel} contains the resulting low-inclination, high eccentricity relative orbit with  observations and residuals presented in Table~\ref{tbl-TR} and orbital parameters listed in Table~\ref{tbl-TSORB}. The second largest residual in Figure~\ref{Rel} (\#18) comes from the February 2020 speckle measurement, secured under very poor weather conditions. We excluded this and one early TRANS observation (\#2, components unresolved along one \FGS axis) in the final orbit determination. 

The relative orbit appears quite plausible. However, see the Appendix  for an arduous and tortuous path traversed, and for a  blind alley  out of which we backed, to eventually arrive at Figure~\ref{Rel}.

\section{Absorption-line Spectroscopy} \label{spec}

\subsection{McDonald Cassegrain Echelle Spectrograph}
We obtained  RV data with the McDonald 2.1m Otto Struve telescope and
Sandiford Cassegrain Echelle spectrograph \citep{McC93}, hereafter CE.
The CE delivers a dispersion equivalent to 2.5 \kms/pix ($R =
\lambda/\Delta\lambda = 60,000$) with a wavelength range of 5500
$\leq$ $\lambda$ $\leq$ 6700 \AA~spread across 26 orders.
The McDonald data were collected during thirty-three observing runs
from 1995 to 2009 and reduced using the standard IRAF \citep{Tod93}
{\tt echelle} package tools. CE RVs (listed in Table~\ref{tbl-CE}) were derived using the IRAF cross-correlation tool {\tt
  fxcorr}.  
  
  \subsection{Center for Astrophysics Digital Speedometers}
vA 351 was monitored spectroscopically at the CfA with the Digital Speedometers (Latham 1992) between January of 1982 and October of 1999, on two telescopes: the 1.5m Tillinghast reflector at the Fred L.\ Whipple Observatory on Mount Hopkins (AZ), and the 4.5m-equivalent Multiple Mirror Telescope (also on Mount Hopkins) before its conversion to a monolithic 6.5m telescope. With a resolving power, $R \approx 35,000$,  the spectra consist of a single echelle order centered on the \ion{Mg}{1}~b triplet (5187~\AA) and spanning 45~\AA. Reductions were carried out with a dedicated pipeline. The average signal-to-noise ratio for the 53 observations is about 12 per resolution element of 8.5~km~s$^{-1}$. We identify this source as CfA, hereafter, and provide the CfA RVs in Table~\ref{tbl-CfA}.

\subsection{IGRINS and other Sources of \vAs RVs}
On UT 2016 February 23 and 25 we obtained two epochs with the R$=$45,000, near-infrared spectrograph IGRINS \citep{Mac16} on the McDonald Observatory Harlan J. Smith 2.7m telescope. An additional observation was taken on UT 2019 April 11 at the 4.3m LDT.  Exposure times were 16--24 minutes, divided into 2 or 4 nodded pairs at two distinct slit positions. The resulting signal to noise was $>$100 per resolution element of $\sim$6 km s$^{-1}$. Data reduction was carried out using standard processing with the IGRINS pipeline \citep{LGK17}\footnote{https://github.com/igrins/plp/tree/v2.1-alpha.3}. The pipeline yields telluric-corrected, one-dimensional spectra with wavelength dispersion solutions determined from OH night sky emission lines and atmospheric absorption lines.  Barycentric velocity corrections were applied to the target and RV standard star data as part of the cross-correlation analysis (Section~\ref{WTF?}).

Hartmann et al. (1987) \nocite{Har87} report a single RV of (presumably) component A with the MMT.  \cite{Sta97} present two RV measurements of vA 351 components A, B, and C obtained with the NOAO 4m Mayall telescope at two epochs separated by $\sim$50 days.  We list dates, RVs, and phases for these three epochs from the literature and the three IGRINS epochs  in Table~\ref{tbl-oRV}.

\subsection{Absorption Line  Analysis} \label{WTF?}
For the CE observations we used the same high-resolution, high signal-to-noise template spectrum of GJ623\,AB 
as in \cite{Ben16}  for all absorption line cross-correlations. The quality of the GJ623\,AB  orbit yields an  absolute RV uncertainty at the epoch of template observation of less than 0.025 \kms \citep[][table 4]{Ben16}. GJ623\,AB has a $\Delta V=5.2$, rendering the effect of the secondary on the cross-correlation
function (CCF) negligible. We visually inspected 
the resulting CCF to select the better of the 26 apertures, typically
using half to obtain an average velocity.  The CCF resulting from the absorption line template had
from one to three peaks (Table~\ref{tbl-CE}), indicating the existence of at least three, not just 
two components (hereafter, components A, B, and C).  We identify component A with the strongest peak in the CCF,
and list all CE component RVs with internal $1-\sigma$ errors
in Table~\ref{tbl-CE}. For components A, B, and C we find average internal per epoch errors; 
0.3, 0.6, and 0.8 \kms. 

Radial velocities from the CfA and IGRINS spectra were measured with algorithms for three-dimensional cross-correlation such as
 TRICOR  \nocite{Zuc95} (Zucker et al. 1995). For the IGRINS data, we used the same template for the three vA 351 components (A, B, and C), a high S/N exposure of the slowly rotating M2 star GJ 725A (Prato 2007; Mann et al. 2015)\nocite{Man15, Pra07}. For the CfA data,
  rotational broadening was applied to achieve the best overall match, resulting in $v \sin i$ estimates of 11$\pm$2 km s$^{-1}$ for stars B and C, and 3$\pm$3 km s$^{-1}$ for the brighter star A.  For the IGRINS observations, the cross-correlation coefficient was optimized
   when we used the GJ 752A template for the A component, with no broadening applied, and the M2.5 dwarf standard GJ 436 for the B and C components with $v \sin i$ of 12$\pm$2 km s$^{-1}$ (Prato 2007; Mann et al. 2015). For both the optical and IR data, these
    $v \sin i$ values  agree with the independent determinations of Stauffer et al. (1997). \nocite{Sta97}


From CfA measurements the light ratio of each of the fainter components relative to star A was found to be $0.53 \pm 0.02$ at the mean wavelength of these observations. This yields a flux ratio BC/AD = 1.06 $\pm$ 0.03, or $\Delta$m = 0.06 $\pm$ 0.03, with BC being marginally brighter at this wavelength.  The bandpass is somewhat close to V though on the blue side and much narrower (45 \AA). It is similar to the D51 filter used for the Kepler Input Catalog \citep{Bro11}, which was modeled after the Dunlap Observatory DDO51 filter. The CfA $\Delta$m = 0.06 $\pm$ 0.03 is consistent with the measurements listed in Table~\ref{tbl-TR}.  The IGRINS $H-$band spectra yielded light ratios of B/A$=$0.66$\pm$0.02 and C/A$=$0.64$\pm$0.08, hence a BC/AD flux ratio of 1.3$\pm$0.1 and $\Delta H=0.28$ (combined components AD fainter than combined components BC), a puzzle given the $V-$ and $K-$band lower $\Delta m$ values (Table~\ref{tbl-TR}).

The two fainter components, B and C, exhibit the largest RV
variation. When fit to an orbit these yield the elements listed in Table~\ref{tbl-RVO}.
We plot component velocities phased to a BC period, $P_{\rm BC}=0.7492425^{\rm d}$
in Figure~\ref{RVabs}. The rms RV residual to these orbits is 2.7 \kms. A ratio of the RV amplitudes yields a mass ratio for components C and B; $K_B / K_C =$ \m$_{\rm C}/$\m$_{\rm B}=0.945\pm0.005$. The average
component A RV is $\langle \rm{V}_{\rm A}\rangle=+40.7\pm2.0$ \kms. An average of the BC RV $\gamma$ values yields 
$\langle \rm{V}_{\rm BC}\rangle=+40.2\pm0.2$ \kms. 
We have consistency with the expected radial velocity of a Hyades member at the location of \vA, V=+39.80 \kms, computed from a convergent point solution with the location of the convergent point as reported by the \cite{Lee17} ($\alpha_{\rm J2000} = 97\fdg73$, $\delta_{\rm J2000} = +6\fdg83$), adopting a space velocity for the cluster of V$_0$ = +47.13 \kms, based on ongoing radial velocity measurements at the CfA. 


In Figure~\ref{fig-WTF?} we plot the RV variation
attributable to the motion of A  about a common center of mass with BC, phased to the period 
(Table~\ref{tbl-RVORB}) determined through a  solution incorporating only RVs. While noisy, the component A velocities show a clear signature of a highly eccentric orbit. We were unable to extract a clear signature of the BC center of mass motions around the AD-BC system barycenter.



\section{POS mode Astrometry}\label{POSast}
\cite{Ben16}  describes the acquisition and reduction of \FGS POS  mode data. Figure \ref{fxy} shows the distribution on the sky of
the four reference stars (numbers 2, 3, 4, 5) for the POS
mode measurements, with number  1 indicating
\vA.    Not all reference stars were
measured at each of the seven epochs of observation that included POS mode.
We acquired all \HST astrometry data sets with \FGSns\,3.  For POS mode we obtained a total of 50 reference star
observations and 19 observations of \vA, secured over 1.8 years.

\subsection{Prior Knowledge and Modeling Constraints} \label{MODCON}
As in our previous astrometry projects, e.g., \cite{Ben01,Ben07,
  Ben11, McA11, Ben16} we include as much prior information as possible in
our \FGS fringe-tracking modeling.  
In contrast to much of our previous relative astrometry
(e.g., Harrison \etal 1999, Benedict \etal 2011)\nocite{Har99,Ben11}, for this project we
obtain  reference star absolute parallaxes and proper motion priors from \Gs DR2 \citep{Bro18}.
A comparison of the Figure~\ref{CCD} color-color diagram with the Table~\ref{tbl-pis} input parallaxes confirms the giant nature of
reference stars 4 and 5, labeled ref-4 and ref-5 in the figure.

We did not simply adopt the \Gs DR2 parallax for \vAs
because the \Gs astrometry does  not yet account for over 80 mas \citep{Fra98b} of orbital motion. Similarly, rather than using the highly precise  \Gs DR2 
proper motion from measures taken over a limited multiple of the AD - BC orbit, we introduce no \vAs proper motion priors.

In a quasi-Bayesian approach we input all priors as observations with
associated errors, not as hardwired quantities known to infinite
precision.  
The lateral color 
calibration (see section
3.4 of Benedict \etal 1999)\nocite{Ben99} and the $B-V$ color indices are also treated as
observations with error.  
i





\subsection{\vAs POS Mode Photocenter Corrections} \label{PCC}
\FGS POS mode measures the zero-crossing of a fringe. 
Thus, for a binary the observed fringe becomes a linear superposition of two fringes.
For a nearly equal brightness component binary (Table~\ref{tbl-TR}) the measured zero crossing can 
be significantly perturbed from a position coincident with either component.
\cite{Ben01}, section 4.2, discusses the generation of such corrections at some length.
Figure~\ref{pcor} demonstrates the determination of corrections along the  x and y axes of \FGSns\,3
for a typical POS mode observation of \vA. Corrections ranged from 4 mas to 80 mas, the latter required
because for that observation the \FGS locked on the other component.

\subsection{The POS Model}
 
From the reference star astrometric data we determine the rotation and offset
``plate constants" relative to an arbitrarily adopted constraint epoch
(the so-called ``master plate") for each observation set. The
\vAs reference frame contains five stars, but only three were observed
at each epoch. Hence, we constrain the scales along $x$ and $y$ to
that provided by the \FGS Optical Field Angle Distortion calibration \citep{Ben99}, and the two axes to orthogonality. 
The consequences of this
choice are minimal. For example, imposing these constraints on the
Barnard's Star astrometry discussed in \cite{Ben99} results in an
unchanged parallax and increases the error by 0.1 mas, compared to a
full 6 parameter model  \citep[e.g.][equations 6,7]{Ben17}.

Our reference frame model becomes, in terms of standard coordinates,
\beq
        x'  =  x + lc_x(\it B-V) 
\eeq
\beq
        y'  =  y + lc_y(\it B-V) 
\eeq
\beq
x''=x'+corrx
\eeq
\beq
y''=y'+corry
\eeq
\beq
\xi = {\rm cos}(A)*x'' +{\rm sin}(A)*y'' + C  \\ - \mu_\alpha \Delta t  - P_\alpha\varpi - ORB_\alpha
\eeq 
\beq
\eta = -{\rm sin}(A)*x'' + {\rm cos}(A)*y'' + F  \\ - \mu_\delta \Delta t  - P_\delta\varpi - ORB_\delta
\eeq

\noindent 
, where $\it x$ and $\it y$ are the measured coordinates from {\it HST}, $corrx$ and $corry$ are
the photocenter corrections derived in Section~\ref{PCC} above,
$\it lc_x$ and $\it lc_y$ are lateral color corrections , and $B-V$ represents the
color of each star, either from SIMBAD or estimated from the
spectral types suggested by Figure~\ref{CCD}.  $A$  is a field rotation in radians, 
$C$ and $F$ are offsets, $\mu_\alpha$ and
$\mu_\delta$ are proper motions, $\Delta$$t$ is the epoch difference from
the mean epoch, $P_\alpha$ and $P_\delta$ are parallax factors, and
$\it \varpi$  is the parallax.  
$\xi$ and $\eta$ are 
relative positions that (once rotation, parallax, the proper motions 
 are determined) 
should not change with time. All Equation 
5 and 6 subscripts are in RA and DEC because the master constraint plate was rolled
into the RA DEC coordinate system before the analysis. We obtain
the parallax factors from a JPL Earth orbit predictor (Standish
1990\nocite{Sta90}), upgraded to version DE405.  We obtain orientation to the
sky for the master plate using ground-based astrometry from
the PPMXL \citep{Roe10} with uncertainties in the field orientation of
$\pm 0\fdg1$.  This orientation also enters the modeling as an
observation with error.    

We derive $A$, $C$, and $F$ for each epoch only from reference star measurements.
Applied as constants, we solve  for a  position
within our reference frame, a parallax, a proper motion, and a major axis measurement, $\alpha_{\rm A}$, for the \vAAs orbit about the system center of mass.   To \vAs we apply $ORB$  as a
correction term that is a function of the traditional astrometric and RV 
orbital elements listed in Tables~\ref{tbl-TSORB} and \ref{tbl-RVORB}.
These include the
orbital period (P), the epoch of passage through periastron in
modified Julian days (T$_0$), the eccentricity ($e$), the inclination, $i$, and the position
angle of the line of nodes ($\Omega$). 
 We also constrain the angle
($\omega$) in the plane of the true orbit between the line of nodes
and the major axis to differ by 180\arcdeg~for the component A and B
orbits.


\subsection{Assessing Reference Frame Residuals}
 From histograms of the astrometric residuals for 50 reference star and
19 \vAs position measurements (Figure~\ref{his}), we conclude that we
have obtained corrections at the $1.2$ mas level in the region
available at all {\it HST} roll angles (an inscribed circle centered
on the pickle-shaped \FGS field of regard).  The resulting reference
frame ``catalog'' in $\xi$ and $\eta$ standard coordinates (Equations
5 and 6) was determined with median absolute errors of $\sigma_\xi=
0.9$ and $\sigma_\eta = 1.0$ mas in $x$ and $y$, respectively. 


\subsection{Results of \vAs  Modeling}  


In Table~\ref{tbl-SUM} we list our final values for the absolute parallax and proper motion  (with $1\sigma$ errors), where
the \Gs parallaxes from both DR2 and EDR3 \citep{Lin20a} are also included for comparison.  We included neither parallax as a prior. Extending the \Gs observational time baseline has brought their parallax closer to our value, now only slightly more than $1\sigma$ different. \cite{Sta21} find that the \Gs $RUWE$ (renormalised unit weight error) robustly predicts unmodeled photocenter motion, even in the nominal "good" range of 1.0--1.4 \citep[see also][]{Bel20}. For \vAs EDR3 lists a value very much larger, $RUWE$=9.33, clearly indicating the {\it Gaia} astrometry was affected. 
The median $RUWE$ value for the 9 stars (with a Gaia $G-$band magnitude $G<16$, not used as astrometric reference stars) in EDR3 within $7\arcmin$ of \vAs is 1.05. Reference stars (Table~\ref{tbl-pis}) have a median $RUWE$ value  of 1.38, a partial explanation for our poorer than average \citep[see the 105 parallax results in][]{Ben17} parallax result for \vA.

Table~\ref{tbl-ORB}
contains the final orbital parameters with formal ($1 \sigma$)
uncertainties.
Figure~\ref{vAorb} illustrates component A and B astrometric
orbits.  
 That the FGS POS measures did not sample the entire 
  orbit and  that the \Gs solution thus far neglects orbital motion could provide a partial explanation for the  proper motion discrepancies seen in Table~\ref{tbl-SUM}. 
 The TRANS and speckle astrometric residuals (Table~\ref{tbl-TR}, RMS=2.8 mas) suggest an unmodeled perturbation to the TRANS measured separation 
  between components A and B.  We discuss further  evidence for this
  perturbation in Section~\ref{BS}. 
  
  In the past \citep[e.g.][]{Ben16} we have employed a relationship between astrometry and RV \citep[c.f.][]{Pou00}
 to narrow the range of possible combinations of parallax, $\varpi$, and orbit semi-major axis, $\alpha$:
\beq
\displaystyle{{\alpha_{\rm A}\,{\rm sin}\,i \over \varpi_{\rm abs}} = {P K_{\rm A} \sqrt{(1 -\epsilon^2)}\over2\pi\times4.7405}} 
\label{PJeq}
\eeq
However, the significant inclination error (Table~\ref{tbl-TSORB}) renders this constraint less useful. The right hand (RHS) and left hand (LHS) quantities calculated from results listed in Tables~\ref{tbl-TSORB}, \ref{tbl-RVORB}, and \ref{tbl-SUM} do agree within their errors
(RHS$=0.267\pm0.154$, LHS$=0.220\pm0.004$).

\subsection{\vAs Component Masses} \label{MT}
Our orbit solution and derived absolute parallax provide an orbital
semimajor axis, $a$ in AU, from which we can determine the system mass
through Kepler's Third Law.  
We find (in solar units) ${\cal M}_{\rm tot} = 2.06 \pm 0.24{\cal M}_{\sun}$.
At each instant in the orbits of the two components around the common
center of mass,

\beq
{\cal M}_{\rm A} / {\cal M}_{\rm B} = \alpha_{\rm B} / \alpha_{\rm A}
\eeq

\noindent a relationship that contains only one observable,
$\alpha_{\rm A}$, the \vAs semi-major axis measurement. Instead, we calculate
the mass fraction

\beq
f = {\cal M}_{\rm B}/({\cal M}_{\rm A}+{\cal M}_{\rm B}) = \alpha_{\rm A} / (\alpha_{\rm A} + \alpha_{\rm B}) = \alpha_{\rm A} /a, \label{frac}
\eeq

\noindent where $\alpha_{\rm B}$ = $a-\alpha_{\rm A}.$ This parameter,
$f$, also given in Table~\ref{tbl-ORB}, creates a ratio of the two quantities
directly obtained from the observations: the \vAAs semi-major axis
 ($\alpha_{\rm A}$ from POS mode, Figure~\ref{vAorb}) and the relative orbit size ($a$
from TRANS mode, Figure~\ref{Rel}),  listed in
Table~\ref{tbl-ORB}. From these we derive a mass fraction of 0.443
$\pm$ 0.022.  Equations 5, 6, and 7 yield ${\cal M}_{\rm A} = 1.14
\pm 0.14{\cal M}_{\sun}$ and ${\cal M}_{\rm B} = 0.91 \pm 0.11
    {\cal M}_{\sun}$. 
    
    The mass errors of $\sim$12\% exceed any of our past  \citep{Ben16,Ben01} determinations.  We ascribe this to a combination of orbit low inclination and high eccentricity, increasing the errors on relative orbit size and perturbation size, and primarily to the parallax error on what is the smallest parallax yet measured for a binary by the FGS. 
The median parallax in our previous MLR work was 113 mas, 6 times larger than 
that measured for \vA. The median percent error in our MLR study was 0.4\%, significantly better 
than the vA 351 3.5\% parallax error. 

Nonetheless, \vABs with \mB $= 0.91{\cal M}_{\sun}$ is consistent with a G5V star, and \vAAs  with 
\mA $= 1.14{\cal M}_{\sun}$  an F8V star \citep{Cox00}. This is not consistent with each component being a single low mass red
    dwarf as suggested by the Figure~\ref{CCD} color-color diagram, which  cleanly and unambiguously identifies \vAs as such. To make up for the excess mass we now identify component A to have a companion, component D. Component B also must have a companion, component C. We now assert that \mA+\mD~= 1.14\msun and \mB+\mC~= 0.91\msune. 

\section{Photometry} \label{allphot}

\subsection{\FGS POS Mode Photometry} \label{Fphot}
\FGSns\,3 has demonstrated relative photometric precision approaching the milli-magnitude level for two other M dwarf stars, Proxima Cen and Barnard's Star \citep{Ben98a}.  The \FGS acquires data at a 40Hz rate with each observation lasting from 80 to 130 seconds. With similar exposures and a brightness two magnitudes fainter than Proxima Cen, measures of \vAs have lower S/N. We flat-fielded  the \vAs observations using a sum of the average counts for astrometric reference stars ref-2 and ref-3 identified in Figure~\ref{fxy}. Figure~\ref{FP} contains flat-fielded average counts as a function of BC orbital phase. 
We observed \vAs three times during each scheduled \HST observation set, hence, the \vAs groupings of three.
The observations used the \FGSns\,3 F583W filter. The bandpass center, 583nm, has a width 234nm, which includes
both the \Has and \Hes emission lines discussed below (Section~\ref{specem}).
Figure~\ref{FP} includes a sine wave with a period constrained to the BC orbital period. We find a F543W ($V$+\Ha) amplitude of 0.021$\pm$0.005 mag. The sampling is too sparse to make definitive statements,
but the fit
to the BC orbital period indicates only one dimming event per cycle, when \vABs is more distant than \vAC. This might
require additional complications (such as eclipses or a dark spot on the back side of component C). Alternatively, because \vAs varies (V* V805 Tau, flare star), we might attribute the  larger signal  at 
$\Phi_{\rm BC} \simeq 1$ to flaring activity. 

 \subsection{$V$-band Photometry} \label{BSP}
To test the eclipse hypothesis, one of us (BS) obtained absolute 
$V$-band photometry in autumn 2019 using the Lowell Observatory
0.7-m robotic telescope.  \vAs  was observed with 100-second exposures, 
typically several visits each on a total of 13 nights, using
 ``Canopus" software \footnote{http://bdwpublishing.com/mpocanopusv10.aspx} to reduce the data, yielding
conventional aperture photometry.  The zero-point was adjusted
to within a few percent of standard V using magnitudes of
four field comparison stars.  

Rather than providing additional evidence to assist in choosing between the interpretations above (Section~\ref{Fphot}),  a periodogram analysis of the newer $V$-band photometry yields a
highly significant peak  near, $P_V=1.7^{\rm d}$ with a lower peak near the BC orbital period, $P_{\rm BC}=0.7^{\rm d}$. Fitting a sine curve to the $V$ photometry yielded $P_V=1.783\pm0.003^{\rm d}$ and an amplitude $\Delta V = 0.011\pm 0.001$ magnitude with $\langle V \rangle= 13.27$.

The left side of Figure~\ref{BSPfig} presents the $V$ measurements phased to  the 1.783$^{\rm d}$
period with the FGS F543W measurements superposed. The fit includes only the Lowell $V$ measurements,
The right side of Figure~\ref{BSPfig} phases
FGS and $V$ photometry to the BC orbital period with the single cycle fit from Figure~\ref{FP} superposed.
The variation in $V$ only does not correlate with the \vAs BC orbit. The bandpass containing \Has does.

\section{Dissecting \vA}\label{BS}
Our astrometry  (Section~\ref{MT}) yields masses for the component A,D and the component B,C pairs. Our RV measurements (Section~\ref{WTF?}) yield a mass ratio for the B,C components. From \mBC=$0.91 \pm 0.11$\msun and 
\mC/\mB=$0.945\pm0.005$ we find \mB=$0.47\pm0.06$\msun and \mC=$0.44\pm0.06$\msune. Assuming for the AD-BC pair a $\Delta K=0$,  that $K_B\simeq K_C$, $A_K\simeq0$, and from our parallax, a distance modulus m-M=3.68, we find 
absolute magnitudes $M_K=6.09$ for each. The \cite{Ben16} $K-$band MLR predicts \m=0.46\msune, assuming equal brightness components B and C. The agreement provides validation for the astrometry and RV results. We now lack only individual masses for components A and D.

To establish their masses 
 we use
the data and sources listed at the top of Table~\ref{tbl-RIP}. This table also
provides a line by line guide to the process resulting in a mass estimate for components A  and component D, the (suspected) white dwarf required to satisfy the
total mass resulting from the astrometry (Section~\ref{MT}).

Table~\ref{tbl-RIP} begins with photometry of the total system, $V$-band from Section~\ref{BSP} and $K$-band from 2MASS \citep{Skr06} on {\bf line 1}.
We transform these into arbitrary intensities ({\bf line 2}), and assert a magnitude difference between components AD and BC on
{\bf line 3}.
In support of the {\bf line 3} data, Table~\ref{tbl-TR} contains magnitude differences in various bandpasses. 
The \FGS TRANS $F583$ measures are close to $V$-band \citep{Hen99}.
The speckle magnitudes are similar to $r$ and $i$ \citep{Hor12}.
Some of the components of \vAs are quite active, but, without the flare events
on mJD 50481 and 56934, the measures 
are consistent with $\Delta V = 0.0\pm0.1$. Typically $\Delta K$ is less than $\Delta V$ for M dwarfs \citep{Ben16},
supporting an estimate of $\Delta K = 0.0\pm 0.03$. Finally we introduce ({\bf line 4}) estimates of interstellar absorption, taking as priors a little less than half the  total absorption of $A_V=1.17$ mag (with $A_K/A_V=0.11$) along this line of sight \citep{Sch11}.
Including an absorption parameter significantly reduced the Table~\ref{tbl-RIP} solution residuals, and brought the $V$ and $K$ results into closer agreement.

Next we introduce ({\bf line 5})  data provided by the RV measures, the \mC/\mB~ mass ratio from the absorption
line results (Figure~\ref{RVabs}, Table~\ref{tbl-RVO}), \m$_{\rm C}/$\m$_{\rm B}=0.945\pm0.005$. {\bf Lines 6-10}  introduce the mass and parallax data from the astrometric modeling results (Table~\ref{tbl-ORB}).

We then appeal to three external relationships, allowing us 
to estimate a mass from an absolute magnitude. The first two relate M dwarf intrinsic luminosity to mass.
Rather than use the 
fifth-order polynomial Luminosity to Mass relations (LMR) in \cite{Ben16}, 
we simplify to a linear LMR for the $V$-band and a parabolic LMR for the $K-$band,
 \beq
{\cal M} = C_0 + C_1 (M_{V,K}) + C_2 (M_{K} ) ^2 \label{LMRf}
\eeq
with coefficients given in Table~\ref{tbl-LMR}. The total dynamical mass of \vAs precludes very low
component masses, where the considerable non-linearity in the LMR requires higher order terms.
The third relation maps  $K-$band absolute magnitude as a function of mass for
white dwarf stars. We obtain this mapping from  Bergeron cooling models 
\citep{Hol06,Kow06,Tre11,Ber11}, assuming a Hyades age of 670 My \citep{Gos18}.
Figure~\ref{WDM} shows this mapping, and a relation  fit with a third order polynomial with offset,
\beq
{\cal M} = C_0 + C_1 (M_{K} - X_0) + C_2 (M_{K}  - X_0) ^2  +C_3 (M_{K}  - X_0) ^3\label{LMRWD}
\eeq 
with the coefficient zero-point  at the bottom of Table~\ref{tbl-LMR}. Figure~\ref{WDM}
shows a multi-valued relation between mass and luminosity  (in a very narrow luminosity range) for the $V-$band. Hence, in the least-squares analysis below
we only use the $K-$band to determine the WD mass.
  
We next search for component intensity values, system parallax, and line of sight absorption ({\bf lines 11-16}) that minimize  the
sum of the squares of the O-C residuals, where O are the {\bf line 1-10} values, and C are the {\bf line 17-24}
final parameter values and residuals for both $V$ and $K$ bandpasses. We transform the {\bf line 11-14} intensities
into the {\bf line 26 - 30} magnitudes (corrected for the estimated absorption) and then with the parallax (yielding a distance modulus, $m-M=3.68$) into the {\bf line 31 - 35} absolute magnitudes. Using 
Equations~\ref{LMRf}~and~\ref{LMRWD} we derive the {\bf line 36 - 39} masses. This sequence iterates 
(separately for both the $V-$ and $K-$bands) until  minima in the 
sums of the squares of the {\bf line 17 - 24} residuals are found ({\bf line 25}). The $K$-band yields a significantly better fit with smaller residuals. 

We find final masses by applying the above least-squares process to both the $V-$ and $K-$band data, then averaging
the results ({\bf lines 36-40}). We find \m$_{\rm A} = 0.53\pm0.10$\msune, \m$_{\rm B} = 0.43\pm0.04$\msune, \m$_{\rm C} = 0.41\pm0.04$\msune, and \m$_{\rm D} = 0.54\pm0.04$\msune, where the errors are from the intrinsic scatter in the $V-$ and $K-$band MLR. The WD mass, \m$_{\rm D}$, comes only from the $K-$band photometry. 
The major mass disagreement between the $V-$ and $K-$band is for component A. None of these M dwarf masses can contribute new
data points for the \cite{Ben16} MLR, having been derived by appeal to that MLR.

 The {\bf line 25} fit quality parameter is much worse for the $V$-band. We find those values are decreased to 1.31 for $V$ and 0.001 for $K$ by including a larger absorption prior, $A_V=0.59$, $A_K=0.05$. This change also improves the A,B,C component mass agreement, yielding \m$_{\rm A} = 0.56\pm0.07$\msune, \m$_{\rm B} = 0.46\pm0.01$\msune, \m$_{\rm C} = 0.44\pm0.01$\msune, and \m$_{\rm D} = 0.53\pm0.04$\msune. 
However, \cite{Tay06} concludes that Hyades interstellar extinction  is negligible. Thus, if real, this extinction must be local to \vA. However, photometry with the Wide-field Infrared Survey Explorer (WISE) \citep{Wri10} shows no excess far-IR flux  indicating heated dust that might increase local absorption. 
We note that the Table~\ref{tbl-RIP} WD $V-K=+1.18$ far exceeds that predicted for a 670My old 0.54\msune WD, $V-K=+0.3$. The fit including a higher absorption yields the same WD mass, but a $V-K=+0.66$ much nearer to that predicted. 

We estimate the WD mass error from the scatter in the Figure~\ref{WDM} LMR, and have ignored any contributions from the WD and Hyades age uncertainties. The inputs to our mass derivation ({\bf lines 1 - 10}) provide very little leverage with which to assign an age to the WD. For example, our assumed WD age of 670My and a mass, \m$_{\rm D} = 0.53$\msune, yields from the Bergeron cooling models  an absolute magnitude M$_K=12.01$. An age of 400My for that same mass WD yields M$_K=11.75$. However the other components have M$_K\sim 6$, over 100$\times$ brighter, rendering the WD component age ineffective at changing the mass results ({\bf lines 36 - 40}).

The IGRINS spectra yielded an AD-BC $\Delta H=0.28\pm0.10$. Relatively normal M dwarf binary stars would evidence $\Delta K \simeq \Delta H$. Redoing the modeling outlined in  Table~\ref{tbl-RIP}, substituting in {\bf line 3} $\Delta K =0.28 \pm 0.1$ for $\Delta K =0.0 \pm 0.03$ yields slightly higher masses for components B, C, and D (each up by 0.02\msune), and a slightly lower mass for component A (down by 0.04\msune). However the $K-$band goodness of fit parameter ({\bf line 25}) increases by a factor of 20. Thus, a lower $\Delta K$ value is a better fit to our aggregate information.

From the Table~\ref{tbl-RIP} component B and C masses
and the RV amplitudes, $K_{\rm A}$ and $K_{\rm B}$ from Table~\ref{tbl-RVO}, we can derive an inclination of $i_{\rm BC} \simeq  23\arcdeg$ and a separation, $a_{\rm BC} \simeq 0.016$ AU. Estimating component radii using the \cite{Boy12} eclipsing binary mass-radius relation, components B and C are separated by $\sim4$ stellar diameters. Furthermore, if we assume that B and C are synchronized in the 0.75-day orbit, and spin-orbit aligned, as it seems very likely, we can predict a 30 \kmse equatorial rotational velocity for a typical radius of ~0.46$R_\odot$ for a 0.46\msun star \citep{Boy12}. Our analysis of the  CfA and IGRINS spectra (Section~\ref{WTF?}) yielded $v\,sin\,i \simeq11$ \kms, implying $ i_{\rm BC} \simeq 20 \arcdeg$. 


We have established that the \vAs system contains four components with components A and D orbiting a common center of mass and
components B and C orbiting a common center of mass.
The AD subsystem and BC subsystem orbit a common center of mass.
Radial velocity measurements easily detect and characterize the BC motion (Figure~\ref{RVabs}) and the AD-BC orbit (Figure~\ref{fig-WTF?}).
The residuals to the BC RVs cannot contain any signature of the
AD RVs. However RV measures of component A might evidence AD motion. Additionally, when measured by FGS TRANS and speckle the separation between A and the BC subsystem
should change because of the AD motion. Component A dominates the TRANS fringe for AD, because
A is $> 3$ magnitudes brighter than component D. Component A will thus be measured to be nearer to or more distant from
BC because it orbits a barycenter with a WD, component D. 

In an attempt to tease out possible effects of component D on 
component A we  appeal to periodograms. Figure~\ref{LSRVT} presents  Lomb-Scargle periodograms
for (top) the time series of x and y TRANS residuals to the relative orbit shown in Figure~\ref{vAorb}, and (bottom) for the time series of residuals to the 
component A RV orbit  (Figure~\ref{fig-WTF?}).
We find one significant peak in the component A RV power spectrum at $P=21^d$ with a false alarm probability  $<0.1$\%. As expected, the Figure~\ref{RVabs}  B and C RV residuals  have no significant power in the range 500$^d > P > 8^d$. 
The TRANS X and Y residuals contain a  peak produced by a slight increase of power in both
axes at $P\sim25^d$ with a false alarm probability  $> 50$\%. That the only peak simultaneously present in both the TRANS astrometric residuals and the component A RV occurs
at near $P\sim 21^d$ suggests an AD period near  that value.

A formal fit of a sine wave to the RV A residuals yields $P_{\rm AD}=20.82\pm 0.01^{\rm d}$, and $K_{\rm A}=0.30\pm0.09$ \kms.
Our \mA, \mD, and a circular orbit with this period yield a \vAAs perturbation of 1.5 mas,  a total separation, $a_{\rm AD} \simeq 0.15$ AU, and would require a nearly face-on orientation, $i_{\rm AD} > 89\arcdeg$, to produce the small RV amplitude detected. Additional evidence for this  orientation includes the near zero \vAAs $Vsin\. i$ (Section~\ref{WTF?}).


  
\section{Emission Lines} \label{specem}

\subsection{\Ha} \label{specHa}
Early in our spectroscopic data collection process it became apparent that \vAs exhibited a strong and quite
variable \Has emission with between one and four components. Given that we obtained the many observations 
under a large range of weather conditions, any inter-comparisons of \Has behavior are best carried out using normalized data.
We transformed all \Has data to normalized counts, allowing the determination of equivalent width (hereafter, EW).

To measure EW we transformed the intensities, I($\lambda$) thusly;
\beq
AW = (1-(I(\lambda)/\langle I(\lambda)\rangle) ) \label{EW}
\eeq
where $\langle I(\lambda)\rangle$  averages  two ranges, I(6547\AA-6556\AA) and I(6572\AA-6584\AA). To ensure high enough S/N for inclusion in subsequent analysis,
the measured background, $\langle I(\lambda)\rangle$, had to exceed 40 counts per pixel (S/N$\sim9$).
We used a Gaussian multi-peak fitting routine in the
GUI-based commercial package {\it IGOR}\footnote{https://www.wavemetrics.com} to derive component wavelength (for some RV determinations we used a Voigt function) and EW from a 6\AA ~span centered on the \Has region of each spectrum. Component FWHM typically ranged from $0.3< {\rm FWHM}< 0.7$\AA.
Figure~\ref{Qew} provides an example of the fitting process. 
We tabulate the resulting \Has EW in \AA ~ in Table~\ref{tbl-EWs} with component 1,4 (\vACs and B, see below) velocities  in Table~\ref{tbl-HaBCRVs} and component 2,3 velocities in Table~\ref{tbl-Ha23RVs}. While the average total \Has EW, $\langle{\rm EW}_{\rm tot} \rangle = -6.9\pm1.0$, indicates  a highly active {\it single} star \citep{New17}, \vAs is not a single star. The individual EW for components B and C ($\langle{\rm EW}_{\rm B} \rangle =-1.1\pm0.2 , \langle{\rm EW}_{\rm C} \rangle = -0.9\pm 0.3$), when placed on the Newton \etal (2017) figure 2,  argue  only slight component over active for their masses. \cite{Alo15} find an EW$_{tot}=-7.0$ on mJD=55935, consistent with our measures in the range 49962$<$mJD$<$54484.

We probed \Has velocity behavior close to syzygy (phases $\Phi_{\rm BC}=0.25~ {\rm and} ~0.75$) on those nights we were able to obtain observations closer to quadrature ($\Phi_{\rm BC}=0 ~{\rm and} ~0.5$). We constrained  peak parameters (amplitude, width, and Voigt shape)  to near quadrature values, solving only for position in wavelength, as demonstrated in Figure~\ref{HaRVprobe}. This approach depends on an assumption of profile stability for EW2, EW3 over hours, and provides no independent EW information at syzygy for any component.

When we first observed four components in \Ha, these were ascribed to components A, B, C, and now D (3, 4, 1, and 2 in Figure~\ref{Qew}). B and C were easily identifiable with peaks 4 and 1,
given their known, short-period RV behavior from absorption lines (Figure~\ref{RVabs}).  Figure~\ref{HaRV} shows component B and C  RVs from \Has emission as a function of BC absorption line orbital phase. 
Table~\ref{tbl-RVO} includes the results of orbit fitting to all resulting component B and C \Has RVs. The  $K_{\rm B}$ and $K_{\rm C}$ amplitudes from \Has significantly exceed those derived from the absorption lines. The strikingly systematic residuals seen in Figure~\ref{HaRV} might be explained by some sort of flow pattern.  For example the  component B positive residuals  in the ranges $0.1<\Phi<0.25$ and $0.75<\Phi <0.9$  indicate outflow from component B. This  flips to negative RV residuals towards  component B (inflow) for $0.25<\Phi<0.35$ and $0.75<\Phi<0.9$. The residuals for component C invert this sequence, indicating that a similar process works on both stars.

The bottom panel of Figure~\ref{BCrules} plots \Has EW values for components B and C, which show strong correlation with BC orbital phase. The top of Figure~\ref{BCrules} demonstrates that the sum of EW2 and EW3 show no such correlation, arguing that these components have little association with \vABs and C. Returning to the lower panel, EW B is slightly larger than EW C at quadrature phase $\Phi_{\rm BC}=1.0$ or 0.0.  EW C is significantly
larger than  EW B at the quadrature phase $\Phi_{\rm BC}=0.5$. These observations are consistent with the leading hemispheres of both stars having increased \Has activity, component C more so than B.  In general component B and C \Has EW is highest near quadrature ($\Phi_{\rm BC} =0.0, 0.5, 1.0$), consistent with the wide-band photometry (Figure~\ref{FP}), showing a brighter total magnitude near those phases. With an estimated $20\arcdeg$ inclination,  
apparently the \Has emission suffers  attenuation at syzygy ($\Phi_{\rm BC} =0.25, 0.75$). While we have no direct \Has EW measures for \vABs and C at syzygy, the patterns in Figure~\ref{BCrules} show a reduction in EW as either component approaches syzygy,  supporting  that hypothesis.

The period derived from the \Has B and C components agrees with that from absorption lines (Figure~\ref{RVabs}). The far larger 
$K_{\rm B}$ and $K_{\rm C}$ amplitudes do not agree. We offer a working hypothesis to explain this inconsistency. 
\Has becomes excited at some point between components B and C (labeled 4 and 1 in Figure~\ref{Qew}). 
If a ring or cloud exists around each star, the hydrogen gas gets excited between the two components, then diffuses around to the opposite side, where shadowing cools
the gas, allowing the hydrogen to emit \Ha. 
To produce the larger $K_{\rm B}$ and $K_{\rm C}$ amplitudes requires cloud/ring radii approximately $1.7\times$ the star radius. Location in each shadow  increases the distance between gas and barycenter. Exhibiting a similar period,
 the emitting gas travels further in the same time, meaning it has a higher velocity. 

We provide an animation (Figure~\ref{Disney}, $HalphaEWbetter.mp4$) illustrating the behavior of the various components of the \Has emission as a function of B-C orbital phase. We construct the animation  using only results of the constrained probing exemplified by Figure~\ref{HaRVprobe}. The traces have been shifted in wavelength to minimize  component 2 position changes (Figure~\ref{Qew}) from frame to frame. Originally produced to highlight the RV behavior of components B and C (4 and 1 in Figure~\ref{Qew}), the animation also shows the positional (RV) and EW constancy of components 2 and 3. For the 129 values in Table~\ref{tbl-Ha23RVs} peaks 2 and 3 have  $\Delta$RV=$35.4\pm3.4$ \kms and for the 140 values in Table~\ref{tbl-EWs}, a $\Delta$EW=$0.02\pm0.29$.
 This consistency argues for their having the same source, presumably associated with either \vAAs and/or D. Figure~\ref{HaRVAD}, showing the average of component 2 and 3 RVs plotted against the AD-BC orbital period phase, $\Phi_{\rm AD-BC}$, lends support to this assertion. The RVs exhibit significant scatter, but  agree with the orbit derived from the Figure~\ref{RVabs} absorption-line orbit. The bifurcation of the central \Has component (labeled 2 and 3 in Figure~\ref{Qew}) could be  either self-absorption \citep[e.g.][]{Sem13, Pav19} in a rotating cloud or emission from a rotating ring, either associated with A, or D, or both. 
 
In Section~\ref{BS} we  provided evidence for a roughly 21$^{\rm d}$ periodicity associated with \vAAs RV residuals to the AD-BC orbit, presumably because of the orbit of \vAAs around the AD barycenter. Figure~\ref{HaRVAD} identifies \Has components V2 and V3 with the AD pair. The periodograms of V2 and V3 in Figure~\ref{LSRVV2V3} solidify that association, showing a very strong signal, $P\sim25^{\rm d}$ in the sum of the individual periodograms.
Fitting  sine waves to the V2 and V3 RVs yield $P_{\rm V2}=25.6\pm 0.1^{\rm d}$, $K_{\rm V2}=2.3\pm0.4$ \kms,
$P_{\rm V3}=25.8\pm 0.1^{\rm d}$, and $K_{\rm V3}=1.4\pm0.4$ \kms. Again, if a consequence of the A-D orbit, the low $K$ values suggest a nearly face-on orbit.

\subsection{\He} \label{specHe}

 
  
In the CE spectra the $\lambda$ 5876.6\AA ~\Hes line appears in emission with from one to three components.
This line typically appears in stellar absorption spectra in  B and hotter stars, requiring T$>$10,000K \citep{Cox00}. Re-inspection of CE M dwarf spectra
acquired in support of our  MLR project \citep{Ben16} shows the \Hes line in emission for systems with primary mass less than 0.27\msun (Gl\,234, Gl\,791.2, Gl\,473, Gl\,65) and absent for systems with primary mass \m$\ge0.33$\msun (Gl\,22, Gl\,469, Gl\,623, Gl\,831, GJ\,1081, G\,250-029). Thus, it is surprising that \vABs and C, with \m$ \simeq 0.50$ \msune, show \Hes emission.
 
 Figure~\ref{HeQew} contains a plot of {\it AW} (Equation~\ref{EW}, with background local to the \Hes line) as a function of wavelength for some of the higher S/N observations (same epochs of observation as for Figure~\ref{Qew}).  We produced \Hes EW
 and RVs (as  for \Ha), listed in 
Table~\ref{tbl-HeEWs}  and Table~\ref{tbl-HeIRV} 
and plotted in Figure~\ref{HeRV} (RVs) and Figure~\ref{HeIABC} (EW). We label the central \Hes component EW2 and   V$_{\rm cen}$. Our \Hes EW and RV measures contain significantly more noise than for \Ha. 

Again, the RV behavior (Figure~\ref{HeRV}) of the higher and lower velocity peaks identify them as coming from components B and C. The lower S/N precludes probing in closer to syzygy as we did for \Has (Figure~\ref{HaRVprobe}). 
However, unlike \Ha, the component B and C RVs from  \Hes more closely track the absorption line results, yielding similar $K_{\rm B}$ and $K_{\rm C}$ RV amplitudes, with results listed in Table~\ref{tbl-RVO}. This argues that excitation and re-emission of \Hes occurs closer to each stellar surface. Residuals indicate that, as for \Ha, simple, zero eccentricity orbital motion is an inadequate model for the motions of the \Hes emission.  We ascribe EW2 and V$_{\rm cen}$ to the \vAs AD components (see below for rationale).
 We derive \Hes EW averages from the   Table~\ref{tbl-HeEWs} values: $\langle {\rm EW}_{\rm B} \rangle$=-0.29$\pm$0.12, $\langle {\rm EW}_{\rm C}\rangle$=-0.34$\pm$0.14, $\langle {\rm EW}_{\rm cen}\rangle$=-0.64$\pm$0.45, $\langle {\rm EW}_{\rm tot}\rangle$=-1.17$\pm$0.48. Similar to \Ha, most of the \Hes emission comes from the central source presumably associated with  components A and D. 

Figure~\ref{HeIRV} shows the Table~\ref{tbl-HeIRV} V$_{\rm cen}$ RVs plotted against the AD-BC orbital phase with the absorbtion line RV orbit overplotted. While quite noisy, the V$_{\rm cen}$ velocities argue for the association of the central \Hes emission with the AD components of \vA.

An initial supposition, that proximity heating (components B and C separated by only 0.016 AU) excites both \Has and \Hes emission can easily be  falsified.  Our Sun with $T_{\rm eff} = 5777$K has a surface emissivity $F_\odot=6.32\times10^{10}$ erg cm$^{-2}$s$^{-1}$ \citep{Cox00}. With masses near 0.5\msun (Table~\ref{tbl-RIP}, {\bf lines 35-36}), components B and C would have $T_{\rm eff} \simeq 3\,800$K \citep{Cox00} and $F_\odot=1.18\times10^{10}$ erg cm$^{-2}$s$^{-1}$, assuming $F\propto T^4$. For a separation r=0.016 AU (Section~\ref{BS}) the energy received by either component from the other is insufficient to raise the surface temperature to the
10\,000K necessary to excite the \Hes emission. This excitation level requires a non-thermal source, perhaps a flare-induced mechanism  
\citep{Kow18} or magnetic fields. 

Relative to the latter, \cite{Don08} and \cite{Mor08b,Mor10} have mapped the magnetic fields of a set of M dwarf stars. They find that most M dwarfs have poloidal fields which increase in intensity with rotation rate. Presuming tidal locking for components B and C, their sub- one day rotation
period and 0.5\msun masses would suggest a magnetic field for each of $\sim 500$ Gauss (e.g. Morin et al. 2011, figure 2). \nocite{Mor11} Perhaps an interaction of the  component B and C magnetic fields excites the \Has and \Hes emission. Additionally, component poloidal fields might serve to focus energy to polar locations, similar to terrestrial aurorae. We have no concrete evidence that the \Hes emission arises near there. Alternatively, perhaps the magnetic field is
a mix of poloidal and toroidal, with \Hes emission scattered near the stellar surface.  
Table~\ref{tbl-HeEWs} has  time-resolved data with which to tease out more hints as to excitation and emission. Doing so is beyond the scope of this paper. We hope that our observational data will stimulate the derivation of details of some magnetohydrodynamic mechanism operating in this system. 

Lastly, we explore some relations between the various emission line EW. Figure~\ref{HaHeCor}, left, shows the strong correlation between \Has component 2 and 3 EW from Table~\ref{tbl-EWs}. We plot the log of the absolute value of EW2, $log |{\rm EW2}|$, against  $log |{\rm EW3}|$. The maximum values in the upper right all occur within 4 days of AD-BC periastron. We have weaker evidence that the same energy source(s)  power both the    \Has and \Hes emission associated with \vAs AD, shown by their correlation (Figure~\ref{HaHeCor}, right). The one strongest \Hes point also occurred near AD-BC periastron.

\subsection{X-ray Emission} 
 \cite{DEl13} catalog transient 0.3-10keV X-ray bursts observed by Swift-XRT \citep{Geh04}. During the seven years of operation documented in the catalog, 
\vAs was detected three times, listed in Table~\ref{tbl-SX}. We tag each observation with both the BC and AD-BC orbital phases. In a $5\arcdeg$-radius subset of the catalog centered on \vA, it is among the brightest
with an average 0.3-10keV flux, $\langle {\rm Flux} \rangle = 1.62 \times 10^{-12}$ mW m$^{-2}$.
Two detections occurred close to BC syzygy, phase $\Phi_{\rm BC} \sim  0.25, 0.75$. 
One detection occurred very close to AD-BC periastron.
The BC syzygy correlation suggests the presence of active (coronal?) regions on the side of each star facing the other 
(i.e., facing the observer at syzygy), perhaps associated with dark spots. 
The BC system is fainter at  phase $\Phi_{\rm BC} = 0.75$ (Section~\ref{Fphot}, 
Figure~\ref{FP}), when \vABs is closest to us. Dark spots on the hemispheres of each star facing the other 
might explain both features. From admittedly small number statistics we abstract that some X-ray behavior  comes from BC, some from periastron interaction of AD with BC.

\section{Discussion} \label{SciFi}
With all of these (sometimes conflicting) results we now search for a coherent description of the entire \vAs system.
To do so, we  summarize measured results, identify inferences drawn from these results, clearly label
conjectures, and end with constructing a provisional picture of  \vA.

 \FGS TRANS mode and speckle camera astrometry confirm \vAs binarity with a high eccentricity, low inclination orbit,  producing a total mass requiring more than just two M dwarfs. TRANS and speckle measures provide a near-zero magnitude difference between the the originally identified components A and B. RVs acquired from two primary sources yield a component A-B orbit agreeing with the TRANS/speckle result. These RVs also identify two M dwarfs in a very short period, circular orbit, $P_{\rm BC}=0.749^{\rm d}$. 
CfA and IGRINS analyses yield $v\,sin\,i \simeq11$ \kmse for each component. The $K$ values for  \vABs and C provide
 \mC/\mB=0.945. \FGS POS mode astrometry yields an independent parallax, a proper motion, and  a mass fraction,  $f$. The mass fraction shows \mA~far too large to be only an M dwarf. To component A we add component D, presumed to be a white dwarf. \mB, too large to be a single M dwarf, now acquires component C, known from RV spectra to be another M dwarf. We  derive individual masses for all components, using the distance, total masses, masses of detected components, AD-BC magnitude difference, an estimate of interstellar extinction,
\mC/\mB~ mass ratio, $V-$ and $K-$band Mass-Luminosity relations, and a $K-$band WD Mass-Luminosity relation (assuming a 670 My age for the Hyades). 

The values for \mB ~and \mC ~and the observed RV amplitudes yield an orbital inclination, $i_{\rm BC} \simeq 25\arcdeg$ and a physical separation,
$a_{\rm BC}\simeq 0.016$ AU. This inclination is consistent with an observed $v\,sin\,i \simeq11$ \kms,  plausible B and C radii, spin-orbit alignment, and tidal locking. Periodograms of \vAAs absorption line RVs and variations in AD-BC separation suggest $P_{AD}\sim22^d$.  \mA, \mD ~and a presumed astrometric detection limit yield an orbital inclination, $i_{\rm AD} \simeq 90\arcdeg$, and for a circular orbit, a physical separation, $a_{\rm AD}\sim 0.2$ AU. 

We observe \Has in emission with  from two to four components. Components B and C are easily identified through RV variation with the same period, $P_{\rm BC}=0.749^{\rm d}$. These variations have  larger velocity amplitudes than for the absorption lines.  The  \Has BC velocity variation with larger amplitude than seen for the absorption line measures likely places the \Has emission
further from the BC system barycenter, behind, in the shadow of each component, where it can cool sufficiently to emit. The 
EW behavior, phased to the BC orbital period and residual patterns seen for the \Has RV BC orbits offers supporting evidence. 
The two central \Has peaks have an RV signature agreeing with that for component AD-BC orbit. Periodograms of their RVs show a strong peak at $P\sim 25^{\rm d}$.

We observe \Hes in emission at $\lambda 5875$\AA ~with from one to three components. Again, identification of components B and C come from RV variation with the same period, $P_{\rm BC}=0.749^{\rm d}$, this time with  velocity amplitudes similar to those from absorption lines. This suggests closer proximity of the \Hes emission regions  to each component  B and C. We ascribe B, C  emission, which requires 
an excitation temperature, T$\sim10,000$K, to non-thermal  heating effects, either near continuous micro-flaring or magnetic field interactions. 
The RVs of the strongest \Hes peak identifies it with either A and/or D.

X-ray emission has been detected when B-C are near syzygy, phase $\Phi_{\rm BC} \sim  0.25, 0.75$. One outburst occurred very near AD-BC periastron.
Based on sparse sampling, FGS photometry (with an F583W bandpass containing \Ha) suggests that the system is fainter at B-C orbital syzygy, B nearer to us than C. Spots on each hemisphere of components B and C facing the other could satisfy the photometry. Spots associated with stellar activity could then explain
the X-ray coincidence with BC syzygy. A remaining photometric mystery is that the $V$-band photometry yields $P_V=1.78^{\rm d}$, a variation unidentifiable with any previously determined activity. 

When initially formed, all components of \vAs  could have had radii significantly larger than at present. For example the 0.4\msun components of PAR 1802 in the Orion Nebula Cluster (age 1 My), have radii $R\simeq1.65R_{\odot}$ \citep{Che12}, four times larger than drawn in Figure~\ref{LooneyTunes2}. This raises the interesting possibility that the AD and BC separations have changed over time. \cite{Nao14} and  \cite{Nao16} demonstrate that the Kozai-Lidov Effect could force BC into a smaller separation, effected by  interactions with the more distant AD components, which might be treated as a single mass. On the other hand \mBC ~acting as a single mass could have rearranged components A and D into a smaller separation orbit. This latter orbital evolution almost certainly had to occur, given that the WD component D, if similar to other known WD Hyads, had a $\sim 3$\msun \citep{Sal18} early A to late B progenitor, whose red giant phase would have engulfed component A at the inferred, present A-D separation.

\section{Conclusions} \label{summ}
These results support a final  picture: a binary subsystem with M dwarf component A and white dwarf component D with $P_{\rm AD}\sim22^{\rm d}$ lies 0.5-4.4 AU from a binary subsystem with components B and C, both M dwarf stars. The AD-BC inclination, $i_{\rm AD-BC}=14\arcdeg\pm8\arcdeg$, compared to our inferred B-C inclination, $i_{\rm BC}\sim20\arcdeg$ suggests coplanarity. 

The cartoon in Figure~\ref{LooneyTunes2} shows the B and C components of the total system as seen from Earth, with an inclination
inferred from the derived masses (Table~\ref{tbl-RIP}) and absorption line $K_{\rm B}$ and $K_{\rm C}$, with parallax (Table~\ref{tbl-SUM}) impressing a scale in AU. In the absence of any measure, we have constrained the position angle of the line of nodes, $\Omega$, to that of the AD-BC orbit. We take the component stellar sizes  from the \cite{Boy12} mass-radius relation for eclipsing binaries, they typically having short orbital periods, as does \vAs BC. 
The \Has   EW$_{\rm B,C}$ trend towards smaller values as $\Phi_{\rm BC}$ approaches $0.25{\rm ~or~} 0.75$ in  Figure~\ref{BCrules}, demonstrating some eclipse (or darker spot) behavior in whatever
excites the \Has  emission. Seen from the other one, both the B and C components subtend only 20$\arcdeg$ on the sky. 

It is difficult to choose among accretion, stellar winds, continuous microflaring,  and magnetic field interactions to stimulate the \Has and \Hes emission lines. 
We illustrate our hypothesis in Figure~\ref{LooneyTunes2}. The \Has emission, once excited anywhere, 
 can sufficiently cool
enough to emit \Has in shadowed regions near the triangles (always close to components B and C, respectively), at the distance from the BC system center of mass required to produce the observed \Has RV $K$ values. We estimate their placement by solving Equation~\ref{PJeq} for $\alpha$ values   constrained  by our now known parallax and period, and the higher \Has  $K$ values  (Figure~\ref{HaRV}, Table~\ref{tbl-RVO}). The fat arrows in Figure~\ref{LooneyTunes2} indicate notional coronal loop-like flows or circulation that we infer from the 
\Has RV residuals in Figure~\ref{HaRV}. We have made no attempt to provide detailed flow patterns. 
The \Hes emission (from EW, Figure~\ref{HeIABC})
varies little and its velocity matches that from absorption lines, so is likely produced near the stellar surface. The \Hes RV residuals in Figure~\ref{HeRV}  hint at a  flow pattern similar to that of \Ha. 
 These data do not constrain an exact \Hes excitation location (other than near the stellar surfaces), but observations of active M stars would suggest the polar regions of each 
component.  We offer no hypothesis explaining the genesis of \Has and \Hes emission associated with the AD pair, other than to point out that a 670 My old, 0.53\msun WD would have a surface temperature capable of producing such emission.

\section{Summary}\label{sum}

\begin{enumerate}
\item Fringe scans (TRANS mode) spanning 4.7 years and a few more recent speckle observations yield a
low inclination, highly eccentric orbit. They also provide component $\Delta V$ measurements,
indicating nearly equal brightness components. 

\item Absorption-line radial velocities from the McDonald Observatory 2.1m
  telescope and Cassegrain Echelle spectrograph and a long series of observations from the CfA Digital Speedometer 
   yield an RV orbit which agrees with that derived from astrometry. They also yield a period
  and mass ratio for two, short orbital period M dwarf components (now identified as B and C) of this system, with $P_{\rm BC}=0.7492425\pm0.0000003$ days. 
   
\item Spanning 1.8 years, fringe tracking (POS) observations of  \vAAs, including
  photocenter corrections, provide an absolute parallax with
  a 4\% error, $\varpi_{\rm abs} = 18.37\pm0.65$ mas, a proper motion,
  and a mass fraction relative to the  astrometric reference frame. This parallax establishes Hyades membership for \vA,
  satisfying our first goal.

\item The astrometric solution yields  period, $P=2.705\pm0.004$ y, and a total mass for the system, ${\cal M}_{tot} = 2.06 \pm 0.24{\cal M}_{\sun}$. The mass fraction provides ${\cal M}_{\rm A} = 1.14
\pm 0.14{\cal M}_{\sun}$ and ${\cal M}_{\rm B} = 0.91 \pm 0.11
    {\cal M}_{\sun}$, masses  inconsistent with M dwarf components. We assume that component A is comprised of components A and D, and component B
    is component B plus component C. The \vAs system is now ${\cal M}_{\rm AD} = 1.14$\msun and  ${\cal M}_{\rm BC} = 0.91$\msune.

\item We have achieved our second goal, that of establishing component masses. Combining $V$ and $K$ photometry with a parallax  derived from our astrometry, \mAD, \mBC, the AD - BC magnitude difference, \mC /\mB ~mass ratio from absorption line RVs,  an estimate for interstellar absorption, M dwarf Luminosity-Mass relations, and a WD Luminosity-Mass relation for an assumed Hyades age of 670 My we solve for individual component masses that would best satisfy all the priors. We find \m$_{\rm A} = 0.53\pm0.10$\msune, \m$_{\rm B} = 0.43\pm0.04$\msune, \m$_{\rm C} = 0.41\pm0.04$\msune, and \m$_{\rm D} = 0.54\pm0.04$\msune. The WD mass, \m$_{\rm D}$, comes only from the $K-$band photometry. Including an extinction, $A_V=0.6$, lowers the Table~\ref{tbl-RIP} fit residuals, improving the consistency between the $V$ and $K$ solutions, yielding \m$_{\rm A} = 0.56\pm0.07$\msune, \m$_{\rm B} = 0.46\pm0.01$\msune, \m$_{\rm C} = 0.44\pm0.01$\msune, and \m$_{\rm D} = 0.53\pm0.04$\msune. 

\item The \mC ~and \mB, the parallax, and the  absorption line RV amplitudes, $K_{\rm B}$ and $K_{\rm C}$, yield an inclination, $i_{\rm BC} \simeq 23\arcdeg$ and separation, $a_{\rm BC}\simeq 0.016$ AU, consistent with an observed $v\,sin\,i =11\pm2$ \kms, and plausible B and C radii.

\item Periodograms of the TRANS mode astrometric residuals provide extremely weak evidence for a $P_{\rm AD} \sim 25^{\rm d}$. The component A RV AD-BC orbit residuals exhibit a  strong peak in the periodogram at $P_{\rm AD} \sim 21^{\rm d}$. 

\item The spectra show  \Has  in emission with from two to four components. 
The centrally located (in RV) components 2 and 3 are associated with the AD components, and may be caused by self absorption in a circumbinary hydrogen cloud. The amplitudes of the B and C RVs place emission further from the BC system barycenter.

\item  We see \Hes ($\lambda$ 5876.6\AA) in emission with from one to three components. We identify component B and C emission from
RV behavior, which in this case more closely tracks the absorption line RVs. Hence, we surmise a location for two of the \Hes components
closer to the surfaces of B and C than the \Has emission. The RV behavior of the central \Hes component supports association with the AD components of the system.

\item Only partially achieving our third goal, a qualitative model explaining all observed phenomena, we propose  hypotheses that will require future observations and theoretical interpretation for completion.

\end{enumerate}

Some future \Gs catalog could contain AD-BC orbital parameters and confirm the total system mass. However,  the very small magnitude difference between the AD and BC pairs could make that determination problematical. Additional low-cadence monitoring of the system with speckle techniques could further improve
the AD-BC orbit. \vAs remains a phenomena-rich laboratory for future magnetohydrodynamic model testing.

\acknowledgments

Support for GFB, OGF, and BEM for this work was provided by NASA through grants GTO-4892, 5657;
and GO-6479, 6881 from the Space Telescope Science Institute,
which is operated by the Association of Universities for Research in
Astronomy, Inc., under NASA contract NAS5-26555.  GT acknowledges partial support from the NSF through grant AST-1509375.
Similarly, EPH acknowledges support from the NSF through grants AST-1517824 and AST-1909560 for the analysis of the speckle data.

This publication makes use of data products from the Two Micron All
Sky Survey, which is a joint project of the University of
Massachusetts and the Infrared Processing and Analysis
Center/California Institute of Technology, funded by NASA and the NSF.
This research has made use of the SIMBAD and Vizier databases and
Aladin, operated at CDS, Strasbourg, France, the NASA/IPAC
Extragalactic Database (NED) which is operated by JPL, California
Institute of Technology, under contract with the NASA, and NASA's
Astrophysics Data System Abstract Service. We acknowledge the essential contribution of the 
WD cooling models Web site (\url{http://www.astro.umontreal.ca/~bergeron/CoolingModels}).

This work has made use of data from the European Space Agency (ESA) mission
{\it Gaia} (\url{https://www.cosmos.esa.int/gaia}), processed by the {\it Gaia}
Data Processing and Analysis Consortium (DPAC,
\url{https://www.cosmos.esa.int/web/gaia/dpac/consortium}). Funding for the DPAC
has been provided by national institutions, in particular the institutions
participating in the {\it Gaia} Multilateral Agreement.

We thank Art Bradley and Linda Abramowicz-Reed for their unflagging and expert \FGS
instrumental support over the last 25 years.  Cassegrain Echelle (CE)
Spectrograph observing and data reduction assistants included
J.~Crawford, Aubra Anthony, Iskra Strateva, Tim Talley, Amber
Armstrong, Robert Hollingsworth, Casey Kyte, and Jacob Bean.  We thank Dave Doss,
John Booth, and many other support personnel at McDonald Observatory
for their cheerful assistance over many years. Thank you Ray Lucas for morale-building
violin playing during a particularly productive observing run!
GFB thanks Chick Woodward and Nancy Morrison
for the suggestion of self-absorption to explain the central \Has peaks.
We thank Team IGRINS, especially Greg Mace and Heeyoung Oh for getting an April 2019 DCT observation.
We thank Gerard van Belle, Catherine Clark, and Zachary Hartman at Lowell who battled difficult weather conditions to secure the February 2020
speckle observation.

The American Astronomical Society supported the preparation of this paper 
as GFB carried out his duties as Society Secretary. For this he is sincerely grateful.
GFB fondly remembers Debbie Winegarten (R.I.P.), whose able assistance with Secretarial matters freed me to devote  time to this analysis. 

We thank an anonymous referee for a careful and useful assessment,  improving the final paper.

\appendix
\begin{center}
{\bf A Cautionary Tale}
\end{center}
There exist many pitfalls between data and definitive orbit determination, some rather entertainingly recounted in \cite{vdB62}. We managed to find a few more in producing our \vAs result. In a few years \Gs will flood us with perturbations crying out for orbit determination. Our experience may prove instructive.

\begin{figure}
\includegraphics[width=6.5in]{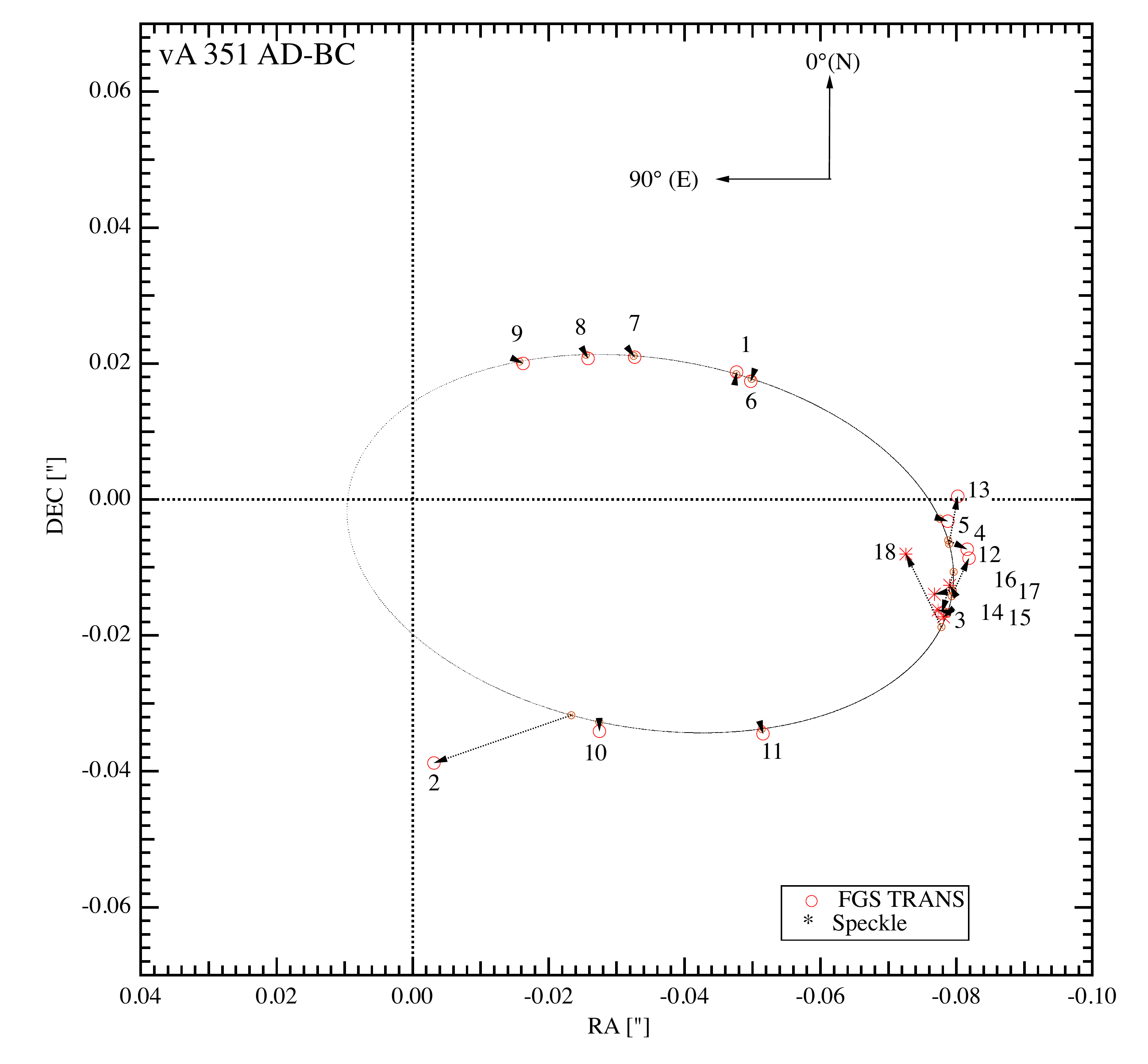}
\caption{ A relative orbit AD-BC for \vAs  derived from FGS TRANS and ground-based speckle observations. We plot observation numbers from Table~\ref{tbl-TR}. We omitted observations 2 and 18 for the final solution. }
\label{Rel}
\end{figure}

\begin{figure}
\includegraphics[width=7in]{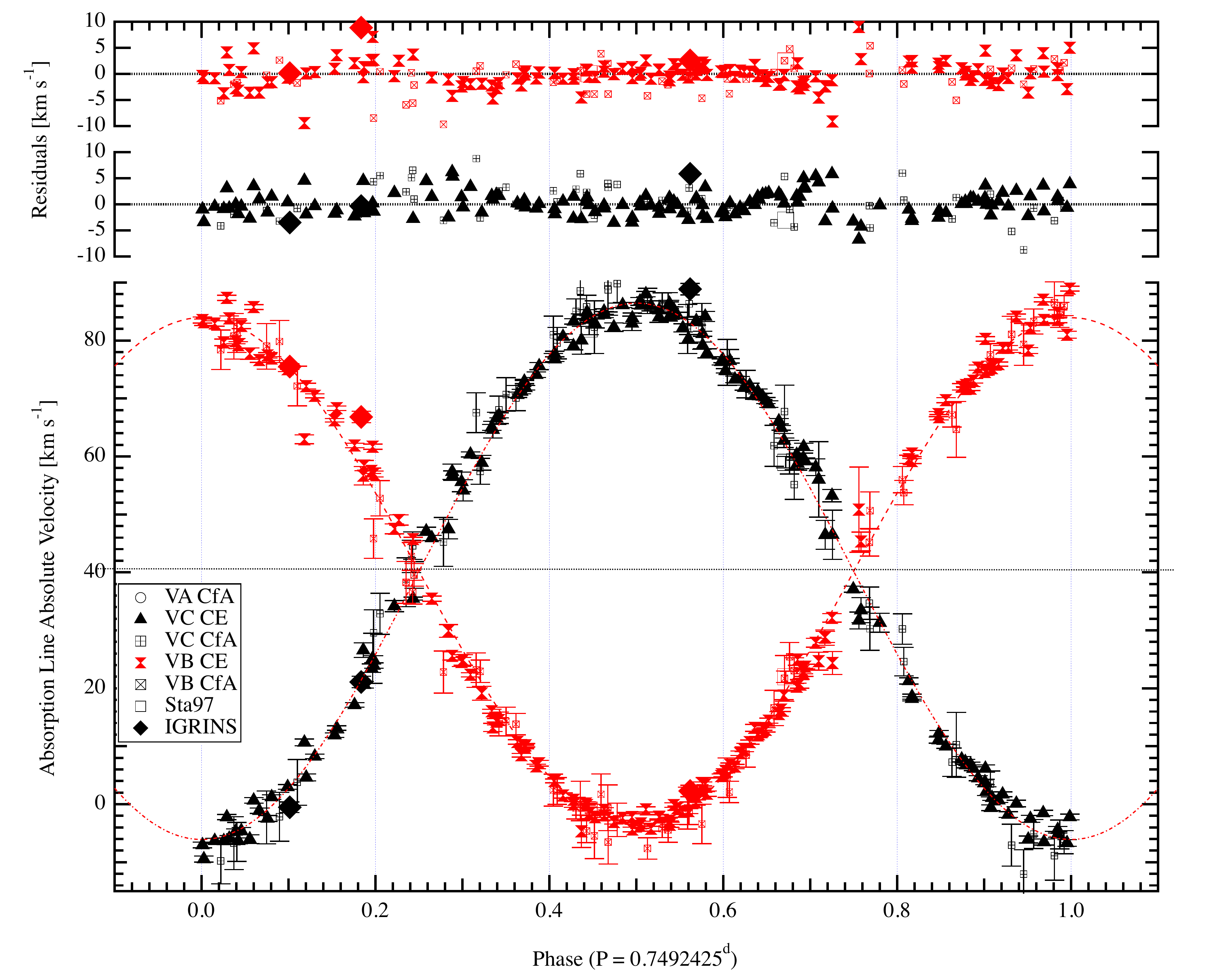}
\caption{ \vAs absorption line radial velocities (Tables~\ref{tbl-CE} and \ref{tbl-CfA}) for components  B and C, phased to the BC period, $P_{\rm BC}$. Orbital elements are given in Table~\ref{tbl-RVO}.
From the BC semi-amplitudes we derive a mass ratio \m$_{\rm C}/$\m$_{\rm B}$ = 0.945$\pm$0.005. Also plotted are independent measures from IGRINS and \cite{Sta97}, where the errors have been estimated at 1 \kms. }
\label{RVabs}
\end{figure}

\begin{figure}
\includegraphics[width=5in]{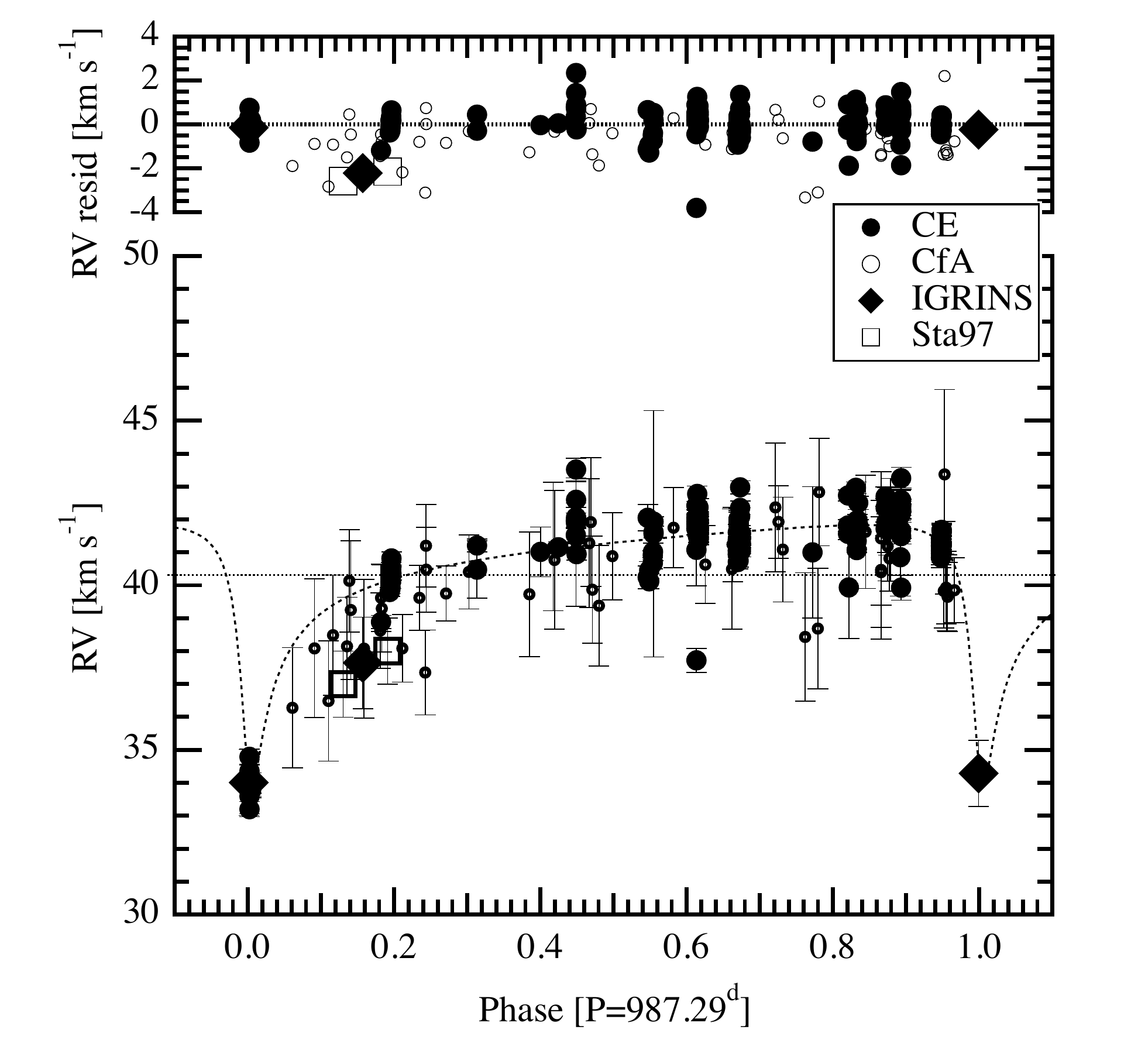}
\caption{Radial velocities for component A  plotted against phase of the AD-BC  orbit (Section~\ref{WTF?}) obtained only from RVs.
 The horizontal line denotes the derived systemic velocity (Table~\ref{tbl-ORB}). 
}
\label{fig-WTF?}
\end{figure}

\begin{figure}
\includegraphics[width=7in]{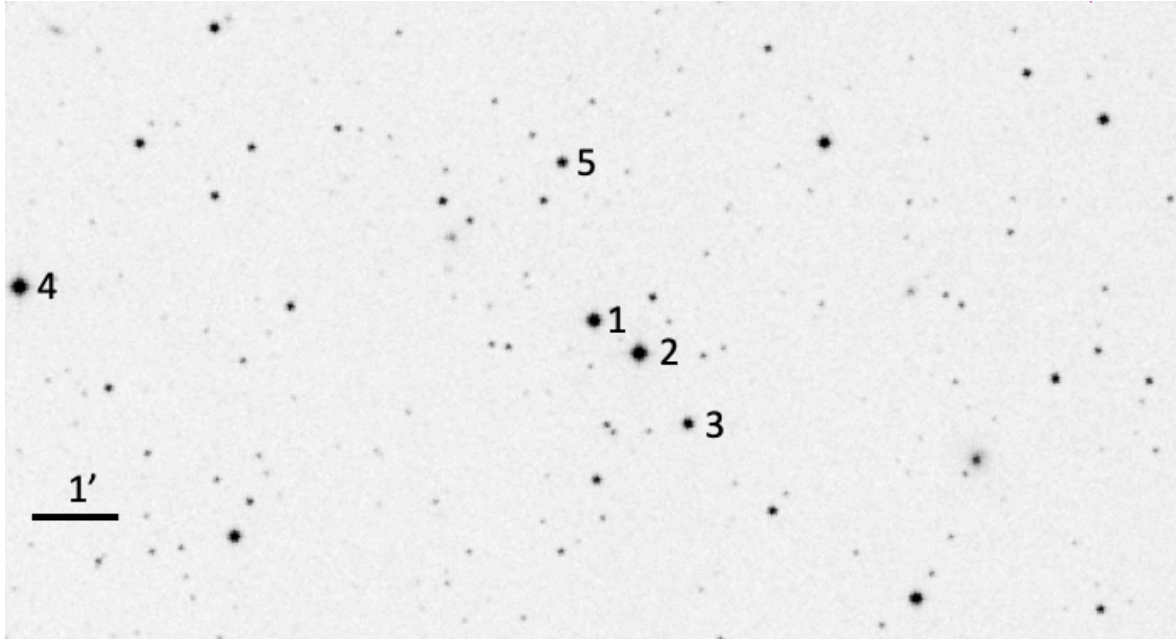}
\caption{ \vAs and reference stars used in the POS mode astrometry; north at top, east at left. ID numbers from Table~\ref{tbl-pis}. The ``1" indicates the location of the \vAs
  photocenter. Image from Aladin.}
\label{fxy}
\end{figure}

\begin{figure}
\includegraphics[width=6.5in]{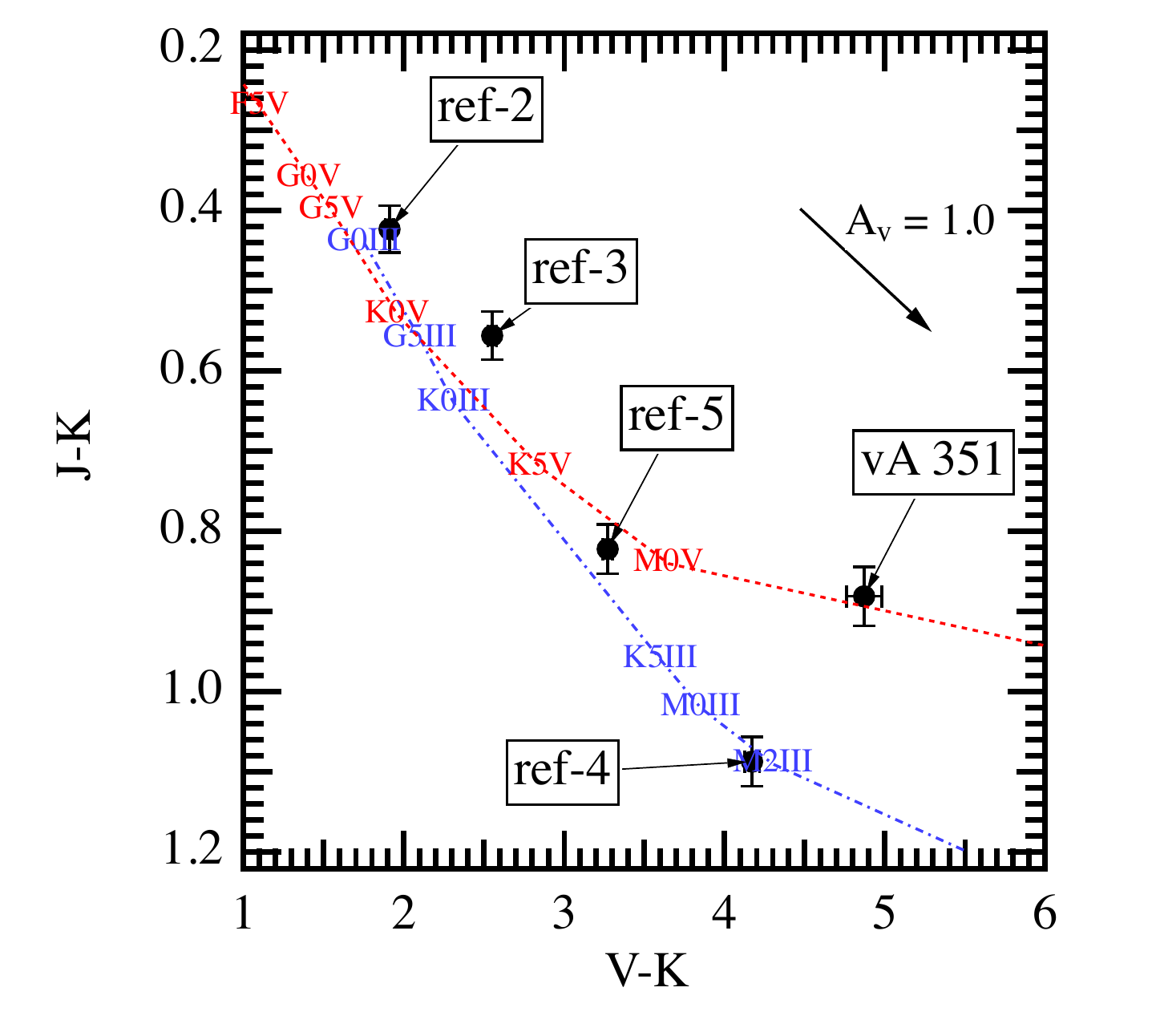}
\caption{$J-K$ vs. $V-K$ color-color diagram for \vA~and reference stars
  listed in Table~\ref{tbl-pis}. The dashed line is the locus of
  dwarf (luminosity class V) stars of various spectral types; the
  dot-dashed line is for giants (luminosity class III) from
  \cite{Cox00}. The reddening vector indicates A$_V$=1.0 for the
  plotted color systems. For this field at Galactic latitude
  $\ell^{II}=-21\arcdeg$ we estimate $\langle A_V\rangle$ = 1.4, midway between 
  redddenings established for vA\,310 and vA\,383 in \cite{McA11}.}
\label{CCD}
\end{figure}

\begin{figure}
\includegraphics[width=7in]{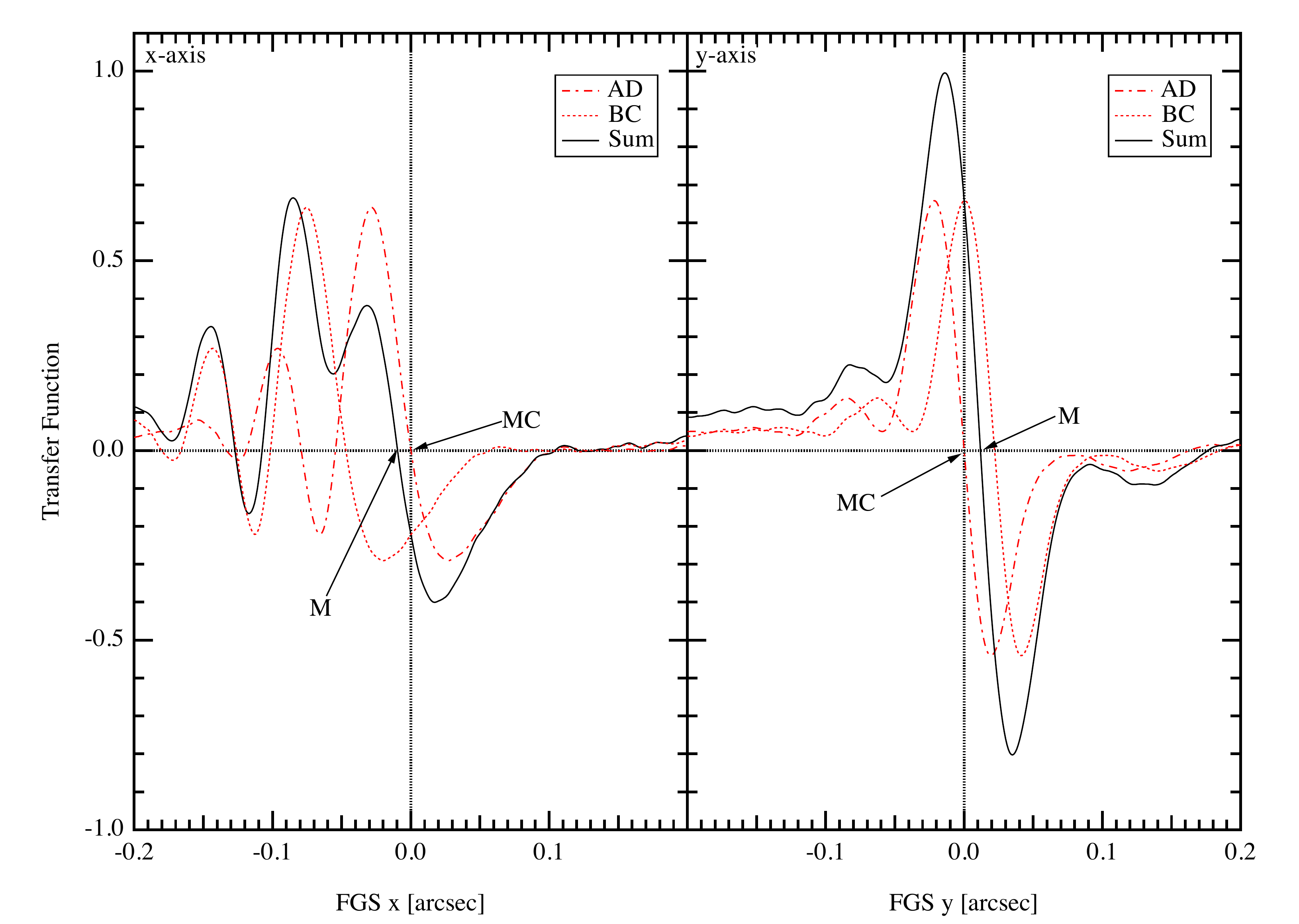}
\caption{Given the very small brightness difference between components AD and BC, we apply photocenter corrections
to the POS mode astrometry, which measures a zero-crossing for a sum of the two interferometric fringes. The arrow labeled M points to the position as measured by \FGSns\,3. The arrow labeled MC points to the actual position of component A. In each case the correction increases the separation between the components. The intrinsic fringe structure in the x-axis adds increased uncertainty to that correction. Corrections are +9 mas in x, -12 mas in y for $\Delta$m=0.0.}
\label{pcor}
\end{figure}

\begin{figure}
\includegraphics[width=5in]{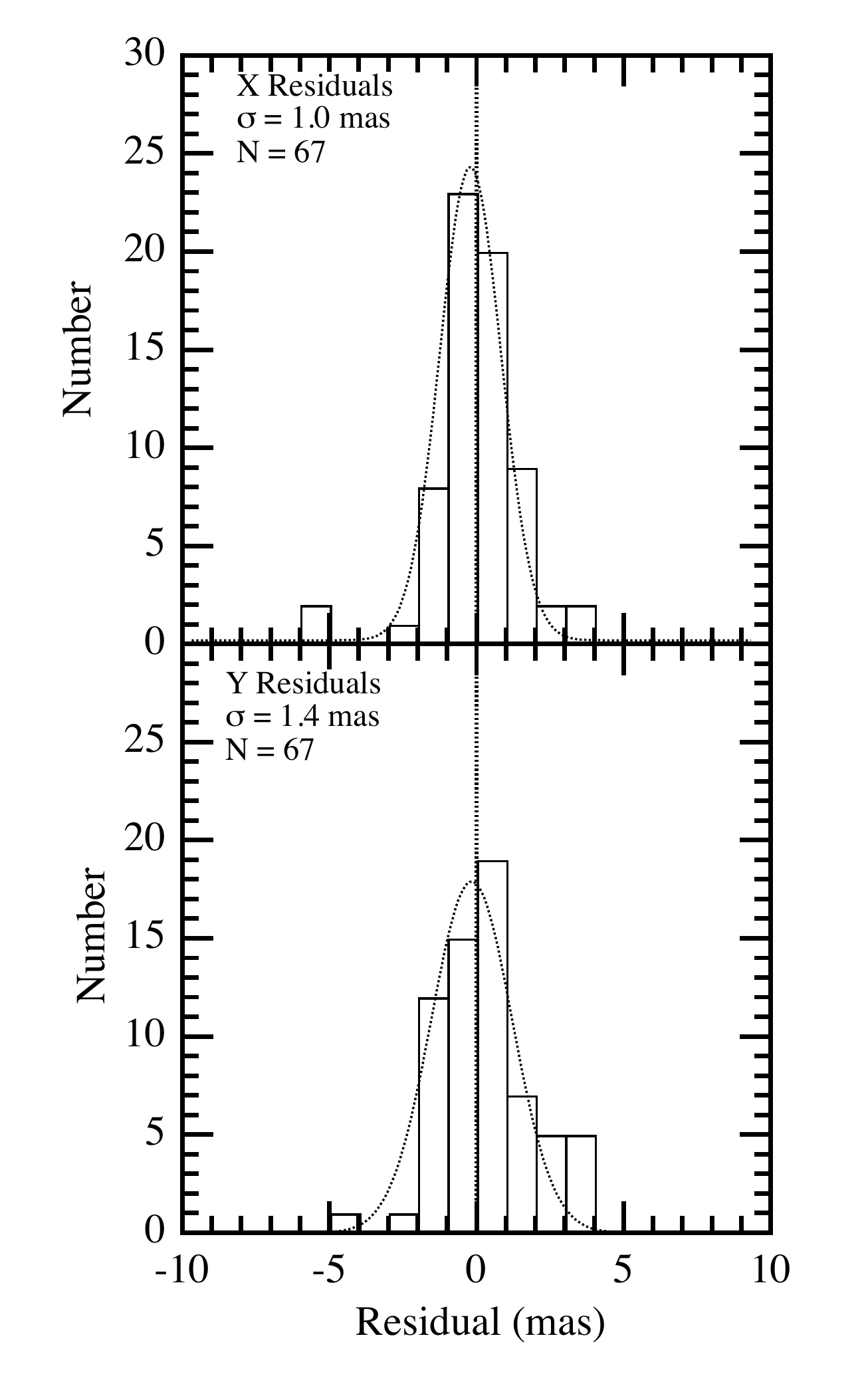}
\caption{Histograms of \vA~ and reference star residuals
  resulting from the application of the astrometric model (Equations 5 and 6).}
\label{his}
\end{figure}

\begin{figure}
\includegraphics[width=7in]{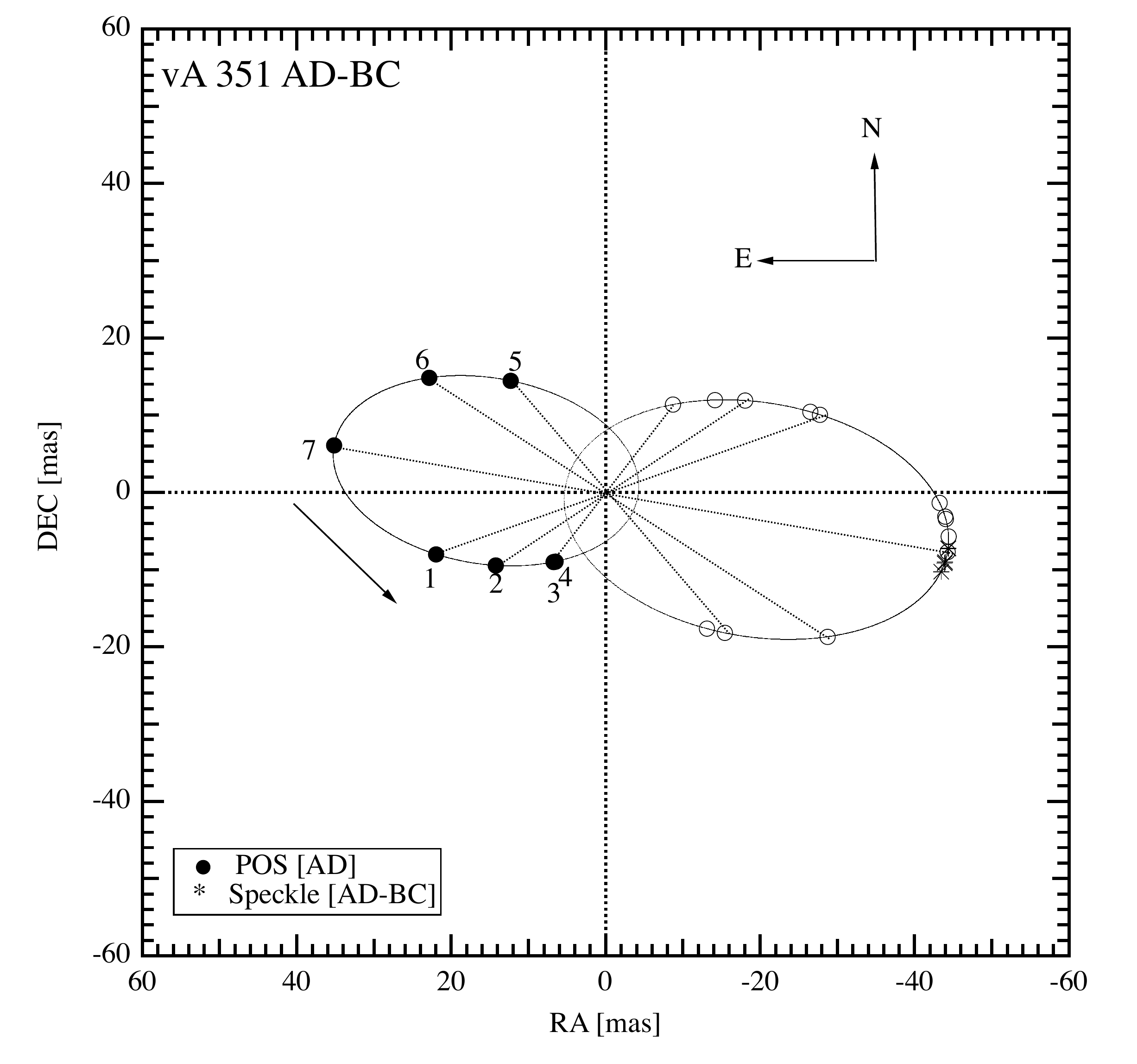}
\caption{\vA~component A (POS orbit calculated positions,\color{black}$\bullet$\color{black}) and
  component B (TRANS calculated, \color{red} $\bigcirc$\color{black}; speckle calculated,*).  POS measures the position of the AD combination relative to reference stars.
 TRANS measures the separations between and the position angles of AD relative to BC. All
  observations, speckle  and TRANS,  were used to derive the orbital elements listed in
  Table~\ref{tbl-ORB}. POS measures yield only parallax, proper motion and mass fraction (Equation~\ref{frac}). 
 The arrow indicates the direction of orbital motion. 
}
\label{vAorb}
\end{figure}

\begin{figure}
\includegraphics[width=5in]{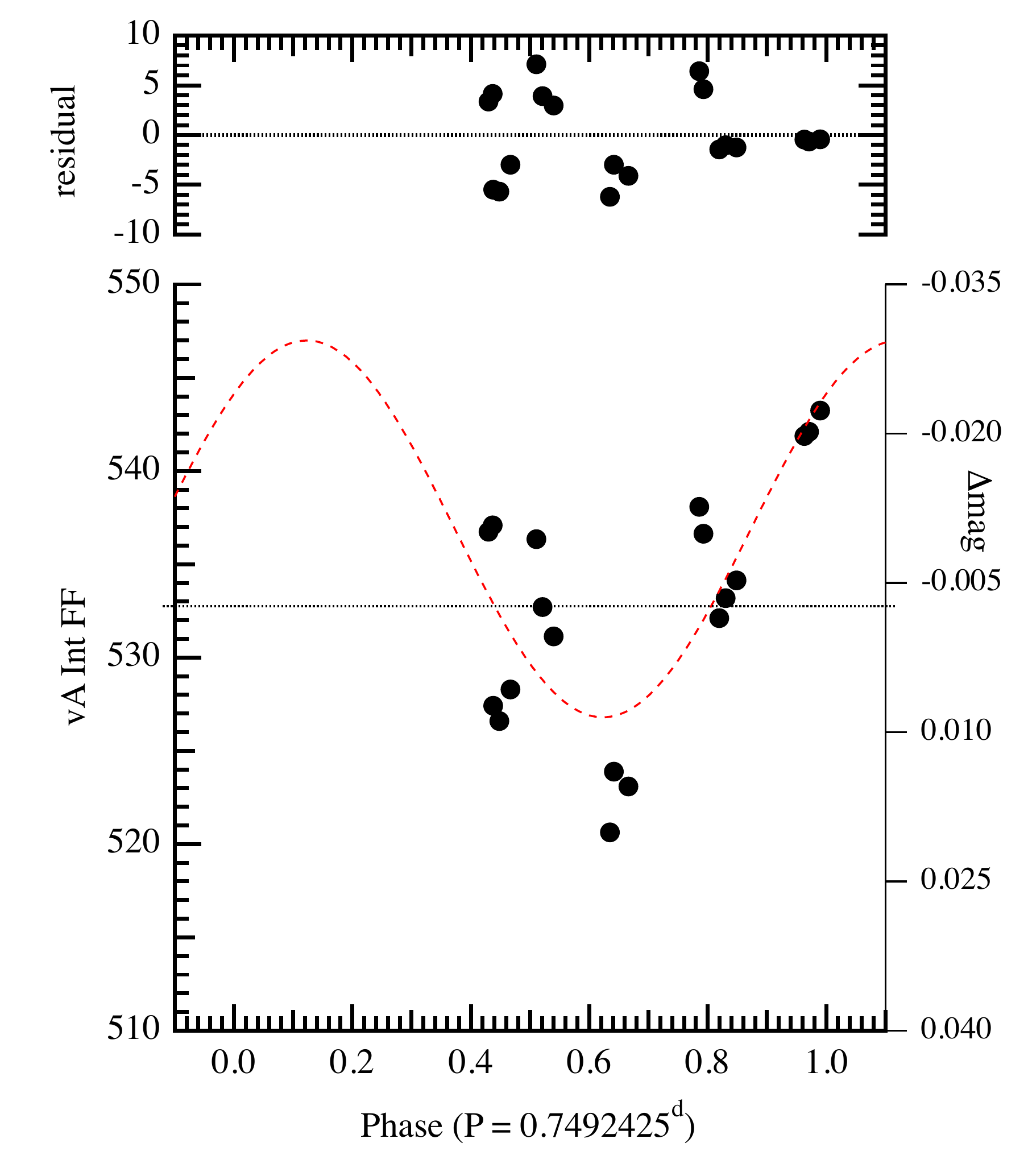}
\caption{Flat-fielded \vAs average counts from POS mode F583W photometry as a function of BC orbital phase. The \HST \FGS F583W filter includes \Ha. The variation in magnitudes ($\Delta$mag) is on the right.
The fit is a sine wave with the period constrained to  the BC orbital period, suggesting that the system is brighter  when  \vACs is closest to us ($\Phi_{\rm BC} \sim 0.25$), and fainter when  \vABs is closest to us (at $\Phi_{\rm BC} \sim 0.75$), consistent with a small amount ($\sim 0.01$ mag) of eclipse-induced dimming. 
}
\label{FP}
\end{figure}

\begin{figure}
\includegraphics[width=7in]{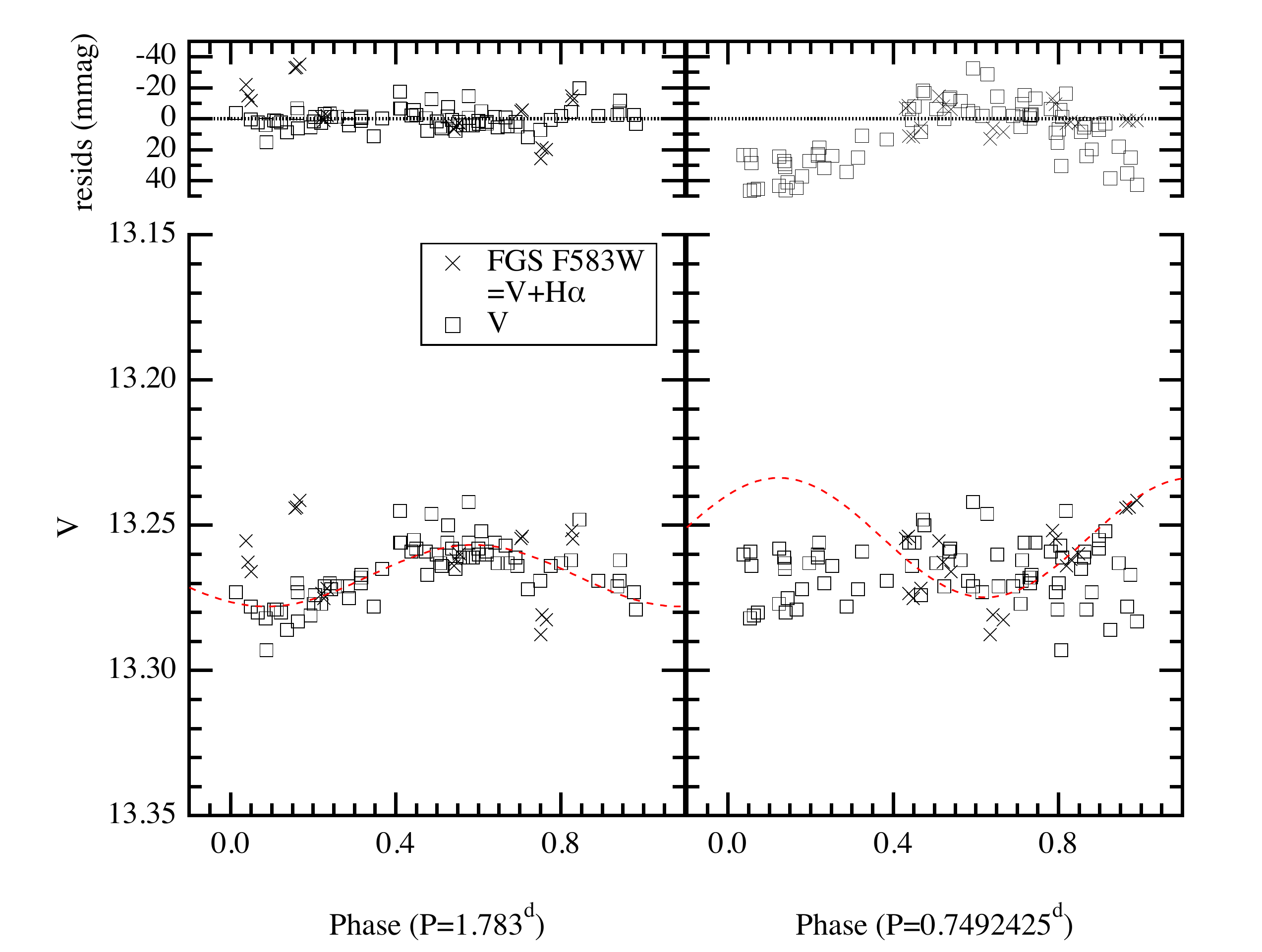}
\caption{Left: $V$-band photometry ($\Box$) and \FGS F583W photometry (\color{black}$\times$\color{black}) phased to a
period, P=1.783$^{\rm d}$.  Right:
the same photometry phased to the BC orbital period with the fit from Figure~\ref{FP}. 
}
\label{BSPfig}
\end{figure}

\begin{figure}
\includegraphics[width=6in]{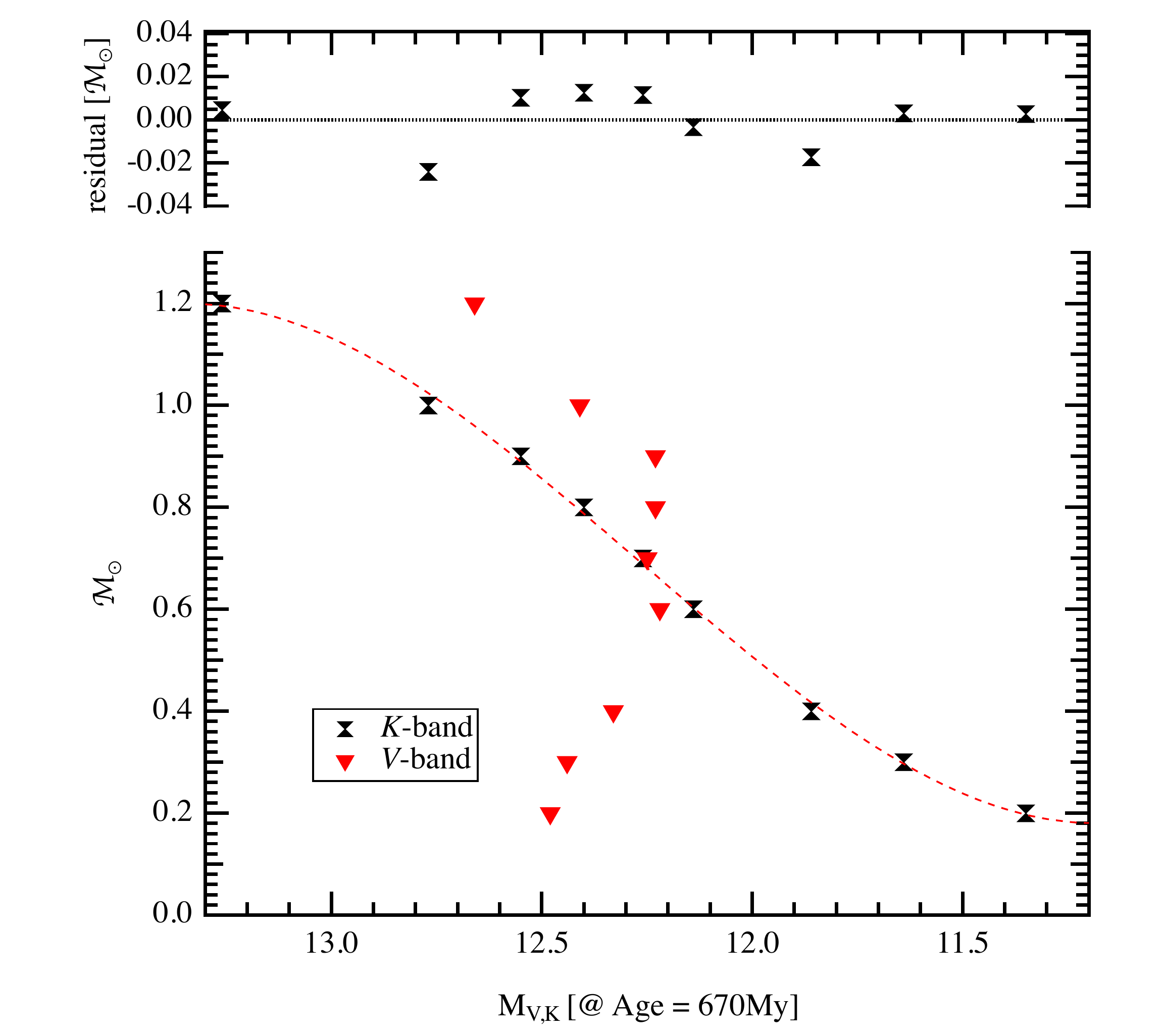}
\caption{White dwarf mass in solar units (\msune) as a function of $K-$ and $V-$band absolute magnitude, derived from Bergeron cooling models,
assuming a Hyades age of 670 My. We do not employ the multi-valued $V-$band relation as a constraint in the Section~\ref{BS}
component mass determination, but do use the  $K-$band relation, fit with a third order polynomial with offset (coefficients  in Table~\ref{tbl-LMR}).}
\label{WDM}
\end{figure}

\begin{figure}
\includegraphics[width=6in]{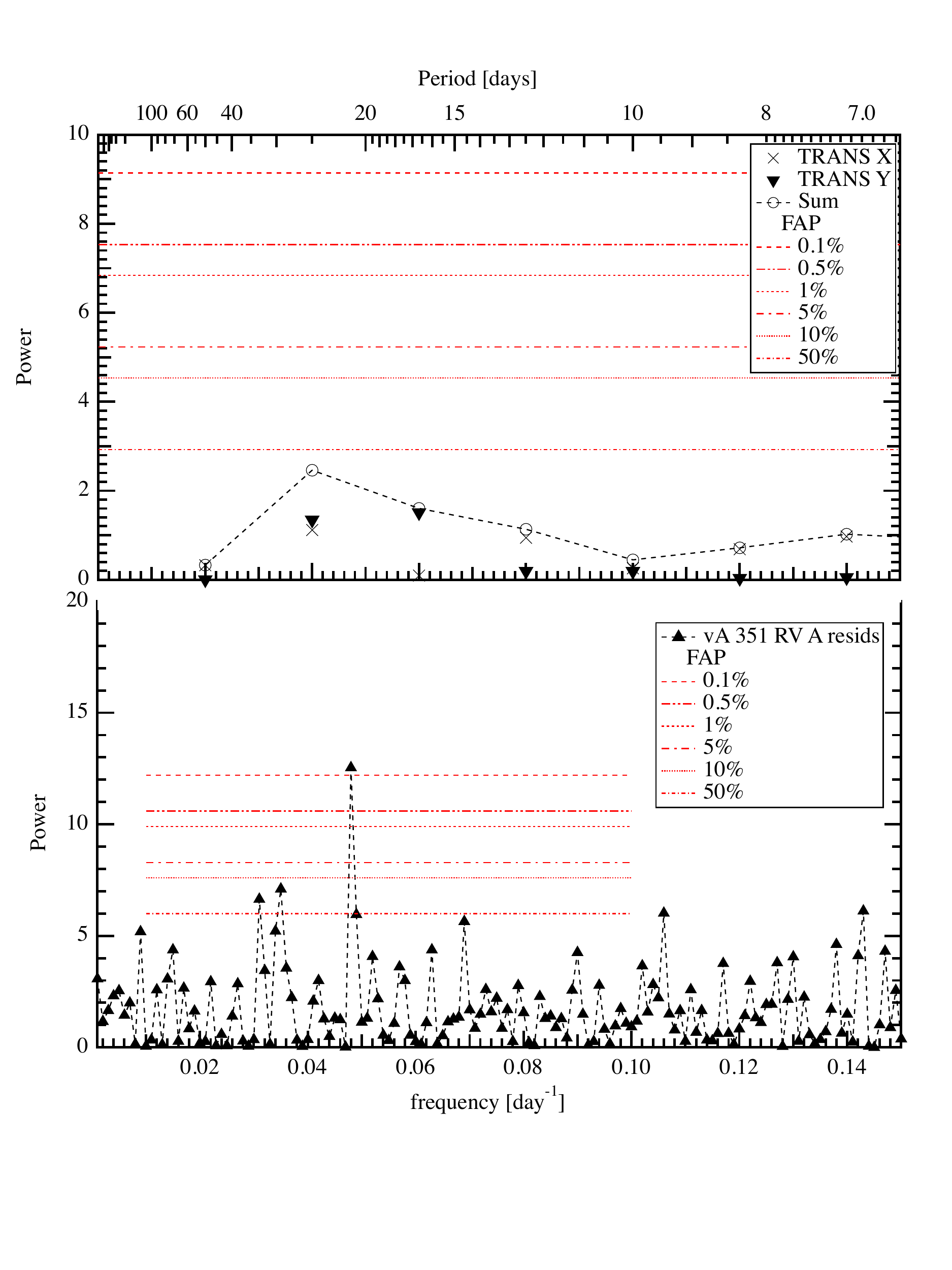}
\caption{Bottom: Lomb-Scargle periodograms for the Figure~\ref{RVabs}  component A measured RV residuals, providing evidence 
(false-alarm probability of $\sim 0.1$\%) for $P_{\rm AD} \sim 21^{\rm d}$. Top: periodograms of the TRANS mode and speckle X and Y residuals to the relative orbit (Figure~\ref{vAorb}). We also plot the sum of the X and Y power. They  provide only weak support for a  detection 
 at $P_{\rm AD} \sim 25^{\rm d}$.}
\label{LSRVT}
\end{figure}

\begin{figure}
\includegraphics[width=5in]{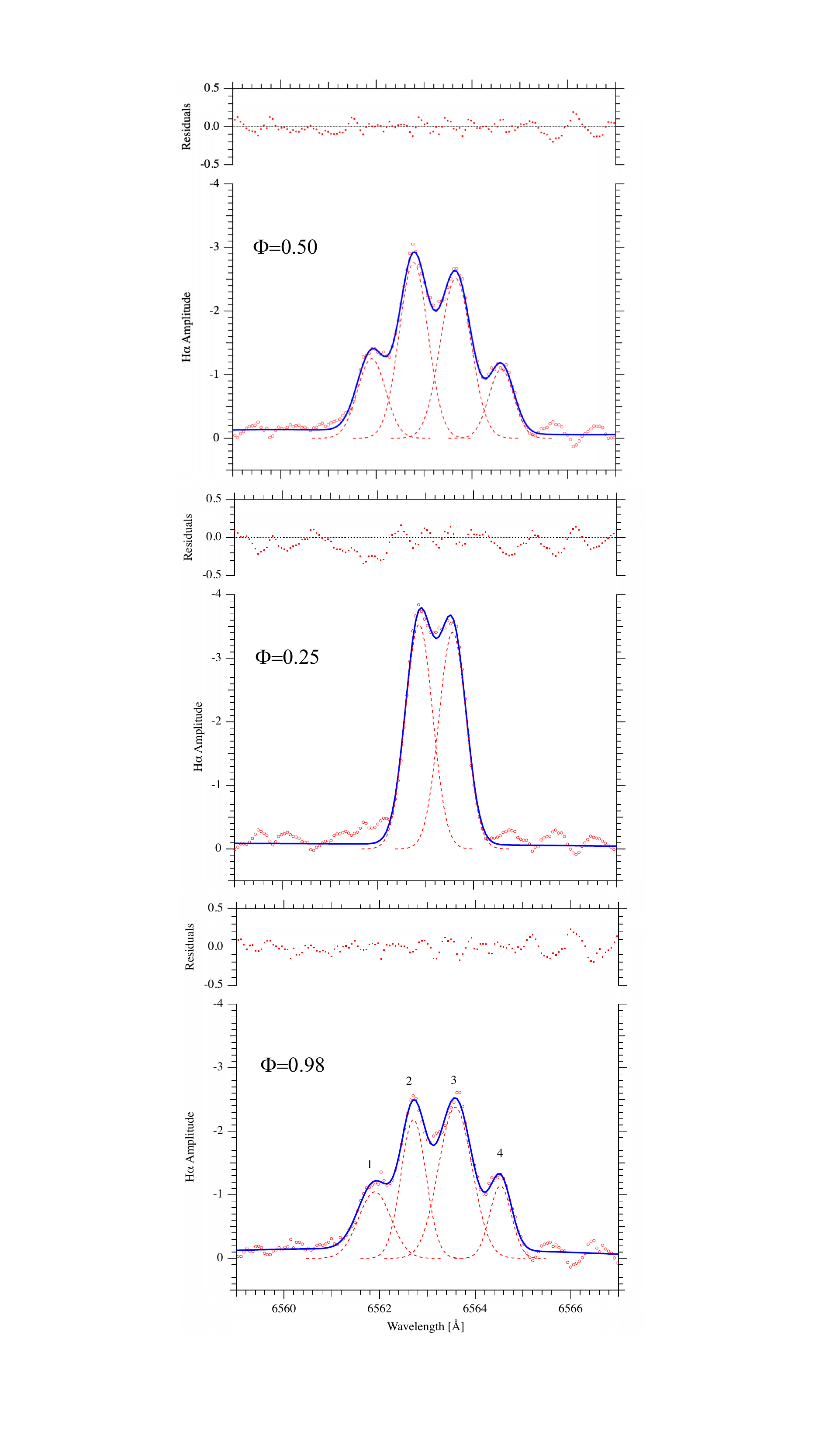}
\caption{Structure in the \Has emission line amplitude ({\it AW}, Equation~\ref{EW}) at $\Phi_{\rm BC}$ = 0.50, 0.25, 0.98; mJD=51467.36, 51466.42, 50724.47. In the bottom panel component B is the right-most peak, labeled 4;
component C the left-most, labeled 1. In the top panel they have swapped sides because of BC orbital motion. In the middle panel components B and C  partially fill-in the possible self absorption. 
}
\label{Qew}
\end{figure}

\begin{figure}
\includegraphics[width=5in]{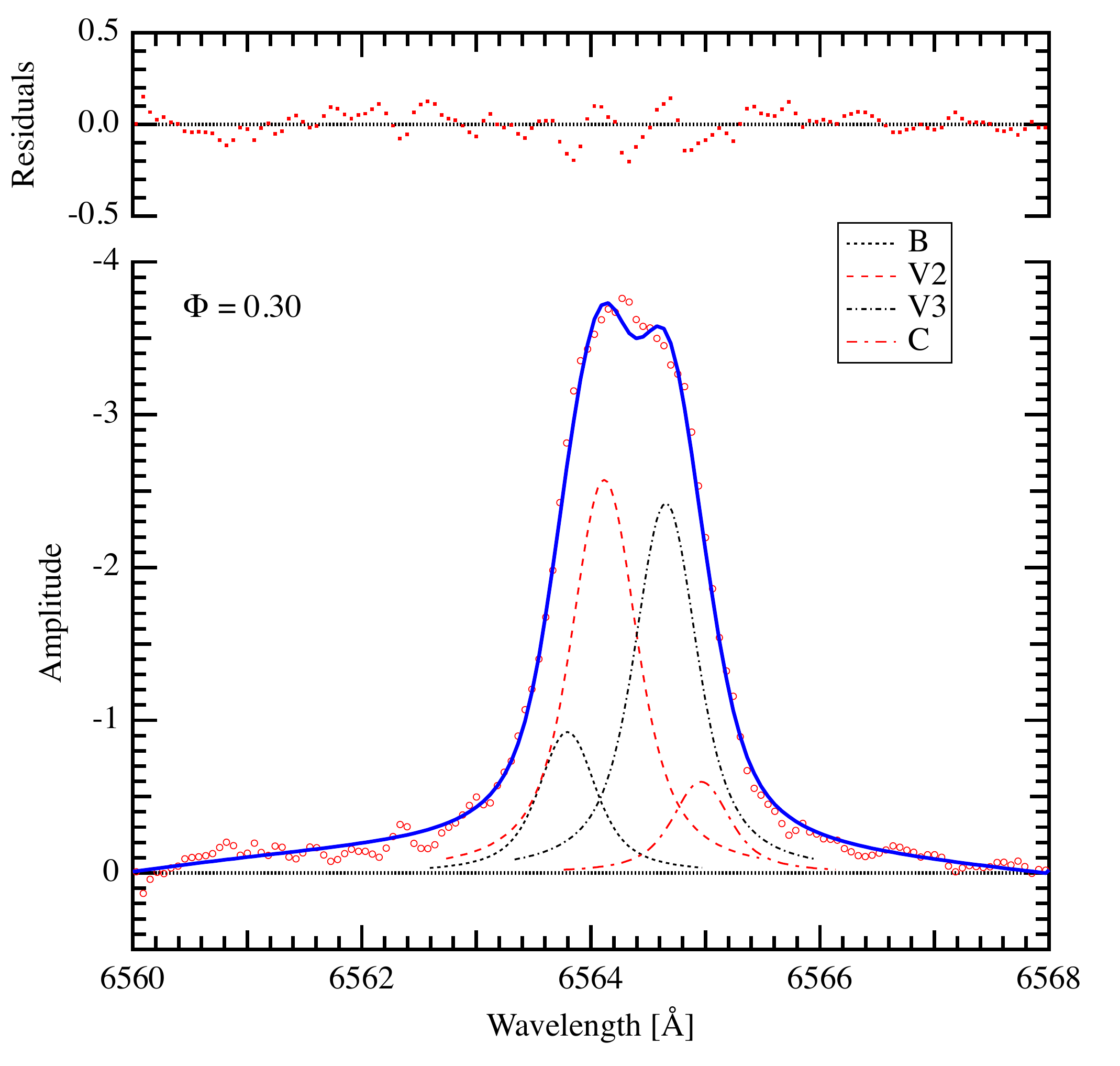}
\caption{An example of extracting component B and C RVs for an observation near syzygy ($\Phi=0.30$) by constraining amplitude and width peak parameters  to values measured for an observation at $\Phi=0.42$, secured approximately two hours later that night. Using Voigt functions, the model solves only for component positions in wavelength.  
}
\label{HaRVprobe}
\end{figure}

\begin{figure}
\includegraphics[width=5in]{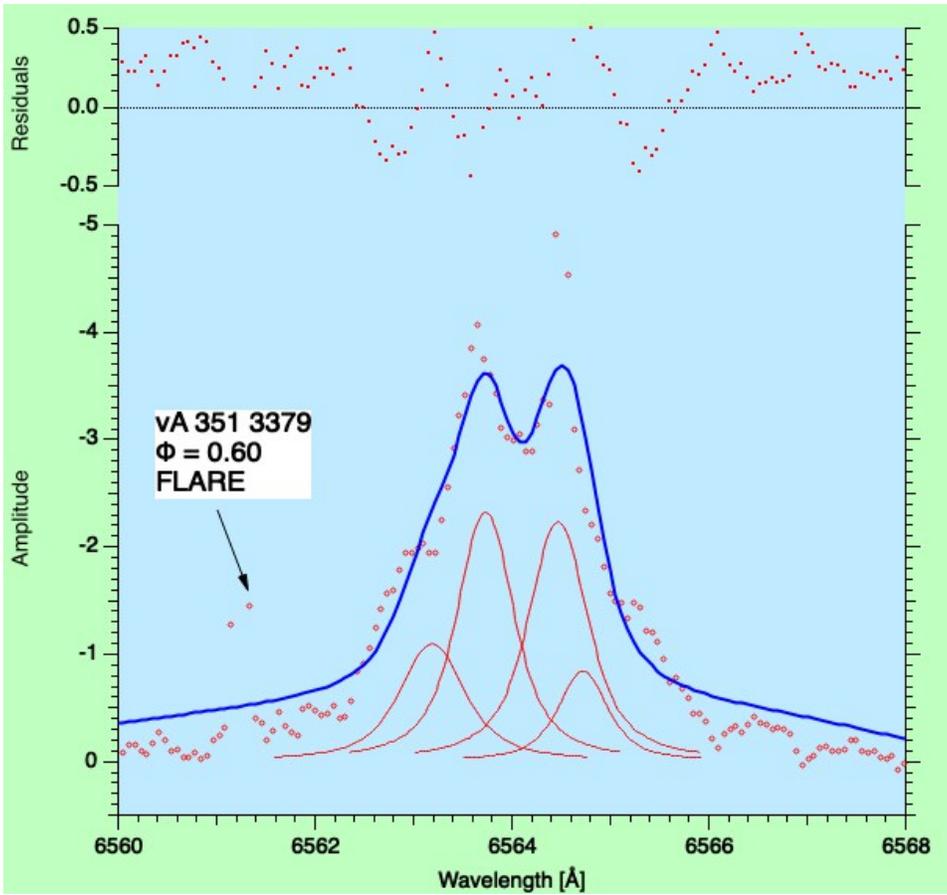}
\caption{An animated version of Figure~\ref{Qew}, showing the behavior of \Has as a function of BC orbital period. This particular frame (observation date mJD=51193.2104) exhibits flaring activity associated with components 2 and 3 identified in Figure~\ref{Qew}. The number next to the vA 351 identification is a running RV observation log number. $\Phi$ indicates BC orbital phase. The animation is constructed from data acquired with the CE, spanning a range $50723<{\rm mJD}<52948$.
}
\label{Disney}
\end{figure}

\begin{figure}
\includegraphics[width=6in]{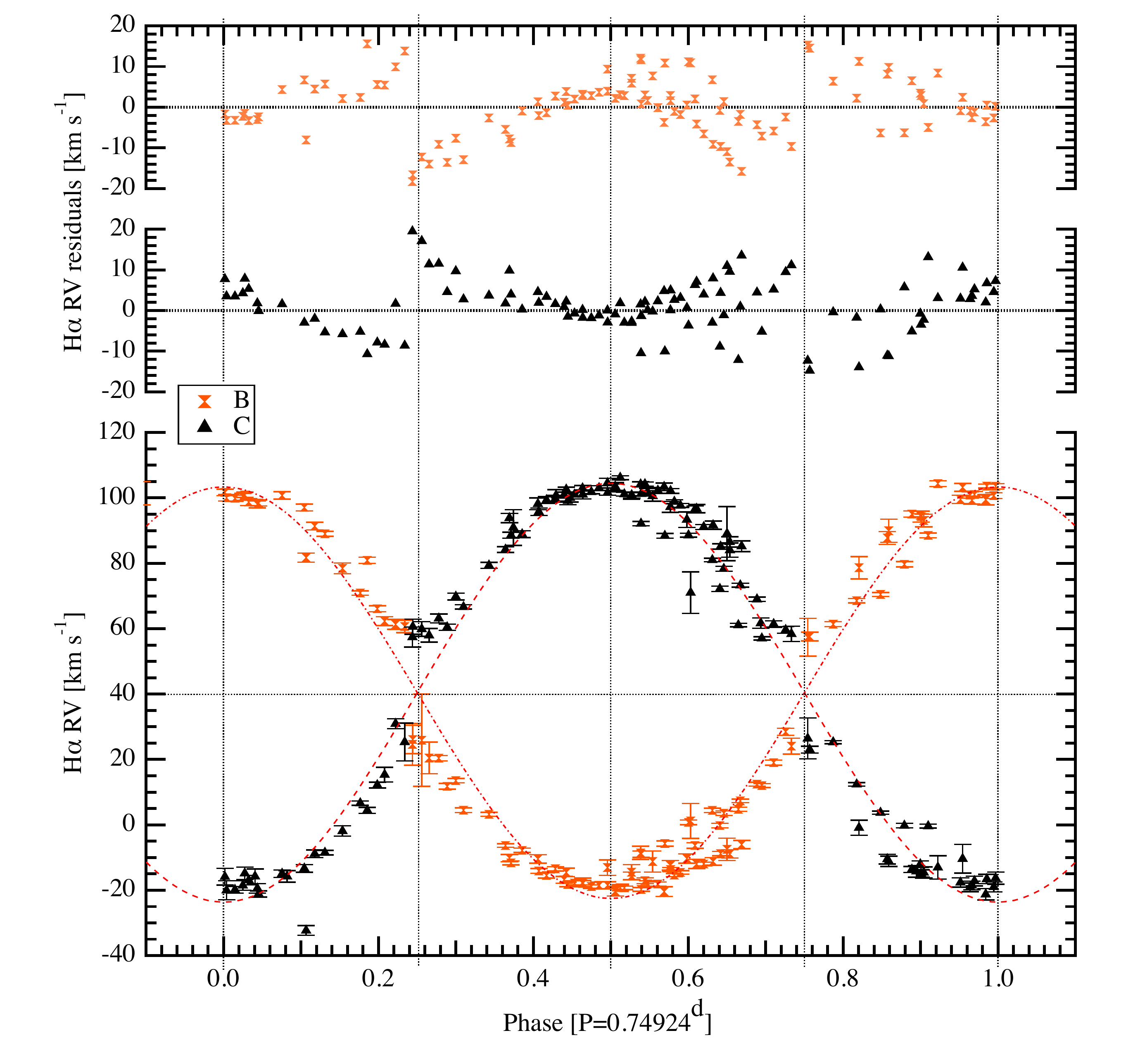}
\caption{\vAs \Has radial velocities for components B and C (peaks 4 and 1 in Figure~\ref{Qew})phased to the period derived from absorption lines (Figure~\ref{RVabs}). 
Vertical lines indicate quadrature (phases $\Phi_{\rm BC}=$0.0, 0.5, 1.0) 
and syzygy (phases $\Phi_{\rm BC}=$0.25, 0.75).  
The amplitude derived from \Has significantly exceeds the amplitude exhibited by absorption lines (Figure~\ref{RVabs}, Table~\ref{tbl-RVO}). 
Residuals indicate unmodeled non-circular motions in  \Ha.}
\label{HaRV}
\end{figure}

\begin{figure}
\includegraphics[width=6.5in]{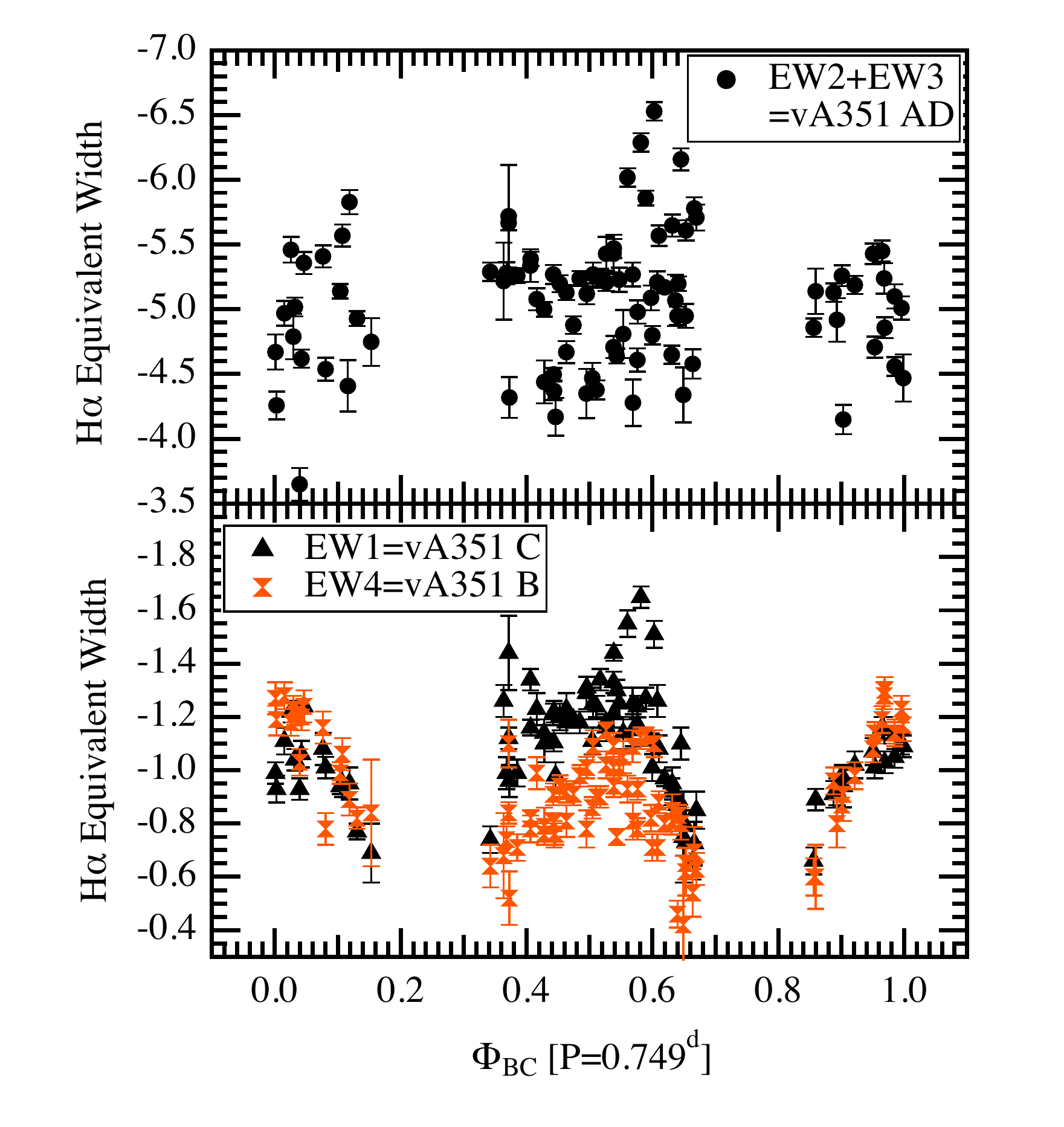}
\caption{\Has equivalent widths for, bottom, components B and C, top, the sum of components 2 and 3, phased  to the orbital period derived from absorption line velocities (Figure~\ref{RVabs}). 
 Values are only for $0.35 \le \Phi_{\rm BC} \le 0.65$,  $0.85 \le \Phi_{\rm BC} \le 1.00$, and $0.00 \le \Phi_{\rm BC} \le 0.15$ to minimize contamination between components. \Has emission from components B and C correlate with the BC period. The sum of components 2 and 3 do not, supporting their association with \vAAs and D.  
 }
\label{BCrules}
\end{figure}

\begin{figure}
\includegraphics[width=6in]{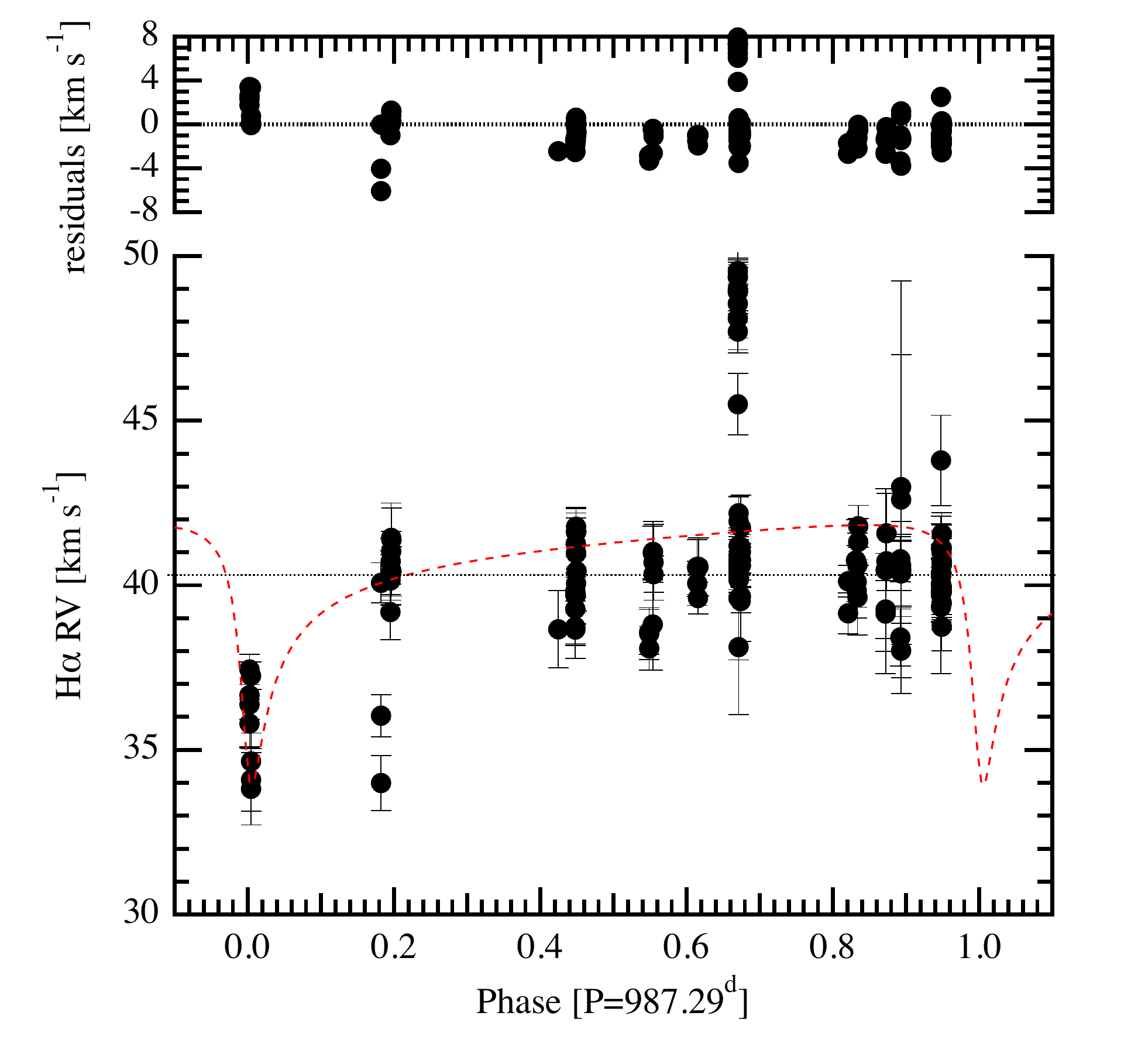}
\caption{\vAs \Has RVs for an average of components 2 and 3 (identified in Figure~\ref{Qew}) phased to the period (Table~\ref{tbl-ORB}) derived from absorption line RVs and the relative orbit. 
The Figure~\ref{fig-WTF?} AD-BC RV orbit is overplotted. While noisy, RVs suggest that components 2 and 3 of the \Has emission have an  association with components AD.}
\label{HaRVAD}
\end{figure}

\begin{figure}
\includegraphics[width=7in]{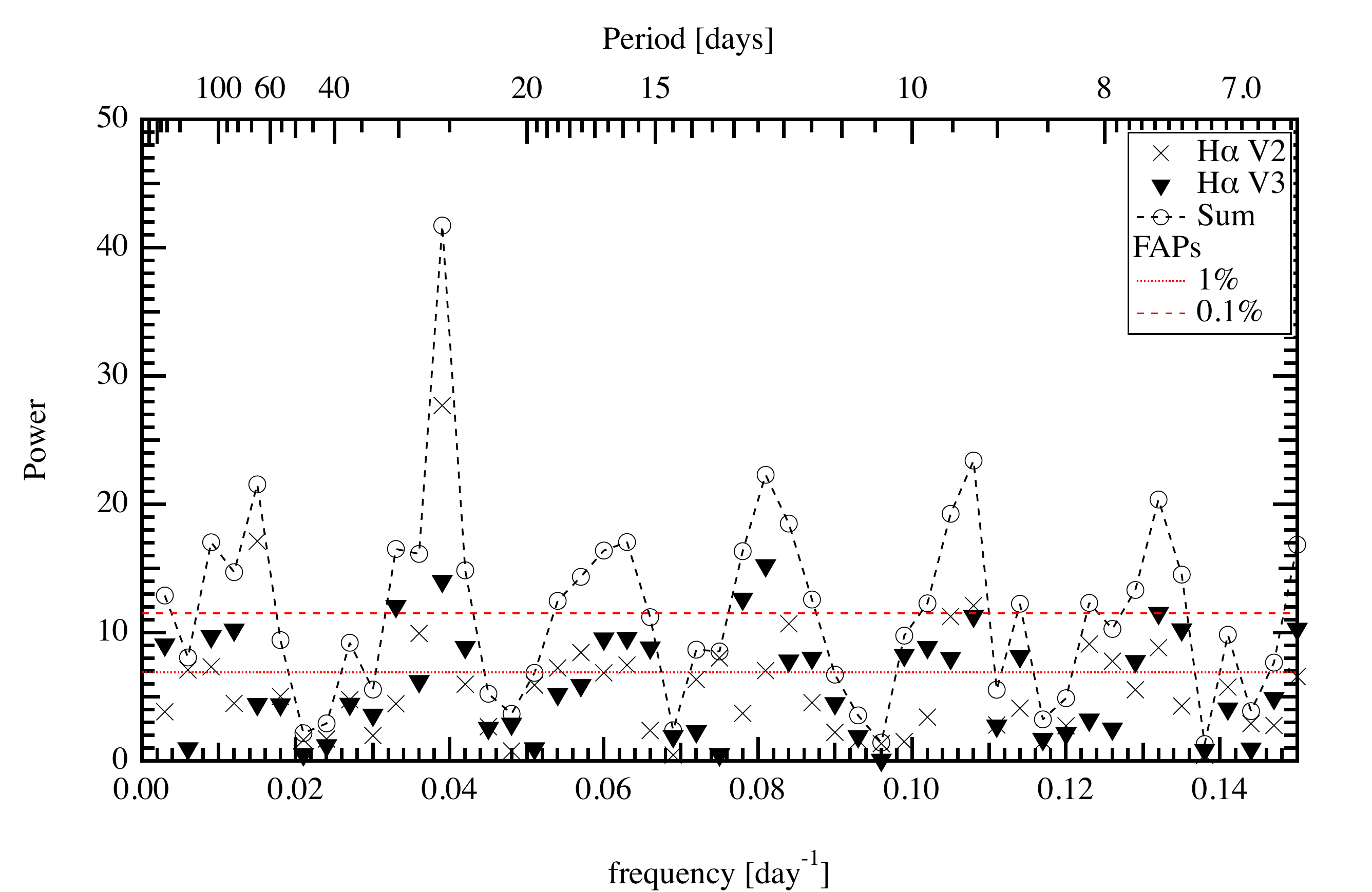}
\caption{Lomb-Scargle periodograms for the Figure~\ref{Qew}  \Has V2 and V3 components and the periodogram sum, providing  significant evidence 
(false-alarm probability $<< 0.1$\%) for activity with $P_{\rm AD} \sim 25^{\rm d}$.  This  provides supporting evidence for  the  reality of the astrometrtric and RV detections 
 at $P_{\rm AD} \sim 21^{\rm d}$ (Figure~\ref{LSRVT}).}
\label{LSRVV2V3}
\end{figure}

\begin{figure}
\includegraphics[width=3in]{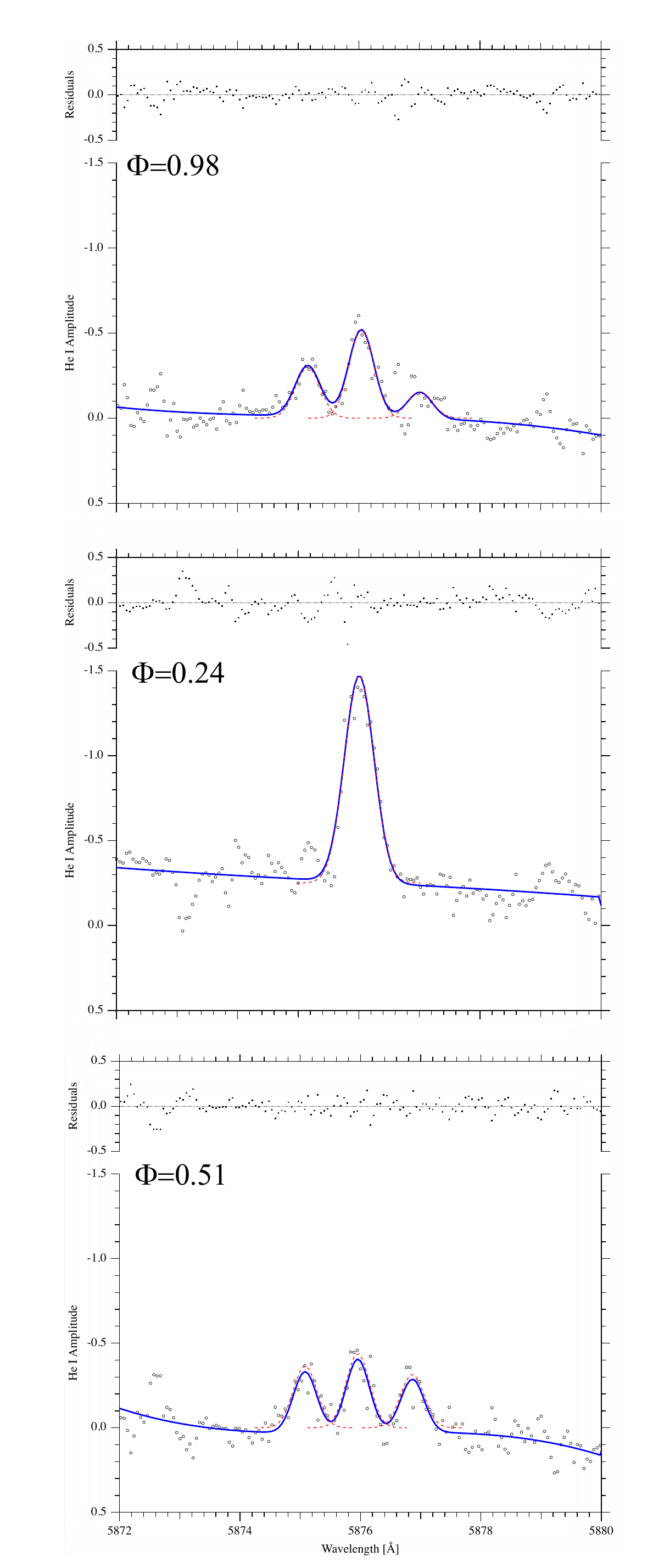}
\caption{Structure in the \Hes emission line ($\lambda 5875$\AA) amplitude ({\it AW}, Equation~\ref{EW}) from bottom to top at $\Phi_{\rm BC}$ =0.51, 0.24, 0.98; mJD=51467.36, 51466.42, 50724.47.  At  phase $\Phi_{\rm BC}$ = 0.51 component C is the right-most peak;
component B the left-most. There is no central peak structure at any phase. Tables~\ref{tbl-HeEWs}  and \ref{tbl-HeIRV} list EW and RVs for these and other observations. The emission is far weaker (factors of 2-6), thus the S/N lower, than for the \Has (Figure~\ref{Qew}) from the same spectra.  The broad Na D absorption line at $\lambda 5889.9$\AA  ~produces the sloping background.}
\label{HeQew}
\end{figure}

\begin{figure}
\includegraphics[width=5in]{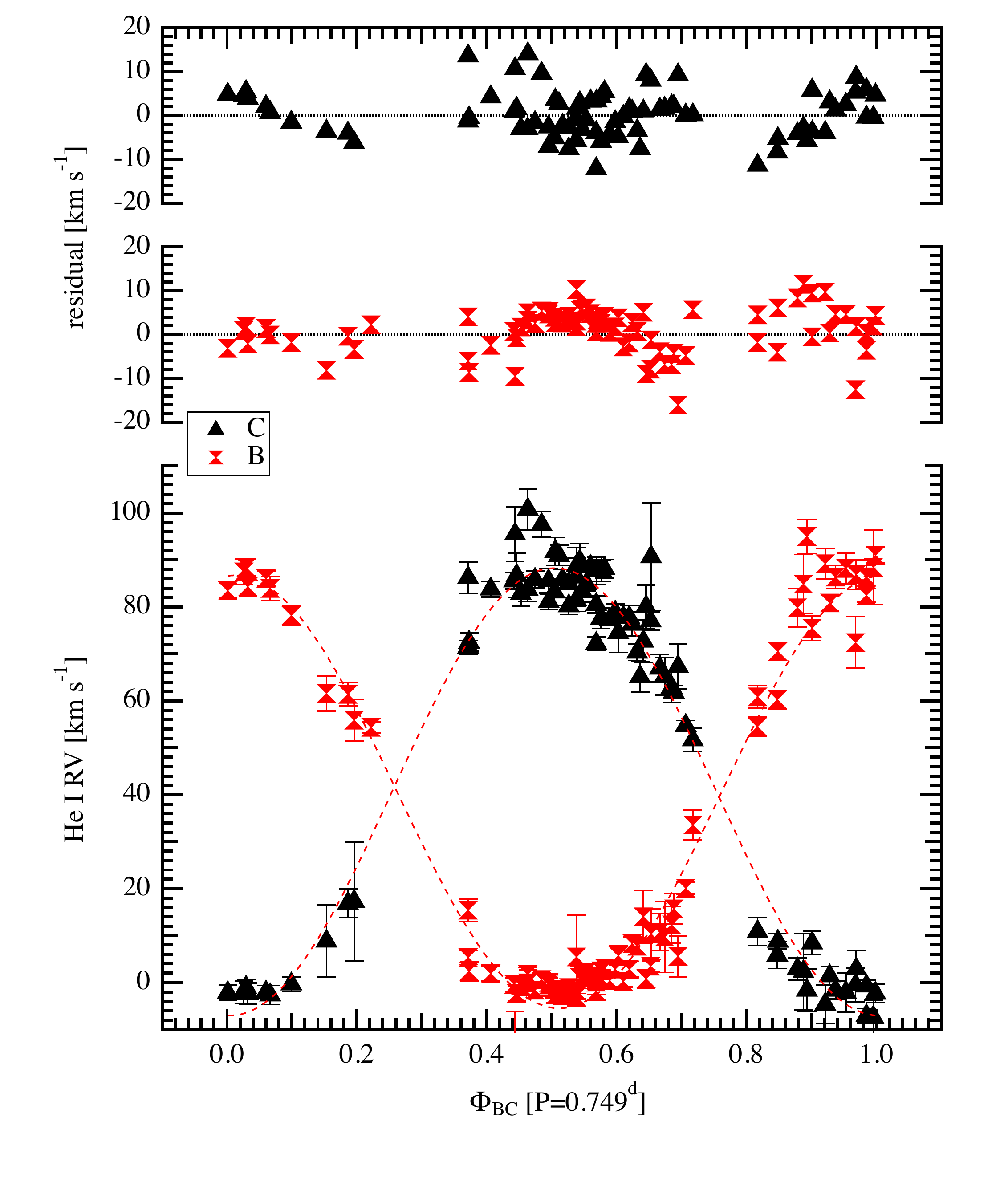}
\caption{\Hes ($\lambda 5875\AA$) emission line radial velocities for components B and C phased to the period derived from absorption lines (Figure~\ref{RVabs}). 
 Emission peak fitting provides all velocities.
Scatter is larger than for absorption or \Has measures. Amplitudes, $K$, and systemic velocity, $\gamma$, (Table~\ref{tbl-RVO}) have values closer to those from absorption line velocities. }
\label{HeRV}
\end{figure}

\begin{figure}
\includegraphics[width=6.5in]{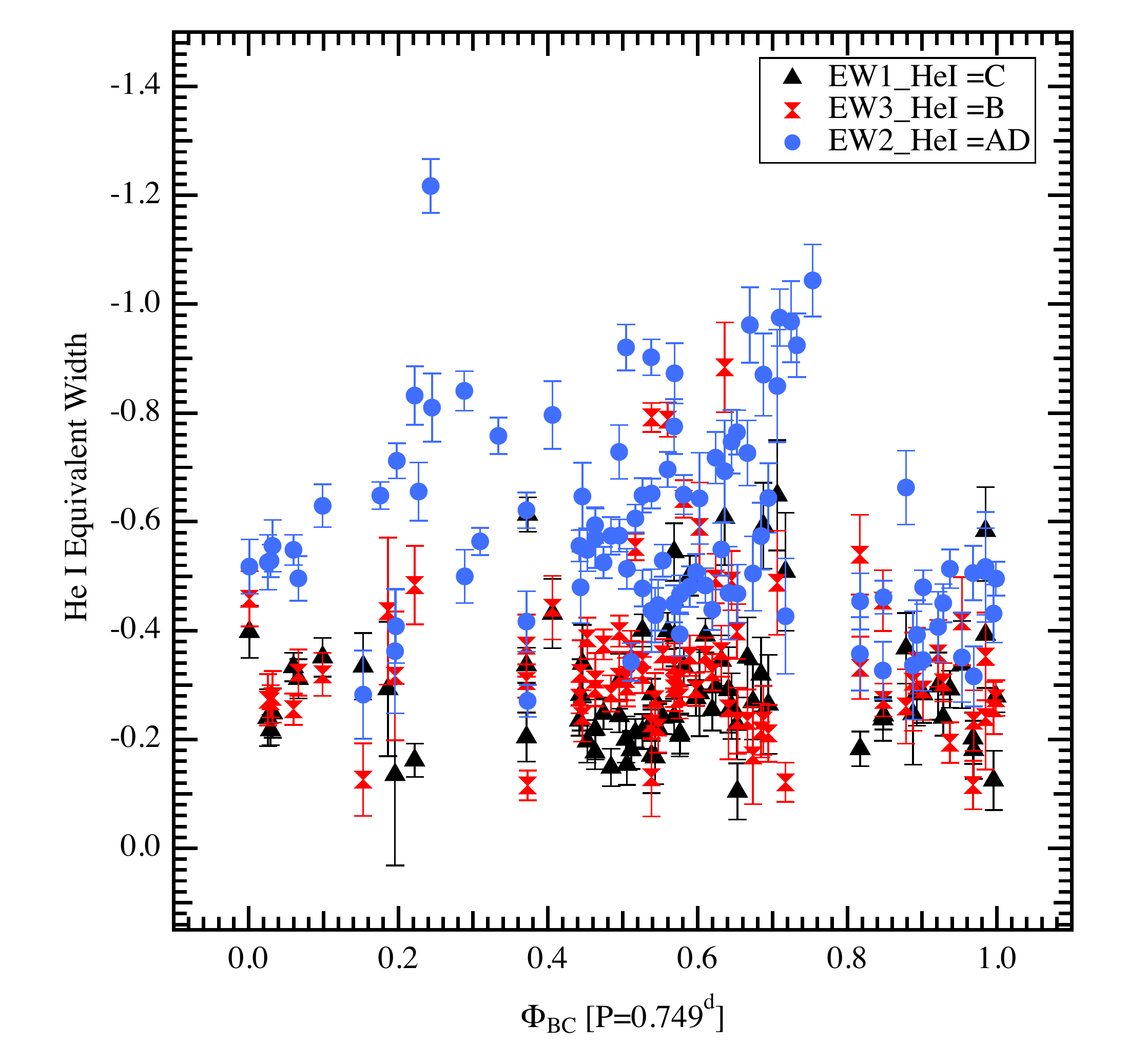}
\caption{\Hes equivalent widths for components A, B and C phased  to the orbital period derived from absorption line velocities (Figure~\ref{RVabs}). 
\Hes emission from components A, B, and C show little correlation  with the BC period. 
 }
\label{HeIABC}
\end{figure}

\begin{figure}
\includegraphics[width=6in]{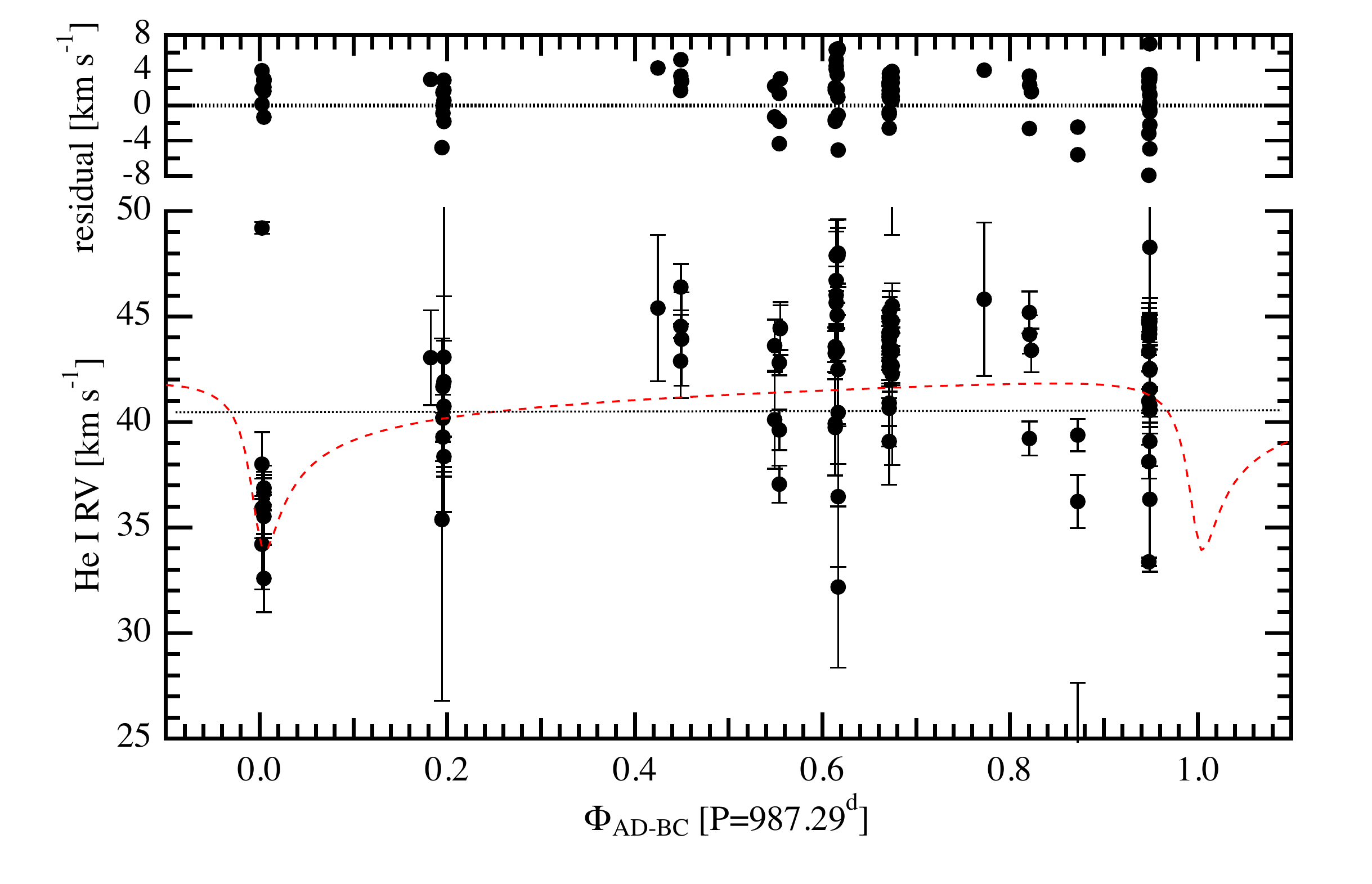}
\caption{\vAs \Hes RVs for the central peak emission (Figure~\ref{HaRVprobe}) phased to the AD-BC period (Table~\ref{tbl-ORB}) derived from absorption line RVs and the relative orbit. 
The Figure~\ref{fig-WTF?} orbit is overplotted. While noisy, velocities suggest that the central component of the \Hes emission is associated with components AD.}
\label{HeIRV}
\end{figure}

\begin{figure}
\includegraphics[width=7in]{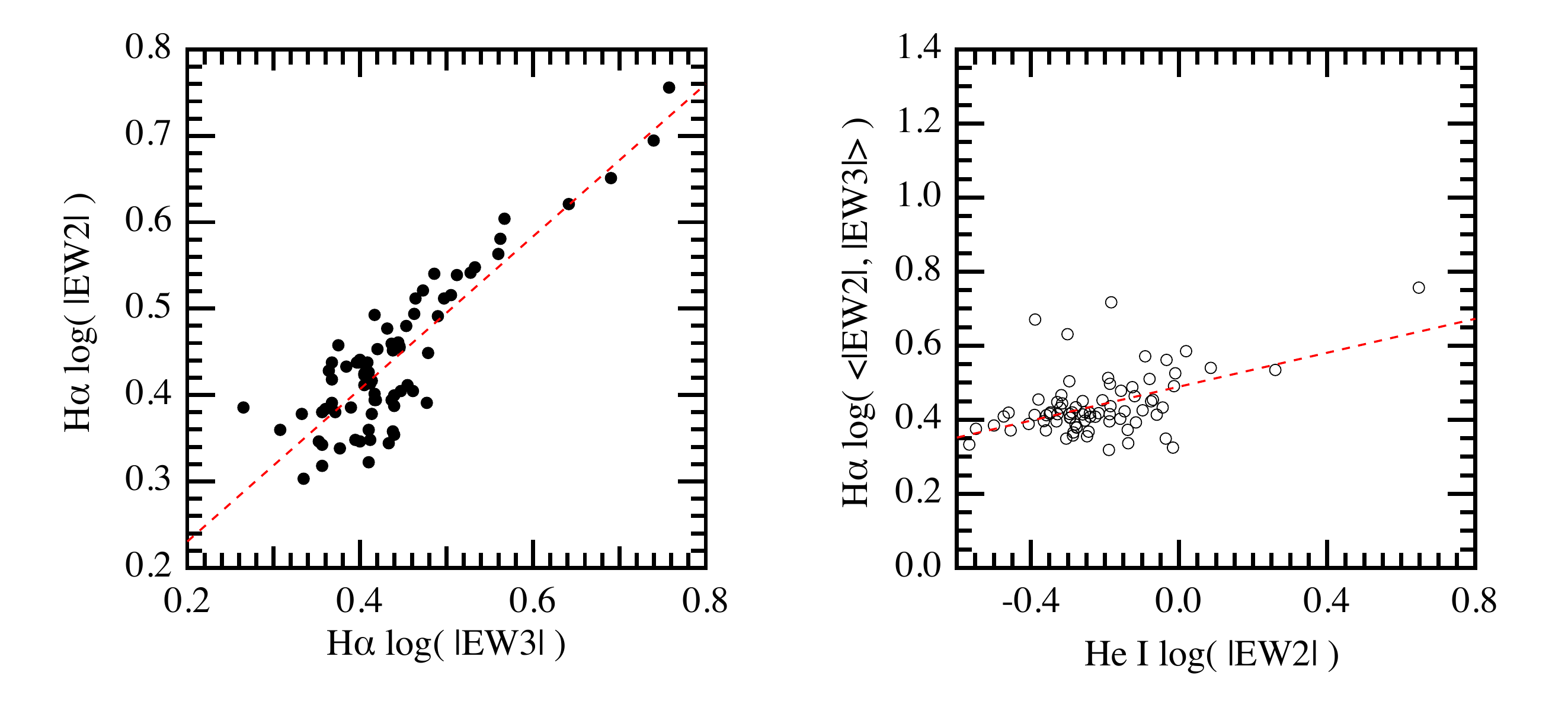}
\caption{Left: log of the absolute values of  EW2 versus EW3 (Table~\ref{tbl-EWs}, peaks 2 and 3 in Figure~\ref{Qew}).The maximum values occur within 4 days of AD-BC orbit periastron. Right: log of the average absolute values of EW2 and EW3 versus the log of the absolute value of the \Hes central peak (EW$_{\rm cen}$, Table~\ref{tbl-HeEWs}; central peak in Figure~\ref{HeQew}). 
The correlation (Pearson's r value, Pr = 0.49) only weakly supports the assertion  that the same energy source powers the \Has and \Hes activity. The strongest values for a few of the \Has EW2, EW3 averages,  but only one  \Hes  epoch, occur within 4 days of AD-BC periastron.
}
\label{HaHeCor}
\end{figure}

\begin{figure}
\includegraphics[width=7in]{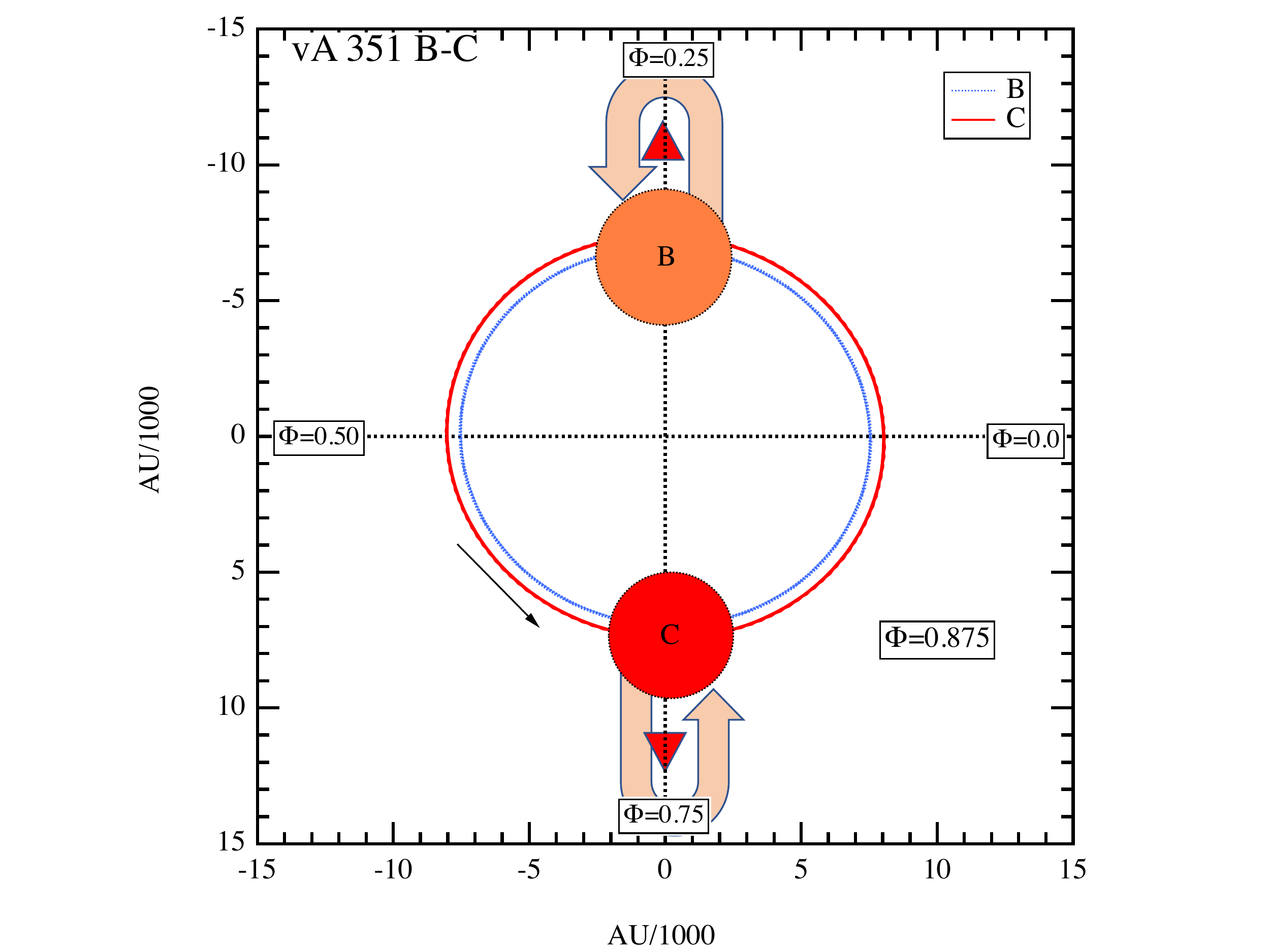}
\caption{
A cartoon of the \vAs BC components, scaled to AU using knowledge of the component masses, absorption line RV amplitudes ($K_{\rm B}$ and $K_{\rm C}$), the system parallax, and Kepler's Laws. We adopt component stellar sizes from the \cite{Boy12} mass-radius relation for eclipsing binaries. The fatter colored tracks represent coronal loop-like flows that could explain the \Has RV residual patterns in Figure~\ref{HaRV}.  The arrows are notional, not to scale, only indicating flow direction. 
\Has emission  takes place in the darker triangles, parts of the flow most shaded from the other component. Those locations are to scale. Phases, $\Phi$, match  Figures~\ref{RVabs} and \ref{HaRV}. The \Hes  $K$ values (Figure~\ref{HeIRV}, Table~\ref{tbl-RVO}) match those from absorption lines, so \Hes emission is likely produced nearer the stellar surface. The \Hes RV residuals (Figure~\ref{HeRV}), while far noisier, suggest a  flow pattern similar to that of \Ha. 
}
\label{LooneyTunes2}
\end{figure}

\begin{figure}
\includegraphics[width=6.5in]{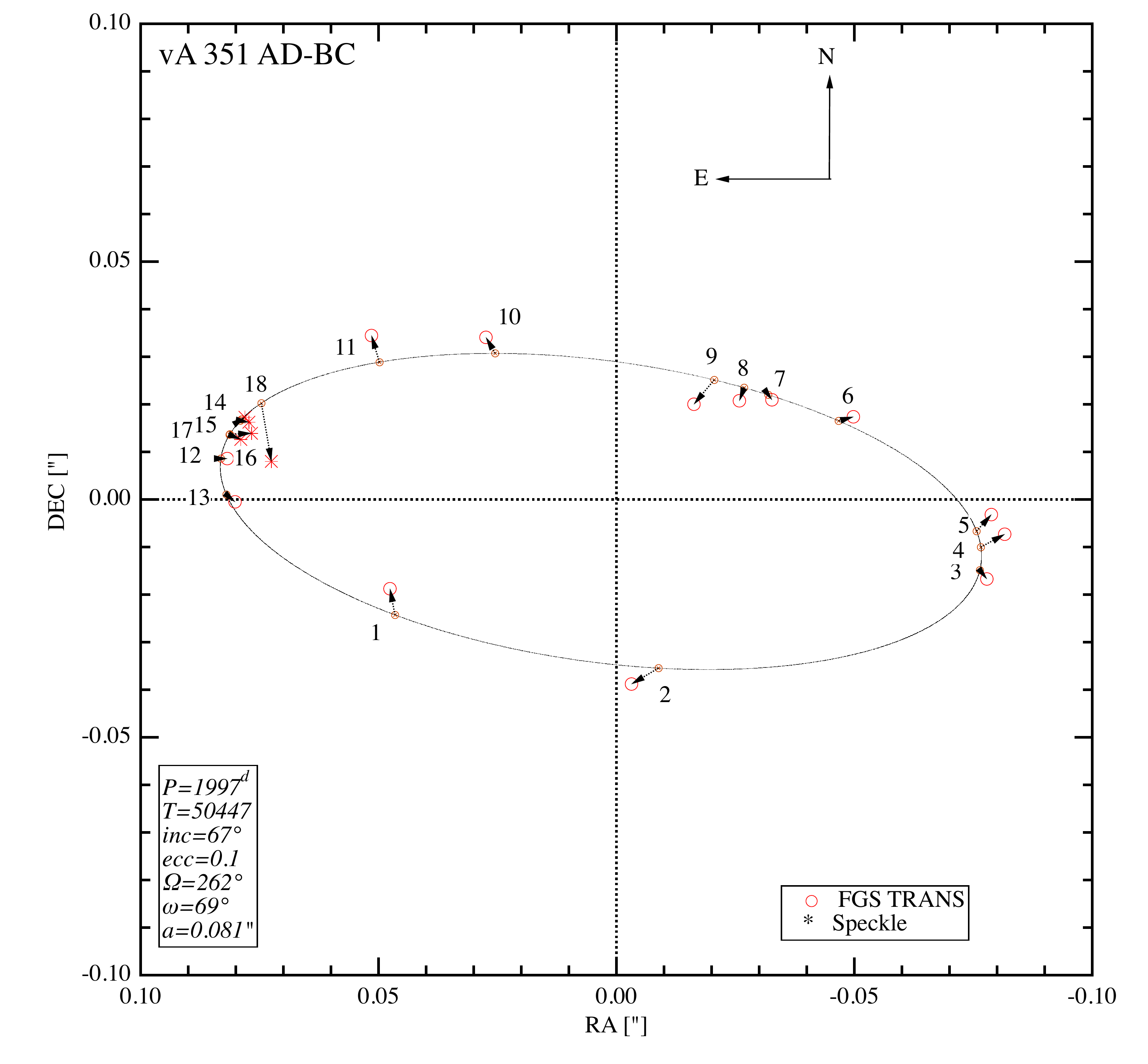}
\caption{ Misidentification of nearly equal and variable brightness  components result in this relative orbit AD-BC for \vAs  derived from FGS TRANS and ground-based speckle observations. Compare with Figure~\ref{Rel}. }
\label{BO}
\end{figure}

For an embarrassingly long period of time, we worked with an AD-BC astrometric orbit solution (Figure~\ref{BO}) that produced severe cognitive dissonance, to the point that two of us (LP, OGF) strongly, and correctly objected to paper submission.
The astrometry and the RVs were inconsistent with each other. As seen in Figure~\ref{fig-WTF?}, the RVs were basically flat line except for a very few epochs. In retrospect this is easily explained by the highly eccentric AD-BC orbit. Even though one of us (GFB) checked and rechecked the RVs, they refused to make sense, given our acceptance of the Figure~\ref{BO} orbit. However, one of us (EH) explored the consequences of component misidentification (for many epochs swapping AD for BC, essentially changing the measured position angle by 180$\arcdeg$) and proposed the Figure~\ref{Rel} orbit, having an orbital period half that previously determined, which the  RVs (Figure~\ref{fig-WTF?}) immediately confirmed (GT). 

Take care with component identification, particularly for near equal brightness systems with component brightness variability. Always explore alternate possibilities. Believe  rechecked RVs. If possible, always appeal to more than one observational technique. RVs and astrometry together made the proper orbit identification possible. Intriguingly, the final component masses obtained from analyses based on either orbit agree within their respective errors. But, Figure~\ref{Rel} and Table~\ref{tbl-RIP} describe the real \vAs system.
  
\bibliography{/Active/myMaster}

\clearpage

\begin{deluxetable}{l r l l r r r r r }
\tablecaption{\vAs Component  B Relative to Component A Separations, Position Angles, Residuals, $\Delta m$, and Sources\label{tbl-TR}}
\tablewidth{0in}
\tablehead{ 
\colhead{}
& \colhead{ mJD }
&  \colhead{ $\rho$}
& \colhead{$\delta \rho$\tablenotemark{a}}
&  \colhead{ $\theta$}
& \colhead{$\delta \theta$}
& \colhead{$\rho \delta \theta$}
& \colhead{$\Delta$m\tablenotemark{b}}
& \colhead{Sources}\\
&&(mas)&(mas)&(deg)&(deg)&(mas)&(mag)
}
\startdata
1&49396.6222&51.2&0.0&291.50&0.41&0.37&-0.04&\FGSns\,3\\
2\tablenotemark{c}&49396.6222&38.9&10.0&184.61&-24.86&-18.02&-0.04&\FGSns\,3\\
3&50082.4521&79.6&-0.2&257.89&-4.51&-6.27&0.06&\FGSns\,3\\
4&50128.8754&81.9&2.9&264.88&-0.67&-0.96&0.08&\FGSns\,3\\
5&50160.0312&78.8&1.2&267.69&-0.17&-0.24&-0.04&\FGSns\,3\\
6&50375.0539&52.7&-0.3&289.28&-0.26&-0.24&0.01&\FGSns\,3\\
7&50437.9500&38.8&-0.1&302.69&-0.22&-0.15&0.00&\FGSns\,3\\
8&50457.8196&33.1&-0.3&308.82&-0.91&-0.53&-0.08&\FGSns\,3\\
9&50481.6339&25.8&0.0&320.95&-1.02&-0.46&-0.54&\FGSns\,3\\
10&50650.4159&43.8&1.2&218.85&-0.96&-0.74&0.13&\FGSns\,3\\
11&50751.4075&62.0&0.6&236.22&-0.38&-0.41&0.00&\FGSns\,3\\
12&51033.4170&82.3&2.1&263.99&3.99&5.75&0.02&\FGSns\,3\\
13&51111.4344&80.2&1.6&270.35&4.96&6.96&&\FGSns\,3\\
14&56933.5560&80.0&-0.4&257.52&-0.41&-0.58&-0.25&speckle\\
15&56936.5145&79.0&-1.4&258.12&-0.01&-0.02&0.00&speckle\\
16&56969.4600&78.0&-2.5&259.72&-0.72&-0.99&-0.01&speckle\\
17&56970.4462&80.0&-0.6&260.92&0.46&0.64&-0.03&speckle\\
18\tablenotemark{c}&58888.1125&73.0&-5.6&263.70&8.61&11.06&-0.04&speckle\\
\hline
&&&&&&&0.06&CfA spectra
\enddata
\tablenotetext{a}{All $\delta$ values are observed minus  calculated (O-C) from the Table~\ref{tbl-TSORB} orbit.}
\tablenotetext{b}{Magnitude difference between components AD and BC. For \FGSns\,3 $\Delta F583W \simeq \Delta V$ \citep{Hen99}.
For speckle, $\Delta\lambda 692$nm \citep{Hor12}. For CfA spectra see Section~\ref{WTF?}. Observations \#9 and \#14 detect flaring activity from one of the M dwarf components. }
\tablenotetext{c}{Observations omitted from final Table~\ref{tbl-TSORB} orbit.} 
\end{deluxetable}

\begin{deluxetable}{clrl}
\tablecaption{Parameters for \vAs AD-BC Relative Orbit \label{tbl-TSORB}}
\tablewidth{3in}
\tablehead
{
\colhead{Parameter} &  
\colhead{Units} &
\colhead{Value} &
\colhead{err} 
}
\startdata
P&days&987.9&1.4\\
P&years&2.705&0.004\\
$T_0$&mJD&51522.4&2.5\\
ecc&&0.790&0.006\\
$\omega_{\rm AD}$&$\arcdeg$&149&13\\
$a$&mas&45.4&0.6\\
inc&$\arcdeg$&14&8\\
$\Omega$&$\arcdeg$&289.5&0.9
\enddata
\end{deluxetable}

\begin{deluxetable*}{l r l r l r l r r }
\tablecaption{CE Absorption Line Radial Velocities\tablenotemark{a} \label{tbl-CE}}
\tablewidth{6in}
\tablehead{ \colhead{mJD\tablenotemark{b}}&
\colhead{VA} &
\colhead{VAerr}&
\colhead{VB}&
\colhead{VBerr} &
\colhead{VC}&
\colhead{VCerr} &
\colhead{$\Phi$(BC)\tablenotemark{c}}&
\colhead{$\Phi$(AD-BC)\tablenotemark{d}}
}
\startdata
49962.4226&41.16&0.17&75.33&0.44&4.51&0.44&0.6424&0.7536\\
50090.1532&40.92&0.14&10.18&0.32&71.84&0.4&0.1227&0.8175\\
50090.2080&40.67&0.17&-1.6&0.39&83.71&0.54&0.1958&0.8175\\
50090.2517&41.02&0.26&-3.73&0.45&86.77&0.68&0.2541&0.8176\\
.&.&.&.&.&.&.&.\\
\enddata
\tablenotetext{a}{Full table available electronically. All velocity units in \kms.}
\tablenotetext{b}{mJD=JD-2400000.5}
\tablenotetext{c}{BC orbit phase}
\tablenotetext{d}{AD-BC orbit phase}
\end{deluxetable*}

\clearpage
\begin{deluxetable*}{l r l r l r l r r}
\tablecaption{CfA Absorption Line Radial Velocities\tablenotemark{a} \label{tbl-CfA}}
\tablewidth{6in}
\tablehead{ \colhead{mJD\tablenotemark{b}}&
\colhead{VA} &
\colhead{VAerr}&
\colhead{VB}&
\colhead{VBerr} &
\colhead{VC}&
\colhead{VCerr} &
\colhead{$\Phi$(BC)\tablenotemark{c}} &
\colhead{$\Phi$(AD-BC)\tablenotemark{d}}
}
\startdata
46428.1524&41.63&1.72&50.69&3.2&30.28&3.72&0.5008&0.9848\\
46449.2532&40.4&1.68&81.07&3.12&-7.05&3.63&0.6638&0.9953\\
47436.3954&40.47&2.11&-5.45&3.92&82.31&4.56&0.1888&0.4894\\
47436.4072&41.43&2.03&-6.53&3.77&88.74&4.38&0.2046&0.4894\\
48281.1881&42.37&1.95&86.51&3.63&-8.87&4.22&0.7219&0.9122\\
.&.&.&.&.&.&.&.\\
\enddata
\tablenotetext{a}{Full table available electronically. All velocity units in \kms.}
\tablenotetext{b}{mJD=JD-2400000.5}
\tablenotetext{c}{BC orbit phase}
\tablenotetext{d}{AD-BC orbit phase}
\end{deluxetable*}
\begin{deluxetable*}{l r l r l r l r r }
\tablecaption{Other Absorption Line Radial Velocities \label{tbl-oRV}}
\tablewidth{6in}
\tablehead{ \colhead{mJD\tablenotemark{a}}&
\colhead{VA} &
\colhead{VAerr}&
\colhead{VB}&
\colhead{VBerr} &
\colhead{VC}&
\colhead{VCerr} &
\colhead{$\Phi$(BC)\tablenotemark{b}}&
\colhead{$\Phi$(AD-BC)\tablenotemark{c}}
}
\startdata
57441.1134\tablenotemark{d}&33&1.5&2.4&1.5&88.4&1.5&0.3476&0.4965\\
57443.0777\tablenotemark{d}&29.2&1.5&61.6&1.5&29.2&1.5&0.9694&0.4975\\
58584.1012\tablenotemark{d}&37.65&1.39&75.54&0.53&0.51&0.9&0.8777&0.0924\\
49672.3000\tablenotemark{e}&37&1&22&1&59&1&0.4198&0.6084\\
49732.0700\tablenotemark{e}&38&1&-1&1&82&1&0.1940&0.6383\\
46388.5\tablenotemark{f}&41.4&1.5&&&&&0.5759&0.9649
\enddata
\tablenotetext{a}{mJD=JD-2400000.5}
\tablenotetext{b}{BC orbit phase}
\tablenotetext{c}{AD-BC orbit phase}
\tablenotetext{d}{from IGRINS}
\tablenotetext{e}{from \cite{Sta97}, errors assumed}
\tablenotetext{f}{from \cite{Har87}}

\end{deluxetable*}
\begin{deluxetable}{l r l r l}
\tablewidth{3in}
\tablecaption{RV BC Orbit\label{tbl-RVO}}
\tablehead{
\colhead{Parameter}
&\colhead{B}
&\colhead{err}
&\colhead{C}
&\colhead{err}
}
\startdata
\underline{Absorption Lines}\tablenotemark{a} &&\\ 
P [days] &0.7492425&0.0000003&&\\
T$_0$ [mJD]&50367.09409 &0.0005&&\\
K [\kms]&43.76&0.15&46.29&0.17\\
ecc &0 &0.002&&\\
$\gamma$ [\kms]&40.2& 0.2&&\\
$\omega [\arcdeg]$&-&-&&\\
Derived &&&&\\
K$_{\rm B}$/K$_{\rm C}$ = \m$_{\rm C}$/\m$_{\rm B}$&0.945 &0.005&&\\
\underline{\Has Emission Lines}\tablenotemark{b}&&&&\\ 
K [\kms]&62.93 &0.64&64.05 &0.69\\
ecc &0 &0.002&&\\
$\omega [\arcdeg]$&-&-&-&-\\
$\gamma$ [\kms]&40.45& 0.40&&\\
Derived &&&& \\
K$_{\rm B}$/K$_{\rm C}$ = \m$_{\rm C}$/\m$_{\rm B}$&0.982 &0.015&&\\
\underline{\Hes Emission Lines}\tablenotemark{c}&&&&\\ 
K [\kms]&44.41 &0.38&45.15 &0.37\\
$\gamma$ [\kms]&41.1 &0.5&&\\
Derived &&&&\\
K$_{\rm B}$/K$_{\rm C}$ = \m$_{\rm C}$/\m$_{\rm B}$&0.984 &0.011&&\\
\enddata
\tablenotetext{a}{Section~\ref{spec}}
\tablenotetext{b}{Section~\ref{specHa}}
\tablenotetext{c}{Section~\ref{specHe}}
\end{deluxetable}
\begin{deluxetable}{clrl}
\tablecaption{RV Orbital Elements for \vAs AD-BC \label{tbl-RVORB}}
\tablewidth{4in}
\tablehead
{
\colhead{Parameter} &  
\colhead{Units} &
\colhead{Value} &
\colhead{err} 
}
\startdata
$P$&days&987.3&1.8\\
$P$&years&2.703&0.005\\
$T_0$&JD-2400000&51518&8\\
$e$&-&0.81&0.01\\
$\omega_{\rm AD}$&$\arcdeg$&149.2&2.8\\
$K_{\rm A}$&\kms&3.98&0.11\\
$\gamma_{\rm AD-BC}$&\kms&40.64&0.05\\
\enddata
\end{deluxetable}

\begin{deluxetable}{lcccccccc}
\tablecaption{\vAs Reference Star \Gs Input and Final Parallaxes\tablenotemark{a} \label{tbl-pis}}
\tablewidth{6in}
\tablehead{\colhead{Ref Star \#}
 & \colhead{$V$}
  & \colhead{$B-V$}
  &  \colhead{ DR2 Source }
 & \colhead{$G$}
&  \colhead{$\varpi$ }
 &  \colhead{err}
 & \colhead{Final $\varpi$}
 & \colhead{err }
 }
\scriptsize
\startdata
2&11.79&0.73&3313958012105540000&11.79&4.74&0.06&4.73&0.03\\
3&13.88&0.98&3313956534636790000&13.84&2.17&0.04&2.17&0.02\\
4&11.84&1.64&3313957423693360000&11.56&0.23&0.07&0.23&0.04\\
5&14.37&1.24&3313958252623700000&14.31&0.23&0.04&0.22&0.02
\enddata
\tablenotetext{a}{Parallax and errors in mas.}
\end{deluxetable}

\begin{deluxetable}{ll}
\tablecaption{Reference Frame Statistics, \vAs Parallax, and Proper Motion\label{tbl-SUM}}
\tablewidth{0in}
\tablehead{\colhead{Parameter} &  \colhead{Value} }
\startdata
TRANS+speckle time span  &26.99 y  \\
number of observation sets    &   18  \\
POS mode time span & 1.8 y \\
number of observation sets & 7 \\
POS reference star $\langle V\rangle$ &  13.31     \\
POS reference star $\langle (B-V) \rangle$ &1.14   \\
{\it HST}~Absolute $\varpi$& 18.37 $\pm$ 0.65    mas \\
~~~~~~~Relative  $\mu_\alpha$& 108.7 $\pm$ 0.9 mas yr$^{-1}$\\
~~~~~~~Relative  $\mu_\delta$&  -21.9 $\pm$ 1.3  mas yr$^{-1}$\\
~~~~~~~$\vec{\mu} = 110.9$ mas  yr$^{-1}$\\
~~~~~~~P.A. = $101\fdg4$\\
\G~DR2 Absolute $\varpi$& 20.10 $\pm$ 0.15     mas \\
~~~~~~~Absolute  $\mu_\alpha$& 114.77 $\pm$ 0.31 mas yr$^{-1}$\\
~~~~~~~Absolute  $\mu_\delta$&  -25.28 $\pm$ 0.20  mas yr$^{-1}$\\
~~~~~~~$\vec{\mu} = 117.52$ mas  yr$^{-1}$\\
~~~~~~~P.A. = $102\fdg4$\\
\G~EDR3 Absolute $\varpi$& 19.31 $\pm$ 0.19     mas \\
~~~~~~~Absolute  $\mu_\alpha$& 107.26 $\pm$ 0.21 mas yr$^{-1}$\\
~~~~~~~Absolute  $\mu_\delta$&  -27.19 $\pm$ 0.20  mas yr$^{-1}$\\
~~~~~~~$\vec{\mu} = 110.65$ mas  yr$^{-1}$\\
~~~~~~~P.A. = $104\fdg2$\\
\enddata
\end{deluxetable}

\begin{deluxetable}{clrl}
\tablecaption{Orbital Elements for \vAs AD-BC \label{tbl-ORB}}
\tablewidth{0in}
\tablehead
{
\colhead{Parameter} &  
\colhead{Units} &
\colhead{Value} &
\colhead{err} 
}
\startdata
$P$&days&988.0&1.5\\
$P$&years&2.705&0.004\\
$T_0$&JD-2400000&51522&3\\
$e$&-&0.790&0.007\\
$\omega$&$\arcdeg$&149&12\\
$i$&$\arcdeg$&14&8\\
$\Omega$\tablenotemark{a}&$\arcdeg$&289.5&1.0\\
$\alpha_{\rm A}$&mas&20.1&0.9\\
a&mas&45.4&0.6\\
\hline&&&\\
Derived Parameters&&&\\
${\cal M}_{tot} $ &\msun & 2.06& 0.24\\
f\tablenotemark{b}&-&0.443&0.022\\
$\alpha_{\rm A}$&AU&1.09&0.07\\
a&AU&2.47&0.09\\
\m$_{\rm AD}$&\msun&1.14&0.14\\
\m$_{\rm BC}$&\msun&0.91&0.11\\
\enddata
\tablenotetext{a}{Equinox 2000.0}
\tablenotetext{b}{Mass fraction from Equation~\ref{frac}.}
\end{deluxetable}


\begin{deluxetable*}{c r c c c c}
\tablecaption{Dissecting \vA \label{tbl-RIP}}
\tablewidth{7in}
\tablehead{\colhead{{\bf line \#}}
& \colhead{ }
&  \colhead{ }
&  \colhead{ }
&  \colhead{ }
&  \colhead{}
}
\startdata
\\
&&{\bf Input Data from Photometry}&&&\\
&&$V$&$K$&$V-K$&\\
{\bf 1}&total apparent magnitude&13.27$\pm$0.03&8.268$\pm$0.029&5.00$\pm$0.04&\\
{\bf 2}&total intensity&0.0492$\pm$0.0012&4.929$\pm$0.017&&\\
{\bf 3}&AD - BC $\Delta$ mag&0$\pm$0.1&0$\pm$0.03&&\\
{\bf 4}&estimated absorption, $A_{V,K}$\tablenotemark{a}&0.0$\pm$0.05&0.0$\pm$0.05&&\\
\hline
\\
&&{\bf Input Data from RVs}&&&\\
{\bf 5}&\mC/\mB\tablenotemark{b}&0.945$\pm$0.005&&&\\
\hline
\\
&&{\bf Input Data from Astrometry}&&&\\
{\bf 6}&\mB + \mC&0.91$\pm$0.11\msune&&&\\
{\bf 7}&(\mA+\mD)/(\mB+\mC)&1.26$\pm$0.22&&&\\
{\bf 8}&\mA+\mD&1.14$\pm$0.14\msune&&&\\
{\bf 9}&$\varpi_{\rm abs}$&18.4$\pm$0.7 mas&&&\\
{\bf 10}&\m$_{\rm tot}$&2.06$\pm$0.24\msune&&&\\
\hline
\\
&&{\bf Parameters}\tablenotemark{c}&&&\\
&Component Intensites&$V$&&$K$&\\
{\bf 11}&A&0.02300&&2.45858&\\
{\bf 12}&B&0.01363&&1.30917&\\
{\bf 13}&C&0.01107&&1.15657&\\
{\bf 14}&D&0.00164&&0.00519&\\
&{System Parallax}&&&&\\
{\bf 15}&$\varpi_{\rm abs}$&18.15 mas&&18.34 mas&\\
&{System Absorption}&&&&\\
{\bf 16}&$A_{V,K}$&0.04 mag&&0.00 mag&\\\hline
\\
&&{\bf Final Values and Residuals} \tablenotemark{d}&&&\\
&&$V$&res&$K$&res\\
{\bf 17}&total intensity&0.0493&0.00001&4.929&+0.00004\\
{\bf 18}&AD - BC $\Delta$ mag&0.0&-0.002&0.00&0.001\\
{\bf 19}&\mC/\mB&0.945&0.000&0.9454&0.000\\
{\bf 20}&\mB + \mC&0.78\msune&-0.13&0.90\msune&-0.014\\
{\bf 21}&(\mA+\mD)/(\mB+\mC)&1.28&0.02&1.27&0.02\\
{\bf 22}&\mA+\mD&1.00\msune&-0.15&1.14\msune&-0.002\\
{\bf 23}&$\varpi_{\rm abs}$&18.15&0.2&18.34&0.04\\
{\bf 24}&\m$_{\rm tot}$&1.78\msune&-0.28&2.04\msune&-0.01\\
{\bf 25}&$\sum({\rm resid}^2)$&&4.227&&0.029\\
\hline
\\
&&{\bf Final Magnitudes}&&&\\
&component&$V$&$K$&&\\
{\bf 26}&A&14.09&9.02&&\\
{\bf 27}&B&14.66&9.71&&\\
{\bf 28}&C&14.89&9.84&&\\
{\bf 29}&D&16.96&15.71&&\\
{\bf 30}&Total&13.26&8.27&&\\
\hline
\\
&&{\bf Final  Absolute  Magnitudes\tablenotemark{e}}&&&\\
&component&$M_V$&$M_K$&$V-K$&\\
{\bf 31}&A&10.35&5.34&5.01&\\
{\bf 32}&B&10.92&6.02&4.89&\\
{\bf 33}&C&11.14&6.16&4.98&\\
{\bf 34}&D&13.21&12.03&+1.18&\\
{\bf 35}&Total&9.52&4.58&4.94&\\
\hline
\\
&&{\bf Final Masses\tablenotemark{d}}&&& \\
&component&$V$-band fit&$K$-band fit&av&\\
{\bf 36}&\mA&0.46\msune&0.60\msune&0.53$\pm$0.10\msune&\\
{\bf 37}&\mB&0.40\msune&0.46\msune&0.43 0.04\msune&\\
{\bf 38}&\mC&0.38\msune&0.44\msune&0.41 0.04\msune&\\
{\bf 39}&\mD& -&0.53\msune&0.54 0.04\msune&\\
{\bf 40}&\m$_{\rm tot}$&1.78\msune&2.04\msune&1.91 0.13\msune&
\enddata
       \tablenotetext{a}{Our initial estimate is near . }
        \tablenotetext{b}{From absorption line RVs, Table~\ref{tbl-RVO}.}
    \tablenotetext{c}{Final  values varied to minimize {\bf line 17-24} residuals.}
  \tablenotetext{d}{After least squares "fit".}
    \tablenotetext{e}{Absorption-corrected.}
\end{deluxetable*}


\begin{center}
\begin{deluxetable}{l c c c}
\tablecaption{Restricted Expressions for Mass as a Function of Absolute Magnitude   \label{tbl-LMR}}
\tablewidth{0in}
\tablehead{\colhead{Param} 
& \colhead{$M_V$\tablenotemark{a}}
& \colhead{$M_K$\tablenotemark{a}}
& \colhead{WD $M_K$\tablenotemark{b}}
}
\startdata
C0&1.4594$\pm$0.0004&2.47$\pm$0.25&0.197$\pm$0.016\\
C1&-0.09693 0.00058&-0.4768 0.0781&0.198 0.075\\
C2&-&0.0237 0.0059& 0.562 0.096\\
C3&-&-&-0.206 0.033\\
X$_0$& - & - & 11.35\\
\enddata
\tablenotetext{a}{Parameters in the Equation~\ref{LMRf} polynomial.}
\tablenotetext{b}{Parameters in the Equation~\ref{LMRWD} polynomial.}
\end{deluxetable}
\end{center}

\begin{deluxetable*}{l l r r r r r r r r r r r}
\tablecaption{\Ha ~Equivalent Widths\tablenotemark{a} 
 \label{tbl-EWs}}
\tablewidth{7in}
\tablehead
{
\colhead{mJD\tablenotemark{b}} &
\colhead{$\Phi_{\rm BC}$}&
\colhead{$\Phi_{\rm AD-BC}$}&
\colhead{EWtot}&
\colhead{err} &
\colhead{EW1} &
\colhead{err}&
\colhead{EW2} &
\colhead{err}&
\colhead{EW3} &
\colhead{err}&
\colhead{EW4} &
\colhead{err}
}
\startdata
49962.4226&0.8930&0.42452&-6.65&0.20&-0.92&0.05&-2.21&0.16&-2.71&0.05&-0.80&0.09\\
50090.1532&0.3726&0.55390&-5.80&0.20&-0.96&0.06&-2.29&0.13&-2.03&0.09&-0.52&0.10\\
50090.208&0.4458&0.55395&-5.92&0.15&-0.98&0.05&-2.01&0.14&-2.16&0.04&-0.76&0.04\\
50090.2517&0.5041&0.55400&-6.43&0.13&-1.11&0.05&-2.22&0.11&-2.25&0.04&-0.86&0.04\\
50091.112&0.6523&0.55487&-6.36&0.14&-0.79&0.06&-2.62&0.06&-2.33&0.07&-0.62&0.09\\
50091.1417&0.6920&0.55490&-7.16&0.13&-1.22&0.07&-2.48&0.06&-2.47&0.07&-0.98&0.06\\
50091.1662&0.7247&0.55492&-6.18&0.10&&&-3.10&0.07&-3.09&0.07&&\\
.&.&.&.&.&.&.&.&.&.&.&.&.
\enddata
\tablenotetext{a}{Full table available electronically. Equivalent widths in  $\mathring{A}$}
\tablenotetext{b}{mJD=JD-2400000.5}
\end{deluxetable*}

\begin{deluxetable*}{l l r r r r r}
\tablecaption{Component B,C Radial Velocities from \Ha\tablenotemark{a} 
 \label{tbl-HaBCRVs}}
\tablewidth{4in}
\tablehead
{
\colhead{mJD}
&\colhead{$\Phi_{\rm BC}$}
&\colhead{$\Phi_{\rm AD-BC}$}
&\colhead{B }
&\colhead{err}
&\colhead{C }
&\colhead{err}
}
\startdata
49962.4226&0.8930&0.42452&101.48&3.58&-14.21&1.83\\
50090.1532&0.3726&0.55390&-10.09&1.70&90.97&5.46\\
50090.2080&0.4458&0.55395&-12.31&3.02&100.07&1.31\\
50090.2517&0.5041&0.55400&-15.35&2.32&103.08&1.15\\
50091.1120&0.6523&0.55487&-7.75&1.25&84.05&2.18\\
50091.1417&0.6920&0.55490&-53.97&1.07&61.73&1.57\\
50364.4174&0.4279&0.83169&-11.90&2.99&100.78&1.75\\
50366.3739&0.0392&0.83367&105.98&0.80&-15.77&2.25\\
.&.&.&.&.&.&.
\enddata
\tablenotetext{a}{Full table available electronically. All velocity units in \kms.}
\end{deluxetable*}
\begin{deluxetable*}{l l r r r r r}
\tablecaption{Component V2, V3 Radial Velocities from \Ha\tablenotemark{a} 
 \label{tbl-Ha23RVs}}
\tablewidth{4in}
\tablehead
{
\colhead{mJD}
&\colhead{$\Phi_{\rm BC}$}
&\colhead{$\Phi_{\rm AD-BC}$}
&\colhead{V2 }
&\colhead{err}
&\colhead{V3 }
&\colhead{err}
}
\startdata
49962.4226&0.8930&0.42452&20.25&0.82&57.08&0.84\\
50090.1532&0.3726&0.55390&22.13&0.91&55.5&1.05\\
50090.2080&0.4458&0.55395&22.36&0.64&59.62&0.51\\
50090.2517&0.5041&0.55400&21.79&0.68&60.24&0.63\\
50091.1120&0.6523&0.55487&25.25&0.62&56.16&0.97\\
50091.1662&0.7247&0.55492&24.49&0.38&56.19&0.39\\
50364.4174&0.4279&0.83169&22&0.95&59.51&0.8\\
50365.3997&0.7390&0.83269&24.26&0.34&55.43&0.38\\
.&.&.&.&.&.&.
\enddata
\tablenotetext{a}{Full table available electronically. All velocity units in \kms.}
\end{deluxetable*}


\begin{deluxetable*}{l l r r r r r r r r r r}
\tablecaption{\He ~Equivalent Widths\tablenotemark{a} 
 \label{tbl-HeEWs}}
\tablewidth{7in}
\tablehead
{
\colhead{mJD\tablenotemark{b}} &
\colhead{$\Phi_{\rm BC}$}&
\colhead{$\Phi_{\rm AD-BC}$}&
\colhead{EWtot}&
\colhead{err} &
\colhead{EW$_{\rm B}$} &
\colhead{err}&
\colhead{EW2} &
\colhead{err}&
\colhead{EW$_{\rm C}$} &
\colhead{err}
}
\startdata
49962.4226&0.8930&0.42452&-1.02&0.11&-0.29&0.06&-0.39&0.06&-0.34&0.06\\
50090.1532&0.3726&0.55390&-1.28&0.05&-0.61&0.03&-0.27&0.03&-0.12&0.03\\
50090.2080&0.4458&0.55395&-1.31&0.11&-0.34&0.07&-0.65&0.06&-0.24&0.05\\
50090.2517&0.5041&0.55400&-1.43&0.07&-0.20&0.04&-0.92&0.04&-0.31&0.04\\
50091.1120&0.6523&0.55487&-1.39&0.09&-0.23&0.06&-0.76&0.04&-0.40&0.05\\
50091.1662&0.7247&0.55492&-0.97&0.07&&&-0.97&0.07&&\\
50404.2830&0.6358&0.87207&-2.19&0.15&-0.61&0.09&-0.69&0.09&-0.88&0.08\\
50404.3080&0.6692&0.87210&-0.96&0.07&&&-0.96&0.07&&\\
50404.3360&0.7065&0.87213&-1.99&0.17&-0.65&0.10&-0.85&0.10&-0.49&0.10\\
.&.&.&.&.&.&.&.&.&.&.
\enddata
\tablenotetext{a}{Full table available electronically. Equivalent widths in  $\mathring{A}$}
\tablenotetext{b}{mJD=JD-2400000.0}
\end{deluxetable*}
\begin{deluxetable*}{c c c c c c c c c}
\tablewidth{7in}
\tablecaption{\Hes Emission Line RVs \label{tbl-HeIRV}\tablenotemark{a}}
\tablehead{
\colhead{mJD\tablenotemark{b}} &
\colhead{$\Phi_{\rm BC}$}&
\colhead{$\Phi_{\rm AD-BC}$}
&\colhead{VB }
&\colhead{err}
&\colhead{V$_{\rm cen}$ }
&\colhead{err}
&\colhead{VC }
&\colhead{err}
}
\startdata
49962.4226&0.8930&0.42452&-1.6&4.5&45.4&3.5&95.0&3.6\\
50090.1532&0.3726&0.55390&72.5&2.0&39.6&1.0&2.4&2.0\\
50090.2080&0.4458&0.55395&86.8&4.8&37.1&0.9&-2.3&1.6\\
50090.2517&0.5041&0.55400&83.2&2.8&42.8&0.6&-1.5&1.7\\
50091.1120&0.6523&0.55487&77.0&1.9&44.5&1.1&3.5&1.2\\
50091.1662&0.7247&0.55492&&&44.4&1.2&&\\
50404.2830&0.6358&0.87207&65.2&3.3&25.0&2.7&-34.2&1.4\\
50404.3080&0.6692&0.87210&&&36.2&1.3&&\\
50404.3360&0.7065&0.87213&54.8&1.0&39.4&0.8&20.1&1.2\\
.&.&.&.&.&.&.&.&.
\enddata
\tablenotetext{a}{Full table available electronically. All velocity units in \kms.}
\end{deluxetable*}


\clearpage


\begin{center}
\begin{deluxetable}{l c c c c}
\tablecaption{Swift-XRT Detections of \vA   \label{tbl-SX}}
\tablewidth{5in}
\tablehead{\colhead{mJD} 
& \colhead{$\Phi_{\rm BC}$}
&\colhead{$\Phi_{\rm AD-BC}$}
& \colhead{X-ray Flux\tablenotemark{a}}
& \colhead{S/N}
}
\startdata
53832.8833&0.790&0.9884&1.78E-12&6.1\\
54162.0951&0.183&0.3218&1.80E-12&11\\
54437.1163&0.248&0.6004&1.27E-12&8.8\\
\enddata
\tablenotetext{a}{0.3-10 keV flux in units mW m$^{-2}$}
\end{deluxetable}
\end{center}

\clearpage


\clearpage

\clearpage

\clearpage


\clearpage


\clearpage


\clearpage

\clearpage


\clearpage




\end{document}